\documentclass[twocolumn,preprintnumbers,amsmath,amssymb,secnumarabic]{revtex4}

\usepackage{graphicx}
\usepackage{dcolumn}
\usepackage{bm}
\usepackage[usenames]{color}
\usepackage{epsfig}
\usepackage{graphicx}
\usepackage{color}
\usepackage{footnote}

\newcommand{\kap}{\boldsymbol{\kappa}}
\newcommand{\rb}{{\bf r}}
\newcommand{\bp}{\boldsymbol{\partial}}
\newcommand{\qb}{{\bf q}}
\newcommand{\kb}{{\bf k}}
\newcommand{\pb}{{\bf p}}
\newcommand{\sig}{\boldsymbol{\sigma}}

\newcommand{\B}{\boldsymbol{B}}

\newcommand{\R}{\boldsymbol{R}}
\newcommand{\E}{\boldsymbol{E}}

\newcommand{\D}{\boldsymbol{D}}

\newcommand{\Finger}{\boldsymbol{B}}

\begin{document}

\title{Nonlinear Rheology of Colloidal Dispersions}

\author{J.M. Brader}

\affiliation{Department of Physics, University of Fribourg, CH-1700 Fribourg, Switzerland }
\begin{abstract}
Colloidal dispersions are commonly encountered in everyday life 
and represent an important class of complex fluid. 
Of particular significance for many commercial products and industrial processes is the 
ability to control and manipulate the macroscopic flow response of a dispersion
by tuning the microscopic interactions between the constituents. 
An important step towards attaining this goal is the development of robust theoretical methods 
for predicting from first-principles the rheology and nonequilibrium microstructure of well 
defined model systems subject to external flow. 
In this review we give an overview of some promising theoretical approaches and the phenomena 
they seek to describe, focusing, for simplicity, on systems for which the colloidal particles 
interact via strongly repulsive, spherically symmetric interactions. 
In presenting the various theories, we will consider first low volume fraction systems, for which a 
number of exact results may be derived, before moving on to consider the intermediate and high 
volume fraction states which present both the most interesting physics and the most demanding 
technical challenges. 
In the high volume fraction regime particular emphasis will be given to the rheology of 
dynamically arrested states.  
\end{abstract}

\maketitle
{\color{blue}
\tableofcontents
}

\section{Introduction and Overview}
Complex fluids exhibit a rich variety of flow behaviour which depends sensitively 
upon the thermodynamic control parameters, details of the 
microscopic interparticle interactions and both the rate and specific geometry of the flow under consideration. 
The highly nonlinear response characteristic of complex fluids may be readily observed 
in a number of familiar household products \cite{mezzenga}.
For example, mayonnaise consists of a stabilized emulsion of oil droplets suspended 
in water and behaves as a soft solid when stored on the shelf but flows like a liquid, 
and is thus easy to spread, when subjected to shear flow with a knife \cite{larson1}. 
This nonlinear viscoelastic flow behaviour, known as shear-thinning, may be manipulated 
on the microscopic level by careful control of the oil droplet size distribution. 
In contrast, a dispersion of corn-starch particles in water, at sufficiently high concentrations, 
exhibits a dramatic increase in shear viscosity with increasing shear rate; 
a phenomenon called shear-thickening \cite{cornstarch,fall}. 
Even the familiar practical problem of extracting tomato ketchup from a glass bottle 
presents a highly nonlinear flow. 
In this case the applied shear stress, generally implemented by shaking, must exceed a critical 
value, the yield-stress, before the ketchup begins to flow as desired.

Colloidal dispersions are a class of complex fluid which display all of the above mentioned 
nonlinear flow responses \cite{coussot}. 
In addition to being of exceptional relevance for many technological processes,  
the considerable research interest in colloidal dispersions owes much to the existence 
of well characterized experimental systems for which the interparticle interactions may be tuned 
to relatively high precision (often possible by simply varying the solvent conditions) \cite{russel}. 
The ability to control the microscopic details of the colloidal interaction   
facilitates comparison of experimental results with theoretical calculations and 
computer simulations based on idealised models (see e.g. \cite{vanmegen1,vanmegen2,vanmegen3}).  
In particular, the size of colloidal particles makes possible light scattering, neutron scattering and 
microscopy experiments which provide information inaccessible to experiments on atomic systems 
and which have enabled various aspects of liquid state theory to be tested in detail
\cite{capillary}.

\begin{figure} 
\includegraphics[width=8.5cm,angle=0]{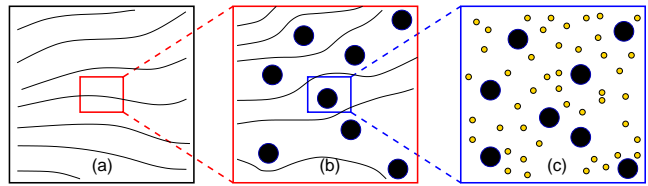}
\caption{
A schematic illustration of coarse graining as applied to colloidal dispersions. Continuum mechanics 
approaches treat the dispersion as a single continuum fluid (panel (a)), whereas a fully 
detailed picture is obtained by treating both colloids and solvent explicitly (panel (c)). 
The theoretical methods considered in this work operate at an intermediate level (panel (b))
in which the colloids are explicitly resolved but the solvent may be 
treated as a continuum.
}
\label{cg}
\end{figure}

The typical size of a colloidal particle lies in the range $10nm$ to $1\mu m$ and thus 
enables a fairly clear separation of length- and time-scales to be made between the 
colloids and the molecules of the solvent in which they are dispersed. 
As a result, a reasonable first approximation is to represent the solvent as a continuum fluid, 
generally taken to be Newtonian and thus characterised by a constant solvent viscosity 
(see Fig.\ref{cg}).  
For suspended particles with a length-scale greater than approximately $1\mu m$ the 
continuum approximation of the solvent is completely appropriate. 
However, this becomes questionable as the average size of the particles is reduced below 
a few nanometers, at which point the discrete nature of the solvent can no longer be ignored.
Colloidal particles occupy an intermediate range of length-scales for which a 
continuum approximation for the solvent must be supplemented by the addition of first 
order Gaussian fluctuations (Brownian motion) about the average hydrodynamic fields describing the 
viscous flow of the continuum solvent. 

The Brownian motion resulting from solvent fluctuations not only plays an important role in 
determining the microscopic dynamics; it is essential for the existence of a unique equilibrium 
microstructure.  
With the important exception of arrested glasses and 
gels, the presence of a stochastic element to the particle motion allows a full exploration of 
the available phase space and thus enables application of Boltzmann-Gibbs statistical mechanics 
to quiescent (and ergodic) colloidal dispersions.  
While the specific nature of the balance between Brownian motion, hydrodynamic 
and potential interactions depends upon both the observable under consideration and 
the range of system parameters under investigation, it is the simultaneous occurrence of these 
competing physical mechanisms which gives rise to the rich and varied rheological behaviour of 
dispersions. 
Unfortunately, the complicated microscopic dynamics presented by dispersions also  
serves to complicate the theoretical description of these systems \cite{dhont}.

The present review has been written with a number of aims in mind. On one hand, we would like 
to present a relatively concise overview of the main phenomenological features of the 
rheology of dispersions of spherical colloidal particles. In order to reduce the parameter 
space of the discussion, emphasis will be placed on the simple hard sphere model for which 
the space of control parameters is restricted to two dimensions (volume fraction and flow rate). 
While both attractive colloids and the response to non-shear flows will be addressed, no 
attempt has been made to be comprehensive in this respect.  
Another primary aim of the present work is to provide an overview, within the context of 
the aforementioned phenomenology, of microscopically motivated approaches to the rheology 
and flow induced microstructure of colloidal dispersions.  
Although we will discuss some less well founded `schematic model' approaches, the focus here 
is upon `first-principles' theories which prescribe a route 
to go from a well defined microscopic dynamics to closed expressions for macroscopically 
measurable quantities. 

The formulation of a robust theory of dispersion rheology from microscopic starting points 
constitutes a formidable problem in nonequilibrium statistical mechanics. 
Although considerable progress has been made in this direction, a comprehensive constitutive 
theory analogous to that of Doi and Edwards for entangled linear polymers 
\cite{doi,larson2,mcleish,cates_fielding} remains to be found. 
At present there exist a number of alternative microscopic theoretical approaches to dispersion
rheology which, despite showing admirable success within limited ranges of the system 
parameters, have so far been unable to provide a unified global picture of the microscopic 
mechanisms underlying the rheology of colloidal dispersions.
Despite common starting points (the many-body Smoluchowski equation) the disparate nature 
of the subsequent approximations, each tailored to capture a particular physical aspect of 
the cooperative particle motion, make it difficult to establish clear relations between 
different theoretical approaches. 
A goal of this work is thus to clarify the range of validity of the various theoretical 
approaches and to identify common ground. 
We note that the present work is well complemented by a number of recent reviews addressing 
dispersion rheology from both experimental \cite{mes6,vermant} and theoretical perspectives 
\cite{lionberger_rev,morris_rev,fuchs_review}.    

The paper is organized as follows: In section \ref{cont} we will discuss briefly some 
traditional continuum mechanics approaches to rheology, both to give a feeling for the spirit of
such work and to put into context some of the microscopic results presented later (in section 
\ref{section:mct}).
In section \ref{section:dynamics} we will introduce and discuss in some detail the Smoluchowski 
equation defining the microscopic dynamics under consideration. 
In section \ref{quiescent} we will consider the equilibrium and non-equilibrium phase behaviour 
of hard-sphere colloids in the absence of flow, which is a necessary pre-requisite to the subsequent
discussions. 
In section \ref{phenom} we will give a brief overview of the relevant basic phenomenology of dispersion 
rheology, including 
the shear-thinning and shear-thickening of colloidal fluids and the yielding of colloidal glasses. 
In section \ref{flowing} we will consider the various theoretical approaches to treating colloidal 
fluids under external flow. 
In particular, exact results for the microstructure and rheology of low volume fraction 
systems and their (approximate) extension to finite fluid volume fractions are discussed in 
subsections \ref{section:brady} and \ref{section:integralequations}, respectively. 
In section \ref{section:mct}, we consider the recently developed mode-coupling based 
approaches to the rheology of dense colloidal suspensions which enable glass rheology to be 
addressed. 
Finally, in section \ref{section:outlook} we will provide an outlook for future work and identify 
possible new avenues for theoretical investigation.

\section{Continuum Mechanics Approaches}\label{cont}
Rheology is primarily an experimental discipline. Indeed, one of the simplest experiments 
imaginable is to exert a force on a material in order to see how it deforms in response. 
More precisely, in a controlled rheological experiment one measures either the stress arising 
from a given strain or, more commonly, the strain 
accumulated following application of an applied stress. In practice, both stress controlled 
and strain controlled experiments are performed and provide complementary information regarding 
the response of a material sample. 
For the purpose of this review we will focus upon situations in which a {\em homogeneous} 
strain field is prescribed from the outset. 
The description of experiments for which macroscopic stress is employed as a control 
parameter poses an enormous challenge for microscopically based theories and demands careful 
consideration of the nontrivial mechanisms by which the applied stress propagates into the 
sample from the boundaries.  

Given the apparent complexity of any microscopic theory, it is quite natural to begin first at 
a more coarse-grained level of description in an effort to establish the general phenomenology 
and mathematical structure of the governing equations at the continuum level. 
Historically, this methodology was pioneered by Maxwell in his 1863 work on viscoelasticity 
and continued to develop into the following century through the efforts of distinguished rheologists 
such as Rivlin and Oldroyd \cite{hassager}. 
While much of this early work aimed to acheive a more fundamental mathematical understanding of viscoelastic 
response, strong additional motivation was provided by experiments on polymeric systems which exposed a
large variety of interesting nonlinear rheological phenomena in need of theoretical explanation.
Theoretical approaches to continuum rheology thus seek to obtain a 
constitutive equation relating the stress, a tensorial quantity describing 
the forces acting on the system \cite{batchelor}, to the deformation history encoded in the 
strain tensor. 

The typical `rational mechanics' approach to this problem is to assume a sufficiently general 
integral or differential constitutive relation between stress and strain and to then constrain 
this as much as possible via the imposition of certain exact or approximate macroscopic symmetry, 
conservation 
and invariance principles \cite{larson2,truesdell_noll,hassager}.  
The clear drawback to this methodology is that the entire particulate system is viewed as a 
single continuum field, thus losing any contact to the underlying colloidal interactions and 
microstructure ultimately responsible for the macroscopic response (see Fig.1). 
As a result, such constitutive theories are neither material-specific nor genuinely predictive in 
character.
Despite these shortcomings, the continuum mechanics approach to rheology has attained a great level of 
refinement and can be applied to fit experimental data from a wide range of physical 
systems \cite{larson2,hassager}. 
Moreover, the experience gained through continuum mechanics modelling may well prove useful in 
guiding the construction of more sophisticated microscopic theories by providing constraints on 
the admissable mathematical form of the constitutive equations.

\subsection{The Lodge equation}\label{sec:lodge}
It is perhaps instructive to give an illustration of the spirit in which phenomenological 
constitutive relations may be constructed using continuum mechanics concepts. 
The example we choose is not only of intrinsic interest, but will also prove relevant 
to the discussion of a recent microscopically based theory of glass rheology 
\cite{joeprl_07,joeprl_08,pnas} to be discussed in section \ref{section:mct}.  
We consider a viscoelastic fluid subject to shear deformation with 
flow in the $x$-direction and shear gradient in the $y$-direction (a convention we will continue
to employ throughout the present work). 
Suppose that we wish to determine the infinitessimal shear stress $d\sigma_{\rm xy}$ at time $t$ arising 
from a small strain increment $d\gamma$ at an earlier time $t'$. As the material is viscoelastic, 
it is reasonable to assume that the influence of the strain increment 
$d\gamma(t')=\dot\gamma(t')dt'$ on the stress at time $t$ must be weighted by a decaying
function of the intervening time $t\!-\!t'$, in order to represent the influence of dissipative 
processes. Adopting a simple exponential form for the relaxation function it is thus 
intuitive to write 
\begin{eqnarray}
d\sigma_{\rm xy}(t) = G_{\infty}\exp\left[-\frac{t-t'}{\tau}\right]\dot\gamma(t')\,dt',
\label{small}
\end{eqnarray}
where $\tau$ is a relaxation time and $G_{\infty}$ is an elastic constant 
(the infinite frequency shear modulus). 
Assuming linearity, the total stress at time $t$ may thus be constructed by summing up all 
of the infinitessimal contributions over the entire flow history, which we take to extend into the
infinite past. We thus arrive at
\begin{eqnarray}\label{pre_boltz}
\sigma_{\rm xy}(t) = \int_{-\infty}^{t}\!\!dt'\,
G_{\infty}\exp\left[-\frac{t-t'}{\tau}\right]\dot\gamma(t').
\end{eqnarray}
Partial integration leads finally to
\begin{eqnarray}
\sigma_{\rm xy}(t) = \frac{1}{\tau}\int_{-\infty}^{t}\!\!dt'\,
G(t-t')\gamma(t,t'),
\label{boltzmann}
\end{eqnarray}
where $G(t)=G_{\infty}\exp[-t/\tau\,]$ is the shear modulus and $\gamma(t,t')$ is the 
accumulated strain $\gamma(t,t')=\int_{t'}^{t}ds\,\dot\gamma(s)$.
The simple integral relation (\ref{boltzmann}) between shear stress and shear strain was 
first considered by Boltzmann. Indeed, the assumption that the stress increments (\ref{small}) 
may be summed linearly to obtain the total stress is often referred to as the `Boltzmann 
superposition principle'. 

In order to extend (\ref{boltzmann}) to a tensorial relation, i.e. a true constitutive 
equation, an appropriate tensorial generalization of the accumulated strain $\gamma(t,t')$ must 
be identified. 
For the spatially homogeneous deformations under consideration the translationally invariant 
deformation gradient tensor $\E(t,t')$ transforms a vector (`material line') at time $t'$ 
to a new vector at later time $t$ via $\rb(t)=\E(t,t')\cdot\rb(t')$, where 
$E_{\alpha\beta}=\partial r_{\alpha}/\partial r_{\beta}$. An alternative nonlinear choice of 
strain measure is the symmetric Finger tensor $\B(t,t')=\E(t,t')\E^{T}(t,t')$. 
The Finger tensor contains information about the stretching of material lines during a deformation 
but is invariant with respect to solid body rotations of the material sample. 
For simple shear the Finger tensor is given explicitly by
\begin{eqnarray}
     \B=\left(
     \begin{array}{ccc}
     1+\gamma^2  & \gamma & 0 \\
     \gamma & 1  & 0 \\
     0 & 0 & 1 
     \end{array}
     \right)
\end{eqnarray}
where $\gamma\equiv\gamma(t,t')$. 
The accumulated strain in the integrand of Eq.(\ref{boltzmann}) can thus be identified as the 
xy element of $\B(t,t')$. 
This suggests that the Boltmann integral form (\ref{boltzmann}) may be extended using the simple ansatz
\begin{eqnarray}
\sig(t) = \int_{-\infty}^{t}\!\!dt'\; \Finger(t,t') \;\frac{G_{\infty}e^{-(t-t')/\tau}}{\tau},
\label{lodge1}
\end{eqnarray}
for the full stress tensor (see subsection \ref{matob} below for more justification of this nontrivial step). 
Equation (\ref{lodge1}) is known as the Lodge equation in the 
rheological literature and is applicable in both the linear and nonlinear viscoelastic regime 
\cite{larson2}. 

\subsection{Upper convected Maxwell equation}
The assumption of an exponentially decaying shear modulus is generally attributed to Maxwell, 
who realized that this choice enabled an interpolation between a purely elastic response to 
deformations rapid on the time scale set by $\tau$ and a viscous, dissipative response in the limit of 
slowly varying strain fields.   
In fact, the Lodge equation derived above is simply the integral form of a nonlinear 
(differential) Maxwell equation. 
In order to show this we first differentiate (\ref{lodge1}) to obtain
\begin{eqnarray}
\frac{D\sig}{Dt} + \frac{1}{\tau}\sig = \frac{G_{\infty}}{\tau}\boldsymbol{ 1},
\label{upper_maxwell}
\end{eqnarray}
where we have introduced the upper-convected derivative \cite{hassager}
\begin{eqnarray}
\frac{D\sig}{Dt}=\dot{\sig}(t) - \kap(t)\,\sig(t) - \sig(t)\,\kap^{T}(t),
\label{upper}
\end{eqnarray}
and where the velocity gradient tensor $\kap(t)$ is defined in terms of the deformation 
gradient tensor via
\begin{eqnarray}
\frac{\partial}{\partial t}\E(t,t')=\kap(t)\E(t,t').
\label{velocity_gradient}
\end{eqnarray}
For an incompressible material the stress is only determined up to a constant 
isotropic term. Eq.(\ref{upper_maxwell}) may thus be expressed in an alternative 
form by first defining a new stress tensor  
\begin{eqnarray}
\boldsymbol{\Sigma}=\sig - G_{\infty}\boldsymbol{1},
\label{dev_def}
\end{eqnarray}
and substituting for $\sig$ in Eq.(\ref{upper_maxwell}). 
This yields
\begin{eqnarray}
\frac{D\boldsymbol{\Sigma}}{Dt} + \frac{1}{\tau}\boldsymbol{\Sigma} 
= G_{\infty}(\kap(t)+\kap^T(t)).
\label{ucm}
\end{eqnarray} 
This differential form of the Lodge equation is known as the 
upper-convected Maxwell equation \cite{larson2} and is a nonlinear generalization 
of Maxwell's original scalar model to the full deviatoric stress tensor. 
Historically, the upper-convected Maxwell equation was first proposed by Oldroyd \cite{hassager} 
directly on the basis of Maxwell's differential form. 

\subsection{Material objectivity}\label{matob}
The assumption that one can go from (\ref{boltzmann}) to (\ref{lodge1}) on the basis of a single 
off-diagonal element appears at first glance to be rather {\em ad hoc}. 
On one hand, this choice can be justified retrospectively, using the fact that the Lodge equation (\ref{lodge1}) is 
derivable from a number of simple molecular models, e.g. the dumbell model for dilute polymer 
solutions \cite{larson2}. 
However, from a continuum mechanics perspective (\ref{lodge1}) is the simplest generalization of 
(\ref{boltzmann}) which satisfies the `principle of material objectivity'. 
This principle expresses the requirement that the constitutive relationship between 
stress and strain tensors should be invariant with respect to rotation of either the material 
body or the observer, thus preventing an unphysical dependence of the stress on the state of 
rotation. 
That this symmetry is an approximation becomes clear when considering the material from a 
microscopic viewpoint: In a noninertial rotating frame the apparent forces clearly lead 
to particle trajectories which depend upon the angular velocity. For many systems the 
neglect of these effects on the macroscopic response of the system is an extremely good 
approximation. For the overdamped colloidal dynamics considered in this work inertia 
plays no role and the principle of material objectivity is exact \cite{degennes}. 

Mathematically, it is straightforward to check whether or not a proposed tensorial  
constitutive equation is material objective.  
When subject to a time-dependent rotation $\R(t)$ the deformation gradient tensor 
transforms as 
\begin{eqnarray}
\hat{\E}(t,t') = \R(t)\E(t,t')\R^{T}(t'),
\label{rotationE}
\end{eqnarray}
where $\hat{\E}$ is the deformation gradient in the rotating frame. 
The dependence of $\hat{\E}$ upon the state of rotation arises 
because $\E$ contains information about both the stretching and rotation of material lines. 
Insertion of the transformed tensor (\ref{rotationE}) into the constitutive equation for the 
stress thus corresponds to a rotation of the material sample. 
Material objectivity is verified if the resulting stress tensor is given by 
\begin{eqnarray}\label{requirement}
\hat{\sig}(t)=\R(t)\sig(t)\R^{T}(t). 
\end{eqnarray}
As noted, the Finger tensor $\Finger$ contains only information about the stretching of 
material lines and transforms under rotation according to
\begin{eqnarray}
\hat{\B}(t,t')=\R(t)\B(t,t')\R^{T}(t).
\label{finger_rotation}
\end{eqnarray}
The material objectivity of the Lodge equation (\ref{lodge1}), and thus the upper convected 
Maxwell equation (\ref{ucm}), follows trivially from the fact that $\sig$ is a linear functional of $\Finger$. 
Many phenomenological rheological models thus start by assuming a general functional dependence 
$\sig(t)=\mathcal{F}[\B]$ in order to guarantee a rotationally invariant theory. 

The vast majority of microscopically motivated theories of dispersion rheology treat 
only a single scalar element of the stress tensor (generally the shear stress $\sigma_{\rm xy}$).  
Indeed, the rarity of microscopic tensorial constitutive theories may well be the primary 
reason for the apparent gap between continuum and statistical mechanical theories aiming to describe
common phenomena. 
We will revisit the concept of material objectivity in section \ref{section:mct} when considering 
a recently proposed tensorial constitutive equation for dense dispersions.

\subsection{Beyond continuum mechanics}
In the last decade, significant progress has been made in understanding the response of 
colloidal dispersions to exernal flow on a level which goes beyond the 
fully coarse-grained phenomenological approaches of traditional continuum rheology.
Important steps towards a more refined picture have been provided by studies based on mesoscopic 
models \cite{mes6,mes1,mes2,mes4}. 
However, while such phenomenological approaches can reveal generic features of the 
rheological response, they 
are not material specific and can therefore address neither the influence of the microscopic 
interactions on the macroscopic rheology nor the underlying microstructure, as encoded in 
the particle correlation functions. 
This deeper level of insight is provided by fully microscopic approaches which 
start from a well defined particle dynamics and, via a sequence of either exact or clearly specified  
approximate steps, lead to closed expressions for macroscopically measureable quantities. 
The symmetry, invariance and conservation principles used as input in the construction of continuum 
theories, such as the material objectivity discussed in section \ref{matob}, should then emerge 
directly as a consequence of the microscopic interactions. 
Such an undertaking clearly requires the machinery of statistical mechanics. 

Theories founded in statistical mechanics provide information regarding the correlated motion of the 
constituent particles and 
are thus capable, at least in principle, of capturing non-trivial and potentially unexpected cooperative 
behaviour as exhibited by equilibrium and nonequilibrium phase transitions. 
This ability to capture emergent phenomena is in clear contrast to continuum approaches where 
such physical mechanisms must be input by hand. 
An additional advantage of a statistical mechanics based approach to rheology over 
the direct application of continuum mechanics is that important additional 
information is provided regarding the microstructure of the system, as encoded in the 
correlation functions. 
It thus becomes possible to connect the constitutive relations to the 
underlying correlations between the colloidal particles and obtain microscopic insight into 
the macroscopic rheological response.  
Additional motivation to theoretically `look inside' the flowing system is provided 
by developments in the direct visualization and tracking of particle motion in 
experiments on colloidal dispersions (confocal microscopy) \cite{crocker,prasad,besseling}, 
together with advances in the computer simulation of model systems under flow 
\cite{brady_sim,varnik1,doliwa}.  

Although beyond the scope of the present work, we note that the influence of steady shear flow on glassy states has been addressed, 
albeit in an abstract setting, by generalized mean-field theories of spin glasses \cite{sg1,sg2}. 
Spin glass approaches have proved useful in describing the dynamical behaviour of 
quiescent systems \cite{bouchaud}. 
In order to mimic the effect of shear flow a nonconservative force is introduced 
to bias the dynamics and break the condition of detailed balance characterizing the 
equilibrium state \cite{riskin}. 
While the abstract nature of these treatments certainly lends them a powerful generality, the 
lack of material specificity makes difficult a direct connection to experiment.  


\section{Microscopic Dynamics}\label{section:dynamics}
Before addressing the phenomenology (section \ref{phenom}) and approximate theories 
(sections \ref{flowing} and \ref{section:mct}) of colloid rheology it is rewarding to first consider in detail the 
microscopic equation of motion determining the overdamped colloidal dynamics. 
By a careful assessment of the fundamental equation of motion a number of general observations and
comments can be made regarding the character of nonequilibrium states, solutions in special limits, important
dimensionless parameters and influence of hydrodynamics, which are independent of the 
specific system or approximation scheme under consideration.  
  
We consider a system consisting of $N$ Brownian colloidal particles interacting via spherically 
symmetric pairwise additive interactions and  homogeneously dispersed in an 
incompressible Newtonian fluid of given viscosity. 
The probability distribution of the $N$-particle configuration is denoted by $\Psi(t)$ 
and satisfies the Smoluchowski equation \cite{dhont} 
\begin{eqnarray}
\hspace*{1.1cm}\frac{\partial \Psi(t)}{\partial t} + 
\sum_{i} \bp_i\cdot {\bf j}_i =0
\label{smol_hydro}
\end{eqnarray} 
where the probability flux of particle $i$ is given by
\begin{eqnarray}
{\bf j}_i={\bf v}_i(t)\Psi(t) 
- \sum_{j} \D_{ij}\cdot(\bp_j - \beta\,{\bf F}_j)\Psi(t),
\label{flux}
\end{eqnarray}
where $\beta=1/k_BT$ is the inverse temperature. 
The hydrodynamic velocity of particle $i$ due to the applied flow is denoted by 
${\bf v}_i(t)$ and 
the diffusion tensor $\D_{ij}$ describes (via the mobility tensor $\boldsymbol{\Gamma}_{ij}=\beta\D_{ij}$) 
the hydrodynamic mobility of particle $i$ resulting from a force 
on particle $j$.
The hydrodynamic velocity can be decomposed into affine and particle induced fluctuation terms 
${\bf v}_i(t)=\kap(t)\cdot{\bf r_i} + {\bf v}^{\rm fl}_i(t)$, where ${\bf v}^{\rm fl}_i(t)$ can be 
expressed in terms of the third rank hydrodynamic resistance tensor \cite{lubrication}.
The force ${\bf F}_j$ on particle $j$ is generated from the total potential energy 
according to ${\bf F}_j=-\bp_j U_N$, where, in the absence of external fields, 
$U_N$ depends solely on the relative particle positions.    
The three terms contributing to the flux thus represent the competing effects of (from left 
to right in (\ref{flux})) external flow, diffusion and interparticle interactions. 

While the Smoluchowski equation (\ref{smol_hydro}) is widely accepted as an appropriate starting 
point for the treatment of colloidal dynamics, alternative approaches based on the Fokker-Planck
equation have also been investigated \cite{jorquera}. 
On the Fokker-Planck level of description the distribution function retains a dependence on 
the particle momenta. Although this makes possible the treatment of systems with a temperature 
gradient (leading to thermophoretic effects), considerable complications arise when 
attempting to treat hydrodynamic interactions which make preferable the Smoluchowski equation.

For the special case of monodisperse hard-spheres at finite volume fractions under steady flow 
Eq.(\ref{smol_hydro}) can be numerically integrated over the entire fluid range using 
computationally intensive Stokesian dynamics simulation \cite{brady_bossis,brady_sim,banchio}.  
This simulation technique includes the full solvent hydrodynamics and provides a useful benchmark for 
theoretical approaches (for an overview of the computer simulation of viscous dispersions we 
refer the reader to \cite{brady_review} and references therein).  
While Stokesian dynamics simulations have focused primarily on simple shear, results have also 
been reported for extensional flow geometries \cite{stokesian_extensional}.

\subsection{Non-interacting particles}
\begin{figure}
\includegraphics[width=8.5cm]{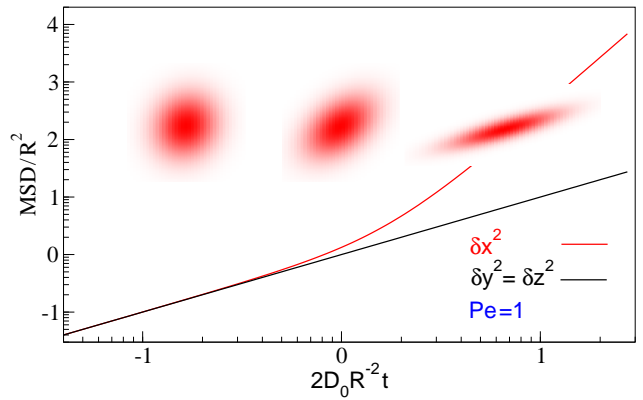}
\caption{
The mean-squared-displacement of non-interacting colloidal particles in flow (x), gradient (y) and 
vorticity (z) directions as a function of time. 
The MSD in flow direction exhibits enhanced diffusion (`Taylor dispersion') for values of the shear 
strain greater than unity. 
Also shown are contour plots of the (non-normalised) probability distribution $P_1(\rb,t)N(t)$ 
(see Eq.\ref{ideal}) in the $z=0$ plane at times $2D_0R^{-2}t=0.15, 1$ and $5$, demonstrating shear induced 
anisotropy for $\gamma>1$ related to the onset of Taylor dispersion. 
}
\label{idealgas}
\end{figure}

For the special case of non-interacting particles (${\bf F}_j=0$) 
equations (\ref{smol_hydro}) and (\ref{flux}) describe the configurational probability 
distribution of an ideal gas under externally applied flow and may be solved analytically 
using the method of characteristics \cite{mathews_walker}. 
For non-interacting particles under steady shear the many-particle distribution function is given by a product 
of single particle functions 
$\Psi(\{ {\bf r_i}\},t)=P_1({\bf r}_1,t)\times\cdots\times P_1({\bf r}_N,t)$, where $P_1$ is 
given by
\begin{eqnarray}\label{ideal}
P_1({\bf r},t) = \!\frac{1}{N(t)}\exp{\Bigg [} 
\frac{-x^2 \!- y^2(1+ \frac{\gamma^2 }{3}) + \gamma x y}{N(t)} \!-\! \frac{z^2}{4D_0t}
{\Bigg ]}\notag
\\ 
\end{eqnarray}
when the initial condition $P_1({\bf r},0)=\delta(0)$ is employed. The Normalization is given by 
$N(t)\!=\!(4\pi D_0t)^3(1+(\dot\gamma t)^2/12)$ and the strain by $\gamma\!=\!\dot\gamma t$. 
Given a suitably localized initial density distribution Eq.(\ref{ideal}) essentially 
describes the dispersion of a colloidal droplet in a solvent (e.g. ink in water) under shear, 
as is apparent from Fig.\ref{idealgas}, which shows contour plots of the probability distribution at 
three different times for a given shear rate. 

Although non-interacting colloids represent a trivial case, it is nevertheless 
instructive to consider the mean-squared-displacement (MSD), characterizing the diffusive 
particle motion, both parallel and orthogonal to the flow direction in simple shear \cite{foister}.  
In both the vorticity and shear gradient directions, flow has no influence and the equilibrium 
result is recovered, 
$\delta z^2\!=\!\delta y^2\!=\!2 D_0 \,t$, with $D_0$ the single particle diffusion coefficient. 
In the flow direction the MSD is enhanced by a coupling between Brownian motion and 
affine advection, yielding $\delta x^2\!=2 D_0 \,t(1+\dot\gamma^2t^2/3)$, where $\dot\gamma$ is the 
shear rate. 
The physical origin of this enhanced diffusion, termed `Taylor dispersion' \cite{taylor}, 
is that the random motion of a given colloid leads to its displacement into planes of laminar flow with 
a velocity different to that of the original point. This constant and random `changing of lanes'
leads, on the average, to a dramatically increased rate of diffusion in the direction of flow. 
The accelerated rate of mixing achieved by stirring a dilute dispersion is thus almost entirely 
attributable to local Taylor dispersion. 
We note also that analogous effects arising from flow-diffusion coupling can also be 
identified in other flow geometries, such as the practically relevant case of Poiseuille 
flow along a cylindrical tube \cite{foister}.

\subsection{Dimensionless parameters}\label{parameters}
The Smoluchowski equation describes the dynamics of spherical colloidal 
particles dispersed in an incompressible Newtonian fluid and provides the 
fundamental starting point for all theoretical work to be described in the following sections. 
An appropriate dimensionless Reynolds number governing the solvent 
flow may be defined as $Re=\rho\dot\gamma R^2/\eta$, with $\dot\gamma$ a 
characteristic flow rate, $\rho$ is the density, $\eta$ the solvent viscosity and $R$ the colloidal 
length scale. Due to the small size of the colloidal particles $Re$ remains 
small for all situations of physical relevance and the Stokes 
equations, rather than the more complicated Navier-Stokes equations, may thus be employed 
in treating the solvent flow.   

Given that $Re$ remains small, two dimensionless parameters are of particular importance 
in determining the equilibrium and nonequilibrium behaviour.  
The first of these is the colloidal volume fraction $\phi\!=\!4\pi n R^3/3$, with number density 
$n$ and particle radius $R$. 
The maximum volume fraction achievable for monodisperse spheres is $0.74$ corresponding to 
an optimally packed face-centered-cubic crystal structure. 
For the purposes of the present work we will find it convenient to divide the physical range of
volume fractions into three subregions: (i) low packing, $\phi\!<\!0.1$, 
(ii) intermediate packing, $0.1\!<\!\phi\!<\!0.494$, and 
(iii) high packing, $0.494\!<\!\phi$. 
While this division is somewhat arbitrary, it will later prove useful in discussing the 
various theoretical approximation schemes currently available. 

The second important dimensionless parameter is the Peclet number $Pe=\dot\gamma R^2/2D_0$ 
\cite{dhont}.   
The Peclet number is a measure of the importance of advection relative to Brownian motion 
and determines the extent to which the microstructure is distorted away from equilibrium 
by the flow field. In the limit $Pe\rightarrow 0$ Brownian motion dominates and the
thermodynamic equilibrium state is recovered. Conversely, in the strong flow limit, 
$Pe\!\rightarrow\!\infty$, solvent mediated hydrodynamic interactions may be expected to dominate 
the particle dynamics, although, in practice, surface roughness and other perturbing effects 
turn out to complicate this limit \cite{brady_morris} 
(see section \ref{section:brady} for more details on this point). 

Finally, we would like to note that there exists a further, nontrivial dimensionless 
quantity implicit in the many-body Smoluchowski equation (\ref{smol_hydro}). 
An increase in either the dispersion volume fraction or attractive coupling between 
particles is accompanied by an increase in the structural relaxation timescale of the 
system $\tau_{\alpha}$ characterizing the temporal decay of certain two-point 
autocorrelation functions. This enables the Weissenberg number to be defined as 
$Wi=\dot\gamma\tau_{\alpha}$. 
For intermediate and high volume fractions, particularly those close to the colloidal 
glass transition, it is the Weissenberg 
number, rather than the `bare' Peclet number $Pe$, which dominates certain aspects of the 
nonlinear rheological response, as has been emphasized in \cite{faraday}. 
For the low volume fraction systems to be considered in 
section \ref{section:brady} the structural relaxation timescale is set by $R^2/2D_0$, leading to 
$Pe=Wi$.

\subsection{Neglecting solvent hydrodynamics}\label{neglect}
In many approximate theories aiming to describe intermediate and high volume-fraction 
dispersions the influence of solvent hydrodynamics beyond trivial advection 
is neglected from the outset. 
For certain situations (e.g. glasses) this approximation is partially motivated by physical intuition, however,
in most cases, the omission of solvent hydrodynamics is an undesirable but unavoidable 
compromise made in order to achieve tractable closed expressions.   
Accordingly, the expression for the probability flux (\ref{flux}) is approximated in two places, 
which we will now discuss in turn. 
 
The first approximation is to set $\D_{ij}=D_0\delta_{ij}$, thus neglecting 
the influence of the configuration of the $N$ colloidal particles on the 
mobility of a given particle. 
For low and intermediate volume fraction fluids this may be reasonable for $Pe\ll 1$ but can be 
expected to break down for $Pe>1$ as hydrodynamics becomes increasingly important in determining 
the particle trajectories. In particular, the near field lubrication forces 
\cite{lubrication} which reduce the mobility when the surfaces of two particles approach contact play  
an important role in strong flow and are responsible for driving cluster formation and 
shear thickening \cite{wagner} (see section \ref{thickening}).
For dense colloidal suspensions close to a glass transition the role of hydrodynamics is 
less clear. 
For certain situations of interest (e.g. glasses close to yield) the relevant value of $Pe$ 
is very small and suggests that hydrodynamic couplings should not be of primary importance.

The second common approximation to (\ref{flux}) arising from the neglect of solvent hydrodynamics 
is the assumption of a translationally invariant linear flow profile 
${\bf v}(\rb,t)=\kap(t)\!\cdot\!\rb$, where $\kap(t)$ is the (traceless) time-dependent velocity gradient 
tensor introduced in Eq.(\ref{velocity_gradient}). 
In an exact calculation the solvent flow field follows from solution of Stokes equations 
with the surfaces of the $N$ colloidal particles in a given configuration providing the boundary 
conditions (essentially what is done in Stokesian dynamics simulation 
\cite{brady_bossis,brady_sim,banchio}). 
By replacing this solvent velocity field with the affine flow, we neglect the need for the 
solvent to flow around the particles and are thus able to fully specify the solvent 
flow profile from the outset, without requiring that this be determined as part of a 
self consistent calculation \cite{stokes_note}. 
If necessary, the assumption of purely affine flow could be corrected to first order. 
For example, under simple shear flow the solvent flow profile around a single spherical 
particle is well known \cite{batchelor} and could form the basis of a superposition-type 
approximation to the full fluctuating velocity field.  

It is important to note that the assumption of a translationally invariant 
velocity gradient $\kap(t)$ is potentially rather severe 
as it excludes from the outset the possibility of inhomogeneous flow, as observed in shear 
banded and shear localized states. 
While physically reasonable for low and intermediate density colloidal fluids, the assumption of 
homogeneity could become questionable when considering the flow response of 
dynamically arrested states, for which brittle fracture may preclude plastic flow 
\cite{banding_note}. 
Moreover, it is implicit in the approximation ${\bf v}(\rb,t)\!=\!\kap(t)\!\cdot\!\rb$ that the imposed 
flow profile acts instantaneously throughout the system. 
In experiments where strain or stress are applied at the sample boundaries 
a finite time is required for transverse momentum diffusion to establish the velocity field.  
Nevertheless, experiments and simulations of the transition from equilibrium to homogeneous steady state flow 
have shown that a linear velocity profile is established long before the steady state regime is
approached, thus suggesting that the assumption of an instantaneous translationally invariant 
flow is acceptable for certain colloidal systems \cite{zausch}.

\subsection{Nonequilibrium states}\label{formal} 
In equilibrium, the principle of detailed balance asserts that the microscopic probability 
flux vanishes, 
${\bf j}_i=\Psi\bp_i(\ln\Psi + \beta U_N)=0$, where $U_N$ is the total interparticle 
potential energy. This balance between conservative and Brownian forces thus yields the 
familiar Boltmann-Gibbs distribution $\Psi_e=\exp(-\beta U_N)/Z_N$, where $Z_N$ is the 
configurational part of the canonical partition function. 
In the presence of flow ($\kap(t)\ne0$) there exists a finite probability current which breaks 
the time reversal symmetry of the equilibrium state and detailed balance no 
longer applies.
A nonvanishing probability current thus serves to distinguish between equilibrium and nonequilibrium 
solutions of (\ref{smol_hydro}) and rules out the possibility of a Boltzmann-Gibbs form for the 
nonequilibrium distribution. 
While such a Boltzmann-Gibbs distribution is clearly inadequate for nonpotential flows 
(e.g. simple shear), for potential flows (e.g. planar elongation) it is perhaps tempting 
to assume such a distribution by employing an effective `flow potential' $U_f$ 
(see e.g. \cite{noneq_ddft}). 
The fundamental error of assuming an `effective equilibrium' description of nonequilibrium 
states is made very clear by the non-normalizability of the assumed distribution 
$\Psi\sim\exp(-\beta (U_N+U_f))$. 
These considerations serve to emphasize the fact that the only true way to determine the 
distribution function for systems under flow is to solve the Smoluchowski equation 
(\ref{smol_hydro}).

For much of the present work we will focus on the response of colloidal dispersions to 
steady flows. While experiment and simulation clearly demonstrate that well defined steady states may be 
achieved following a period of transient relaxation, it is interesting to note that there exists 
no mathematical proof of a Boltzmann H-theorem for Eq.(\ref{smol_hydro}) which would guarantee 
a unique long-time solution for the distribution function. 
The absence of a H-theorem for colloidal dispersions under steady flow is a consequence of the 
hard repulsive core of the particles which invalidates the standard methods of proof generally 
applied to Fokker-Planck-type equations \cite{riskin,fuchs_review}.  

A further nontrivial aspect of Eq.(\ref{smol_hydro}) emerges when considering the translational 
invariance properties of the time dependent distribution function 
$\Psi(t)\equiv \Psi(t,\{ {\bf r}_i \})$, achieved by shifting 
all particle coordinates by a constant vector ${\bf r}'_i={\bf r}_i+{\bf a}$ 
(see subsection $7.3$ for more details). 
For an arbitrary incompressible flow it has been proven that a translationally invariant 
initial distribution function leads to a translationally invariant, but anisotropic distribution 
function $\Psi(t)$, despite the fact that the Smoluchowski operator \cite{dhont} generating the 
dynamics is itself not translationally invariant \cite{joeprl_08} . 
Although the proof outlined in \cite{joeprl_08} omitted hydrodynamic interactions, it may be expected 
that the same result holds in the presence of hydrodynamics due to the dependence of the diffusion
tensors on relative particle coordinates.

\section{Quiescent States}\label{quiescent}
\subsection{Hard-spheres}\label{hs}
\begin{figure} 
\includegraphics[width=8.cm,angle=0]{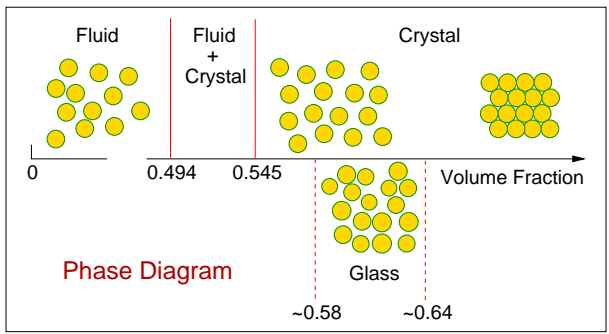}
\caption{
A schematic illustration of the phase diagram of hard-spheres as a function of volume fraction. 
Monodisperse systems undergo a freezing transition to an FCC crystal with coexisting densities 
$\phi=0.494$ and $0.545$. Polydispersity supresses the freezing transition resulting in a glass 
transition at $\phi\sim 0.58$, which lies below the random close packing value of 
$\phi\sim 0.64$.  
}
\label{phase}
\end{figure}

Theoretical and simulation studies based on Eq.(\ref{smol_hydro}) have focused largely 
on the hard-sphere model. 
In addition to being mathematically convenient, the focus on this simple model is motivated 
largely by the availability of well characterized hard-sphere-like experimental colloidal systems 
\cite{vanmegen1}.  
In the absence of flow, a system of monodisperse hard-sphere colloids remain in a disordered 
fluid phase up to a volume fraction of $\phi=0.494$, beyond which they undergo a 
first-order phase transition to a solid phase of $\phi=0.545$ with face-centered-cubic order 
(see Figure \ref{phase}). 
This unexpected, entropically driven, ordering transition was first observed using molecular 
dynamics computer simulation in the late 1950's \cite{alder} and remains a current topic of 
both experimental and theoretical research (for a recent review see \cite{gasser}). 

Making the system slightly polydisperse frustrates 
crystalline ordering and suppresses the freezing transition. 
In sufficiently polydisperse systems \cite{poly_note}
a disordered fluid remains the equilibrium state up to a volume fraction $\phi\thickapprox0.58$, 
at which point the dynamics becomes arrested and a colloidal glass state is formed. 
This dynamical transition to a non-ergodic solid is characterized by a non-decaying 
intermediate scattering function at long times for which dynamic light scattering 
results \cite{vanmegen1,vanmegen2} are well described by the 
mode coupling theory (MCT) \cite{sjoegren}. 
The standard quiescent MCT consists of a nonlinear integro-differential equation for the 
transient density correlator which exhibits a bifurcation, identified as a dynamic glass transition, 
for certain values of the system parameters \cite{sjoegren}. 
One of the appealing aspects of MCT is the absence of adjustable parameters: All information 
regarding both the particle interaction potential and thermodynamic state point enter via the static 
structure factor, which is assumed to be available from either independent measurements or 
equilibrium statisitical mechanical calculations. 
For monodisperse hard-spheres, MCT predicts a dynamic glass transition at $\phi\thickapprox0.516$ 
when the Percus-Yevick \cite{hansen} approximation is used to generate the structure factor, 
although other values may be obtained using either alternative theories, simulation or experiment
to determine the static equilibrium structure \cite{thomas_alternative}.   
We note that using MCT together with Percus-Yevick structure factors enables a glass transition to be
studied for monodisperse hard-spheres at volume fractions above freezing. 
Neither MCT nor PY theory is capable of incorporating crystalline ordering effects and both 
implicitly assume an amorphous microstructure. 

A shortcoming of the quiescent MCT is that it predicts an idealized 
glass transition with a divergent structural relaxation time and 
does not incorporate the activated processes which in experiment and simulation studies 
are found to truncate the divergence. 
While extensions of MCT aiming to incorporate additional relaxation channels have been proposed 
\cite{extended_mct1,extended_mct2}, the underlying microscopic mechanisms remain unclear. 
Despite its mean-field character, the MCT does capture some aspects of the heterogeneous 
dynamics \cite{het2,het3,trappe_timeresolved,trappe_qdependent} which have been observed using 
confocal microscopy \cite{weeks}. 

Finally, we note that a similar scenario of crystallization and dynamical arrest may be observed also 
in two-dimensional systems \cite{gasser,bayer}. 
Despite the reduced dimensionality and new physical mechanisms associated with melting in 
two-dimensions (where the hexatic phase plays an important role) the phase diagram for both 
monodisperse and polydisperse hard-disc systems is qualitatively identical to the 
three-dimensional case illustrated in Figure \ref{phase}. 
The close analogies between two- and three-dimensional systems may be exploited when considering 
nonequilibrium situations for which numerical calculations in $3D$ may prove prohibitively time
consuming \cite{weysser}.  
Viewing a binary mixture as the simplest form of polydispersity, MCT has been employed to study 
the influence of `mixing' 
(variations in composition and size ratio) on the glass transition of three dimensional hard-sphere 
\cite{voigtmann_goetze} and two-dimensional hard-disc \cite{hajnal_brader} systems. 
These studies have revealed intriguing connections between glassy arrest and random close packing.

\subsection{Attractive spheres}\label{attractive}
The addition of an attractive component to the hard-sphere potential can lead to an alternative 
form of dynamical arrest to either a gel at low volume fraction \cite{gel2,gel1} or an attractive 
glass state 
at higher volume fractions \cite{attglass1,attglass2} when the interparticle attraction becomes 
sufficiently strong. 
The origins of the attractive interaction are various e.g. van der Waals forces 
\cite{russel} or  
the depletion effect when nonadsorbing polymer is added to a dispersion 
\cite{dijkstra,brader1d,laurati_jcp}.  
This form of dynamical arrest has been investigated experimentally using both dynamic light 
scattering (see e.g.  \cite{gel1,attglass1,attglass2}) and confocal microscopy 
(see e.g. \cite{solomon}). 
There is now compelling evidence both from experiment \cite{manley} and simulation \cite{foffi} 
that for finite densities gellation occurs via a process of arrested phase separation and that 
only for very dilute, strongly attractive, suspensions does this mechanism cross over to one of 
diffusion limited aggregation.  

When applied to attractive colloidal systems the MCT predicts a nonequilibrium `phase diagram' 
which is in good agreement with the results of experiment and qualitatively describes 
the phase boundary seperating fluid from arrested states as a function of volume fraction 
and attraction strength \cite{attglass1,attglass2}.   
Recent studies of systems in which the depletion attraction between particles is complemented by the 
addition of a competing long range electrostatic repulsion \cite{barrier1} have revealed a rich and 
unexpected phase behaviour, including stable inhomogeneous phases \cite{barrier2} and metastable 
arrested states \cite{klix}. 
In addition, impressive new developments in colloid chemistry have enabled the construction of 
`colloidal molecules' in which the particle surface is decorated with a prescribed number of 
attractive sites, thus rendering the total interaction potential anisotropic \cite{patchy}. 
For a review of these more recent developments we refer the reader to \cite{zacc_rev}.

\section{Rheological Phenomenology}\label{phenom}
\begin{figure}\label{launpic}
\hspace*{-0.3cm}\includegraphics[width=7.5cm]{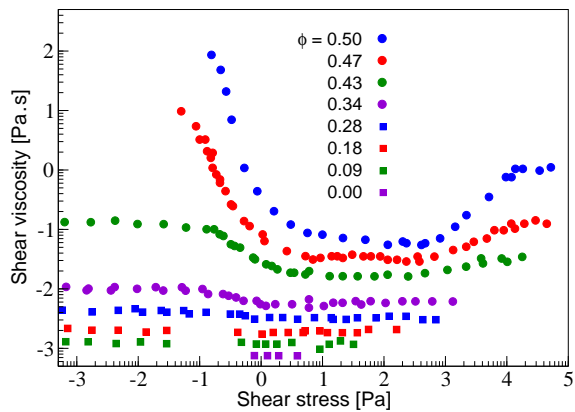}
\caption{
The shear viscosity of an aqueous dispersion of colloidal latex as a function of the externally applied 
shear stress. Data for a range of volume fractions are shown, from dilute up to a dense colloidal 
liquid at $\phi=0.50$. Shear thinning is evident at intermediate stress values as the viscosity 
of the dispersion decreases due to ordering of the particles by the flow. At larger applied 
stresses, for sufficently high volume fraction, the dispersion shear thickens as hydrodynamic 
lubrication forces lead to cluster formation and increased disorder. 
(Figure adapted from \cite{laun_pic})
}
\label{fig:thinning_thickening}
\end{figure}

As noted in the introduction, dispersions of spherically symmetric colloidal particles exhibit 
a diverse range of response to externally applied flow. Much, although not all \cite{att_yielding}, 
of the generic rheological behaviour of colloidal dispersions is captured by the hard-sphere 
model introduced in subsection \ref{hs}. 
In order to focus the discussion we will consider the special case of hard-spheres subject to a 
steady shear flow. 
In Fig.\ref{fig:thinning_thickening} we show the results of stress controlled experiments 
performed on a dispersion of spherical latex particles dispersed in water at various volume 
fractions, ranging from a dilute `colloidal gas' up to $\phi=0.5$,  corresponding to a dense colloidal 
liquid state close to the freezing transition \cite{laun_pic}. 
We note that for the experimental steady shear flow data shown in Fig.\ref{fig:thinning_thickening} 
it is not significant that the shear stress is employed as the control parameter dictating the 
flow. 
The quiescent system is ergodic at all considered state points and qualitatively identical 
results may thus be expected in an analogous strain controlled experiment, 
provided that the flow remains homogeneous. 

\subsection{Zero-shear viscosity}\label{zero_shear}
For each of the volume fractions shown in Fig.\ref{fig:thinning_thickening}  the shear viscosity 
$\eta\equiv\sigma_{\rm xy}/\dot\gamma$ is constant for small applied stresses 
(corresponding to small shear rates) and defines the zero-shear viscosity $\eta_0$. 
The data shown in Fig.\ref{fig:thinning_thickening} clearly demonstrates that the addition 
of colloidal particles leads to a dramatic increase of $\eta_0$ above that of the 
pure solvent (note the logarithmic scale in Fig.\ref{fig:thinning_thickening}). 

From a theoretical perspective, there are two alternative ways to understand the increase of 
$\eta_0$ as a function of $\phi$. The first is to relate the viscosity to the flow distorted pair
correlation functions in the limit of vanishing flow rate (see section \ref{pairsmol}). The 
leading order anisotropy of $g({\bf r},Pe\rightarrow0)$ captures the perturbing effect of weak flow 
on the microstructure and thus describes the increase of $\eta$ in terms of temporally local 
and physically intuitive correlation functions.  
The second method, referred to as either the `time correlation' or `Green-Kubo' approach, provides 
an equally rigorous method in which the viscosity is expressed as a time integral over a transverse 
stress autocorrelation function (see section \ref{section:mct}). 
Although the two approaches are formally equivalent, it is the latter which enables a direct connection to be
made between $\eta_0$ and the timescale describing the collective relaxation of the microstructure. 
  
Within the Green-Kubo formalism the thermodynamic colloidal contribution to the zero-shear viscosity is 
given by \cite{evans_morriss}
\begin{equation}
\eta_0\equiv\frac{\sigma_{\rm xy}}{\dot\gamma} = \int_0^{\infty}\!\!\!dt\, G_{\rm eq}(t),
\label{lr}
\end{equation}
where the equilibrium shear modulus is formally defined as a stress autocorrelation function 
\begin{equation}
G_{\rm eq}(t)=\frac{1}{k_BTV}\langle\, \hat\sigma_{xy}\, e^{\Omega_{\rm eq}^{\dagger}t}\, 
\hat\sigma_{xy}\,\rangle,
\label{exact_mod}
\end{equation}
where $V$ is the system volume and $\Omega_{\rm eq}^{\dagger}$ is the equilibrium adjoint Smoluchowski
operator \cite{dhont}. 
The fluctuating stress tensor element is given by 
$\hat\sigma_{\rm xy}\!\equiv\!-\sum_i\,  F_i^{\rm x}r_i^{\rm y}$, and the 
average is taken using the equilibrium Boltzmann-Gibbs distribution.  

\begin{figure}
\includegraphics[width=7.5cm,angle=0]{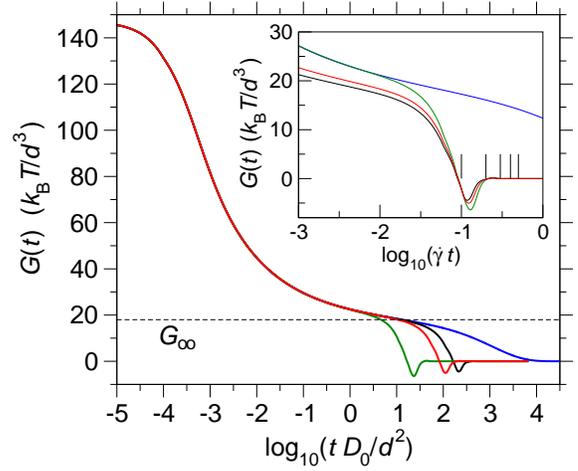}
\caption{
The generalized shear modulus $G(t)\equiv G(t,Pe)$ at a volume fraction 
relative to the glass transition $\phi-\phi_g=-1.16\times 10^{-3}$ calculated using the extended 
mode-coupling approach \cite{zausch}.  
The final relaxation of the equilibrium modulus (blue curve) serves to define the alpha relaxation
timescale. As the applied shear rate is increased the relaxation timescale is reduced. 
Curves are shown for $Pe=5.5\times 10^{-3}$ (green). $1.1\times 10^{-3}$ (red) and 
$5.5\times 10^{-4}$ (black). 
The plateau value which develops for volume fractions approaching the glass transition is indicated
by the broken line. The inset shows the same data as a function of strain $\dot\gamma t$. 
(Reproduced from \cite{zausch}).  
}
\label{fig:nonlinear_mods}
\end{figure}

Eq.(\ref{lr}) is an exact Green-Kubo relation which expresses a linear transport coefficient, 
in this case the shear viscosity, as an integral over a microscopic autocorrelation function. For dense colloidal dispersions the shear modulus starts from a well 
defined initial value \cite{div_note} from which it rapidly decays on a time scale set by 
$d^2/D_0$ to a plateau. 
For much later times the modulus decays further from the plateau to zero, thus 
identifying the timescale of structural relaxation $\tau_{\alpha}$ (see Fig.\ref{fig:nonlinear_mods}). 
The `two step' decay of the time dependent shear modulus is a generic feature of interacting 
systems exhibiting both a rapid microscopic dynamics and a slower, interaction induced, structural 
relaxation and is familiar from experiments and simulations of both colloidal and polymeric 
systems (where the Fourier transform $G^*(\omega)$ is typically considered, rather than $G(t)$ 
directly). 
 
Within the idealized mode-coupling theory (MCT) the equilibrium shear modulus (\ref{exact_mod}) is 
approximated by \cite{naegele}
\begin{eqnarray}
G_{\rm eq}(t) = \frac{k_BT}{60\pi^2d^3}\int_0^{\infty} \!\!\!dk\, k^4 \left(\frac{S'_k}{S_k}\right)^2 
\Phi_k^2(t), 
\label{lrmodulus}
\end{eqnarray}
where $T$ is the temperature, $S_k$ and $S'_k$ are the static structure 
factor and its derivative, respectively, and $\Phi_k(t)$ is the 
transient density correlator defined by 
\begin{equation}
\Phi_k(t)=\frac{1}{S_k}\langle \,\rho_k^*(t)\rho_k(0)\, \rangle,
\label{tc}
\end{equation}
where $\rho_k=\sum_j\exp(i{\bf k}\cdot{\bf r}_j)$. 
The collective coordinates $\rho_k$ are the central quantity within mode-coupling 
approaches and their autocorrelation (\ref{tc}) describes the temporal decay of density 
fluctuations which slow and ultimately arrest as the glass transition is approached. 
The mode-coupling approximation (\ref{lrmodulus}) arises from projection of the dynamics 
onto density-pair modes and thus expresses the relaxation of stress fluctuations 
in terms of density fluctuations. 
Within MCT the correlator is approximated by the solution of a nonlinear integro-differential 
equation
\begin{eqnarray}
\dot\Phi_q(t) + \Gamma_q\left[ \Phi_q(t) + \int _0^{\infty}dt'm_q(t-t')\dot\Phi_q(t')\right]=0,
\label{eom_quiescent}
\end{eqnarray}
where $\Gamma_q=q^2/S_q$ and the memory function is a quadratic functional of $\Phi_k(t)$ which 
depends upon both volume fraction and the static structure factor (which serves as proxy for the pair 
interactions). 
Explicit expressions for the quiescent memory function may be found in \cite{goetze_leshouches}. 
Equation (\ref{eom_quiescent}) predicts that $\tau_{\alpha}$ diverges at the glass transition 
volume fraction which then leads, via Eqs.(\ref{lr}) and (\ref{lrmodulus}), to a corresponding 
divergence of $\eta_0$. 
While in many cases the quiescent MCT give a good account of experimental data 
\cite{goetze_review} the precise nature and location of this apparent divergence remains a 
matter of debate (see e.g. \cite{russel_zero}). 


\subsection{Shear thinning}\label{thinning}
Turning again to Fig.\ref{fig:thinning_thickening} it is evident that for a given volume 
fraction the viscosity decreases as a function of shear rate. 
This shear thinning behaviour typically sets in when the shear rate begins to exceed the inverse of 
the timescale governing structural relaxation, that is for values of the Weissenberg number 
$Wi\equiv\dot\gamma\tau_{\alpha}\!>\!1$, where 
$\dot\gamma$ is the characteristic rate of strain and $\tau_{\alpha}$ is the structural relaxation time.  
Within the range $0<Wi<1$ the system is within the linear response regime and the flow 
rate is sufficently slow that the collective relaxation of the microstructure, characterized by, e.g. 
the decay of the transient density correlator (\ref{tc}), is not influenced.  

For $Wi>1$ the rate of structural relaxation is enhanced by the flow field.  
The modulus thus becomes a function of the shear 
rate and the viscosity shear thins. 
To incorporate this nonlinear response Eq.(\ref{lr}) may be generalized to 
\begin{equation}
\eta(Pe)\equiv\frac{\sigma_{\rm xy}(Pe)}{\dot\gamma} = \int_0^{\infty}\!\!\!dt\, G(t,Pe),
\label{nlr}
\end{equation}
where the functional dependence on $Pe$ has been made explicit. 
The nonlinear modulus is thus defined as 
\begin{equation}
G(t,Pe)=\frac{1}{k_BTV}\langle\, \hat\sigma_{xy} e^{\Omega^{\dagger}t}\,\hat\sigma_{xy}\,\rangle,
\label{exact_mod_nonlin}
\end{equation}
where $\Omega^{\dagger}$ is the adjoint Smoluchowski operator generating the particle dynamics 
\cite{dhont}. 
Despite the equilibrium averaging employed in (\ref{exact_mod}), it is important to note that 
an initial stress fluctuation $\sigma_{\rm xy}$ evolves to a fluctuation at later time 
$t$ under the full dynamics, including the effects of flow.
This serves to distinguish the {\em transient} stress correlator (\ref{exact_mod}) from that which 
would naturally be measured in a computer simulation, where all averaging is performed with respect 
to the full nonequilibrium distribution function. 
In the absence of flow $\Omega^{\dagger}=\Omega^{\dagger}_{\rm eq}$ and the equilibrium result 
(\ref{exact_mod}) is recovered.

Recent generalizations of the mode-coupling theory \cite{fc_jrheol} provide 
approximate expressions for the nonlinear modulus $G(t,Pe)$, see section \ref{section:mct}. 
These approaches incorporate the effects of shear flow into the memory kernel responsible for  
slow structural relaxation and describe the speeding up of the relaxational dynamics. 
In Fig.\ref{fig:nonlinear_mods} we show $G(t,Pe)$ calculated using the generalized MCT 
\cite{zausch} for a volume fractions close to (but below) 
the glass transition at various values of the shear rate. 
For $Wi<1$ the equilibrium result (blue curve) is not influenced by the flow. However, for 
$Wi>1$ the longest relaxation time becomes dictated by the flow and
$\tau_{\alpha}\sim\dot\gamma^{-1}$, as is demonstrated by the inset to Fig.\ref{fig:nonlinear_mods} which
shows the same data as a function of strain.
From Eq.(\ref{nlr}) it is clear that within this generalized Green-Kubo approach the decrease 
of $\tau_{\alpha}$ with increasing $\dot\gamma$ results in shear thinning of the viscosity.

An alternative, although equally valid, viewpoint is provided by approaches focusing on 
the flow distorted pair correlations.  
Exact results for low volume fraction dispersions based on the pair Smoluchowski equation 
(see section \ref{pairsmol}) have shown that shear 
thinning results from a decrease in the Brownian contribution to the shear stress \cite{brady_morris}.    
In dispersions at higher volume fraction the reduction in the Brownian stress is manifest in an 
ordering of the particles in the direction of flow which serves to reduce the frequency of particle 
collisions. 
Within the shear thinning regime evidence for layered or string-like ordering in intermediate volume 
fraction systems has been provided by Brownian dynamics simulations \cite{phung,rastogi} and, 
albeit with different characteristics, by Stokesian dynamics simulations which include hydrodynamic 
interactions \cite{brady_bossis0,brady_bossis,mb_a,mb_b}. 
Interestingly, simulations have also shown that the flow induced order continues to develop 
following its initial onset. 
This `ripening' of the ordered phase leads to a time-dependence of the viscosity known 
as thixotropy \cite{mb_a,mb_b,kulk,mewis_wagner} and complicates the determination of flow curves
in both simulation and experiment. 
For each shear rate the measurement time must be sufficiently long that the viscosity saturates to
a plateau value before the shear rate is updated. 

We note that the same ordering mechanism discussed here for colloidal dispersions would also lead 
to an analogous shear thinning scenario for atomic liquids 
(e.g. liquid argon). In this case, however, the shear rates required to observe such non-Newtonian 
rheology are several orders of magnitude larger than those readily accessible in experiment. 
For this reason, non-Newtonian effects in atomic systems remain largely a matter of academic interest.

\subsection{Shear thickening}\label{thickening}
Following the regime of shear thinning, a second Newtonian plateau develops for which 
the viscosity attains an approximately constant value as a function of shear rate and the 
flow induced ordering of the system continues to develop.  
At higher shear rates the viscosity undergoes a rapid increase once a critical value of the 
shear stress is exceeded. 
Such shear thickening behaviour can be either continuous \cite{wagner} or discontinuous 
\cite{dhaene} in character and is somewhat counterintuitive in light of the discussion 
presented in \ref{thinning} regarding flow induced microstructural ordering and its connection 
to shear thinning. 
Suspensions of nonaggregating particles at intermediate volume fractions generally show 
reversible shear thickening, however 
the details of the increase in viscosity depend upon the details of the system 
(particle type, solvent etc.) as well as the thermodynamic control parameters 
\cite{barnes}. 

In section \ref{thinning} we noted that the onset of shear thinning occurs for values of the 
Weissenberg number $Wi>1$, reflecting the essential competition between flow and 
structural relaxation. In contrast, the onset of shear-thickening behaviour is determined 
by the value of the bare Peclet number $Pe$, thus serving to highlight the different 
mechanisms dominating the physics of thinning and thickening states. In experiment, the 
difference in scaling of $Wi$ and $Pe$ with particle diameter $d$ enables the extent 
of the second Newtonian plateau separating shear thinning and shear thickening regimes to be 
controlled as a function of particle size. When the microstructure under flow is also of 
interest, the contraints of instrumental resolution (in e.g. confocal microscopy) place 
additional limits on the particle size which have also to be taken into consideration.   

The earliest theoretical explanations of shear thickening in colloidal dispersions proposed that the observed 
viscosity increase is the consequence of an order-to-disorder transition \cite{hoffman72,boersma}. 
Within this picture, the ordered planes of particles which form within the shear thinning 
regime, and which persist throughout the second Newtonian viscosity plateau, begin to 
interact via hydrodynamic coupling at sufficiently high shear rates. 
This interaction pulls particles out of the layers, leading to increased particle collisions, 
disorder, and a consequent increase in viscosity. The onset of shear thickening is thus 
identified with the hydrodynamic instability of a layered microstructure (see \cite{laun_92} 
for a discussion of this issue). 

Despite the intuitive appeal of interpreting shear thickening as an order-to-disorder 
transition, questions were raised by the experiments of \cite{laun_91,laun_92} in which a 
specific system (electrostatically stabilized Latex particles in glycols) was found to 
display shear thickening in the absence of an ordered phase. 
These results suggested that while an ordered phase may well precede the shear thickening 
regime as the shear rate is increased, it is not a necessary prerequisite. According to these 
findings, shear thickening occurs via an independent physical mechanism and is not simply 
related to a loss of microstuctural order. 
Further insight into the microscopic mechanism underlying shear thickening was provided by 
Stokesian dynamics simulations \cite{phung,brady_bossis1} which identified the formation of 
hydrodynamically bound particle clusters at high shear rates. 
Such `hydroclusters' form when the shear flow is sufficiently strong that the particle surfaces 
are driven closely together. At such small separations the hydrodynamic lubrication forces 
dramatically reduce the relative mobility of the particles such that they remain trapped together 
in a bound orbit 
(a point which we will later revisit in section \ref{section:brady}). 
Transient shear-driven hydroclusters would appear to be the defining feature of shear thickened 
states and experimental evidence for their importance is accumulating \cite{kalman}. 
Nevertheless, opinion remains divided regarding the fundamental mechanisms at work 
\cite{hoffman}. 

\begin{figure}
\hspace*{-0.15cm}\includegraphics[width=8.5cm]{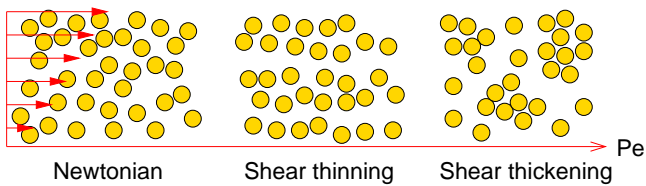}
\caption{
Schematic illustration of the microstructural order and disorder induced in a 
dense colloidal dispersion by shear flow. 
At low values of $Pe$ (leftmost configuration) the viscosity remains constant as diffusion is able to restore 
the equilibrium microstructure more rapidly than the shear flow can disrupt it. 
At intermediate shear rates (central configuration) the rate of shear exceeds the rate of 
structural relaxation, $Wi\!>\!1$ leading to microstructural ordering and shear 
thinning. 
At high shear rates (right configuration) hydrodynamic lubrication forces lead to particle 
clustering which strongly enhances the hydrodynamic contribution to the viscosity and result 
in shear thickening. 
}
\label{thick_fig}
\end{figure}

The hydrodynamic mechanisms described above give rise to a continuous, albeit rapid, rise 
in the viscosity as a function of shear rate. An alternative scenario may arise when, at some 
critical value of the shear rate, the viscosity exhibits a discontinuous jump as the system 
becomes jammed \cite{hoffman72,laun,frith,wagner,egres}. 
While it is anticipated that hydrodynamics will be relevant for the description of shear thickening  
at intermediate volume fractions (e.g. $0<\phi<0.5$, as considered in Fig.\ref{fig:thinning_thickening}) 
alternative mechanisms may become important upon approaching the glass transition. 
In \cite{holmes,holmes1,holmes2} a `schematic' mode coupling theory similar to those to be discussed in 
section \ref{section:mct} was developed, in which a coupling to stress was introduced into the 
nonlinear equations determining the decay of the transient density correlator. 
Upon varying the model parameters a range of rheological behaviour was revealed, including 
both continuous and discontinuous shear thickening, as well as a jamming transition to a 
non-ergodic solid state.  
Within this picture, shear thickening and jamming are viewed as a type of stress induced 
glass transition, for which the applied stress inhibits particle motion, even in the absence 
of hydrodynamics. 
A number of works have suggested a relationship between shear thickening and jamming 
\cite{cornstarch,mb_b,delhommelle,fall} although details of the connection between  
hydrodynamic cluster formation and jamming transitions of the kind more familiar 
from studies of granular media \cite{mehta} remains unclear. 

The addition of an attractive component to the strongly repulsive colloidal core can lead to 
gel formation and irreversible flocculation (see section \ref{attractive}). 
For such systems shear thickening is generally not observed, as the increase in the hydrodynamic 
contribution to the viscosity with increasing shear rate is more than compensated for by the 
decrease in the thermodynamic contribution arising from the attraction (see e.g. \cite{zukoski}). 
The generic behaviour of gel and floc states is thus monotonic shear thinning as a function of shear 
rate \cite{barnes}. 
It is therefore surprising that recent experiments using attractive carbon black particles 
\cite{negi} have identified a rich shear thickening behaviour for which the viscosity increases 
beyond a critical value of the applied shear stress. 
In this case an additional physical mechanism has been proposed by which the forces exerted 
by shear flow cause flocs to break apart, leading to an increased surface area and thus greater 
hydrodynamic dissipation \cite{negi}. 

As pointed out in section \ref{thinning} shear-thinning is not unique to colloidal system 
and can also be observed, albeit at high shear rates, in simple atomic liquids. 
In contrast, shear thickening of the type discussed above is not found in atomic systems 
(for which the `solvent' is a vacuum) and demonstrates clearly the breakdown of the 
correspondence between colloidal and simple liquids for strongly nonequilibrium states. 
Indeed it is quite clear that the view of colloids as `big atoms' \cite{bigatoms} 
will only hold in situations 
for which the influence of the solvent is negligable and that new physics may emerge when 
the role of hydrodynamic interactions becomes significant. 
We note that an alternative type of shear thickening has been observed at 
high shear rates in molecular dynamics simulations of simple liquids 
\cite{evans_morriss_thick,evansextra,matin,daivis}.
In these simulations a profile unbiased thermostat was employed to remove artifacts which may arise 
when a linear flow profile is assumed. 
Shear thickening was observed in simulations performed at constant volume, but not in those performed
at constant pressure.

Finally, we would like to note that the onset of shear thickening at high flow rates has been  
associated with unexpected behaviour of the first normal stress difference 
$N_1=\sigma_{xx}\!-\sigma_{yy}$. 
Typically, dispersions at low or moderate shear rate exhibit a positive value of $N_1$, 
indicating that a Weissenberg (or `rod climbing') effect would be observed in shear 
experiments performed in a Couette geometry \cite{larson1}.   
Experiments on dense colloidal dispersions with repulsive interactions 
\cite{laun,lootens,hebraud} have revealed 
that $N_1$ can change sign from positive to negative upon increasing the flow rate into the regime 
where the viscosity shear thickens. 
Similar behaviour has been observed in Stokesian dynamics simulations \cite{foss} and in 
numerical solutions of the Smoluchowski equation for dilute systems \cite{bergenholtz_brady}. 
In contrast to these results for purely repulsive interactions, recent experiments on 
attractive flocculated colloidal dispersions display a monotonically increasing $N_1$  
throughout the shear thickening regime \cite{negi}. 

\begin{figure}
\hspace*{0.0cm}\includegraphics[width=7.2cm]{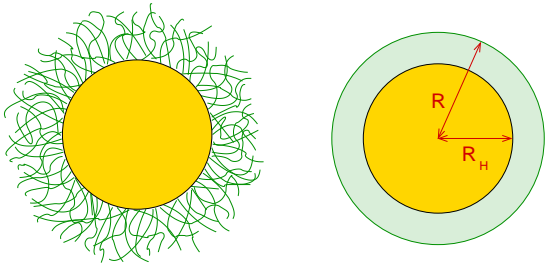}
\caption{
Many commonly studied colloidal particles (e.g. PNIPAM) consist of a polymeric 
core grafted with a layer of polymer (schematically represented on the left) which serves to 
stabilize against flocculation. Given a sufficiently dense and
crosslinked polymer brush the particles exhibit a strongly repulsive effective potential 
interaction approximating that of hard-spheres with radius $R$. 
The ability of the solvent to penetrate into the brush results in a hydrodynamic radius 
$R_H\!<\!R$. These effects may be mimiced by a simple hard-sphere model (sketch on the right) 
in which $R/R_H$ can be used to control the influence of hydrodynamic interactions.    
}
\label{coreshell}
\end{figure}

\subsection{Yield stress}\label{yieldstress}
For colloidal fluid states ($0<\phi<0.494$) the data presented in 
Fig.\ref{fig:thinning_thickening} represent the generic phenomenology of dispersions of strongly 
repulsive colloids under shear flow. In fact, this behaviour is not limited to simple shear.  
Qualitatively identical behaviour is found 
in Stokesian dynamics simulations for the extensional viscosity \cite{larson2} of dispersions 
of hard-sphere colloids under steady extensional flow, with the shear rate replaced by the rate 
of Hencky strain \cite{stokesian_extensional}. 

\begin{figure}
\hspace*{-0.3cm}\includegraphics[width=7.2cm]{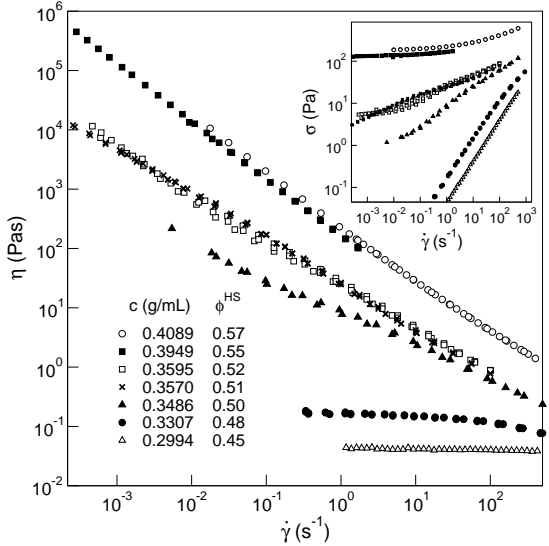}
\caption{Shear thinning and the dynamic yield stress of a concentrated aqueous dispersion of
poly(ethylene glycol)-grafted polystyrene colloidal particles which, to a good approximation, behave 
as hard spheres.  
The main figure and inset show the viscosity and shear stress, respectively, as a function 
of shear rate.
(Reprinted with permission from \cite{zackrisson}. Copyright: American Physical Society)
}
\label{fig:yielding}
\end{figure}

As already noted in section \ref{zero_shear}, increasing the volume fraction of a
colloidal liquid leads to a strong increase in the zero-shear viscosity. 
Assuming that crystallization has been suppressed by polydispersity, the volume fraction 
can then be further increased, eventually resulting in an apparent divergence of the zero-shear 
viscosity, 
either at the glass transition volume fraction (according to mode-coupling theory 
\cite{sjoegren,naegele}), or some higher volume fraction approaching random close packing 
\cite{russel_zero}. 
The variation of the viscosity as a function of shear rate for volume fractions ranging 
from $0.45$ to $0.57$ is demonstrated in more detail by the data shown in 
Fig.\ref{fig:yielding}.  
These experiments were performed on a system of poly(ethylene glycol)-grafted polystyrene 
colloidal particles dispersed in water \cite{zackrisson}. 
For the two lowest volume fractions considered ($\phi\!=\!0.45$ and $0.48$) a clear zero-shear 
viscosity may be be identified from the low shear rate plateau, with shear thinning evident 
at higher shear rates for $\phi\!=\!0.48$. 
As the volume fraction is increased above $0.48$ the low shear rate plateau moves to smaller rates, out 
of the experimental window of resolution, and the dispersion shows shear thinning over the 
entire range.  
Analysis of the intensity correlation function (related to the transient density correlator 
(\ref{tc})) measured using dynamic light scattering leads to an estimate of the glass transition 
for this system of $0.53<\phi_g<0.55$, somewhat lower than the typical value 
$\phi_g\sim 0.58$ obtained for PMMA hard-sphere-like colloids. 
The viscosity data for the two highest volume fractions ($\phi=0.55$ and $0.57$) are 
consistent with a divergence in the zero-shear viscosity at the glass transition, as predicted 
by the MCT (see section \ref{zero_shear}). 

According to the extended MCT \cite{fuchs_cates,faraday,fc_jrheol}, for glassy 
states the slowest relaxation time is $\tau_{\alpha}\sim\dot\gamma^{-1}$, which leads, via 
Eq.(\ref{nlr}), to $\eta\sim G(t\!\rightarrow\!\infty)\dot\gamma^{-1}$, where 
$G(t\!\rightarrow\!\infty)$ is the plateau 
modulus (see Fig.\ref{fig:nonlinear_mods}), thus reproducing the power law decay of the viscosity 
demonstrated by the data in Fig.\ref{fig:yielding}. 
For the idealized glassy states considered by MCT, where $\tau_{\alpha}$ 
is infinite in the absence of flow, this power law dependence extends to the limit 
$\dot\gamma\!\rightarrow\! 0$, resulting in a true divergence. 
In real colloidal experiments, higher order relaxation processes will always endow the 
quiescent system with a finite value of $\tau_{\alpha}$ and the viscosity divergence will be
truncated. 
The low shear divergence of the viscosity and power law shear thinning $\eta\sim\dot\gamma^{-1}$ 
suggested by Fig.\ref{fig:yielding} are supported by independent experiments performed on 
thermosensitive core-shell particles \cite{crassous,winter}. 
However, some recent experiments on sterically stabilized PMMA particles provide contradictory evidence 
and have suggested a nontrivial dependence of the relaxation time on shear rate, namely 
$\tau_{\alpha}\sim\dot\gamma^{-0.8}$, which remains to be understood \cite{expts1}. 

The inset to Fig.\ref{fig:yielding} shows the shear stress as a function of shear rate 
and provides an alternative representation of the viscosity data shown in the main panel. 
For the two highest volume fraction samples ($\phi=0.55$ and $0.57$) the shear stress becomes 
constant for the lowest shear rates considered, thus identifying a {\em dynamic} yield stress 
for glassy states. 
Both the shear thinning as a function of $Pe$ and appearance of a dynamic yield stress as a function 
of $\phi$ evidenced by Fig.\ref{fig:yielding} are well described by the extended MCT
\cite{crassous,winter}. 

The relationship between the dynamic yield stress and the more familiar static yield stress 
mirrors that between stick and slip friction in engineering applications 
($\sigma_y^{\rm stat}>\sigma_y^{\rm dyn}$ is thus to be expected). 
Indeed, it may be argued that the dynamic yield stress is, in fact, a more well defined quantity than 
the static yield stress. 
The latter is typically defined as the step stress amplitude which must be exceeded such that the system 
will flow at long times \cite{creepnote}. 
The point of static yield may therefore be dependent upon details of the system preparation, with the consequence that 
nonstationary properties, such as sample age in colloidal glasses, could influence the outcome of a given 
experiment \cite{creepnote1}. 
Moreover, the existence of creep motion, for which the strain increases sublinearly with time, 
makes difficult an unambiguous identification of the static yield stress. 
In contrast, the dynamic yield stress is defined as the limiting stress within a sequence of ergodic, 
fluidized steady states and is thus independent of prior sample history. 

It is apparent from Eq.(\ref{nlr}) that a dynamic yield stress can only exist in the event that 
$\tau_{\alpha}\sim\dot\gamma^{-\nu}$, with $\nu=1$. 
Values of  $\nu$ less than unity result in a shear stress 
$\sigma_{\rm xy}(\dot\gamma\rightarrow 0)=0$, despite the fact that the viscosity diverges. 
We thus note that a low shear rate divergence of the viscosity is a necessary but not sufficient
condition for the existence of a yield stress. 
While the results presented in \cite{expts1} apparently cast doubts on the existence of 
a dynamic yield stress for certain colloidal glasses, complications due to inhomogeneous, shear localized 
flow make this a subject of ongoing debate \cite{ovarlez_banding}. 

The interplay between static and dynamic yield has been investigated in simulation studies of a glass
forming binary Lennard-Jones mixture (the Kob-Anderson model) using molecular dynamics 
simulations \cite{varnik,varnik_bocquet}. 
In these simulations the mixture was confined between two atomistic walls, one of which was then
subjected to either a constant stress or constant strain in order to induce shear flow. 
It is important to note that due to the application of shear through the boundaries, the flow 
profile within the confined fluid/glass is an output of the numerical calculation and is not constrained
to be linear.    

\begin{figure}
\hspace*{-0.3cm}\includegraphics[width=7.5cm]{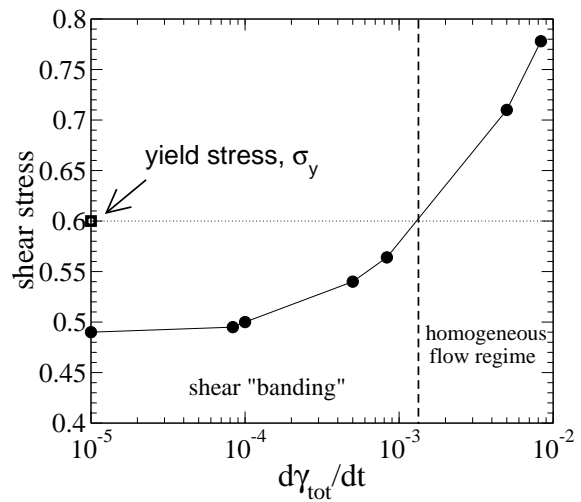}
\caption{Molecular dynamics simulation results for a glass forming binary Lennard-Jones mixture. 
The full line shows the shear stress as a function of the total shear rate 
under conditions for which a constant rate of strain is applied to one of the bounding walls. 
The square (labelled $\sigma_y$) indicates the static yield stress obtained by sequentially 
increasing the applied stress until the system begins to exhibit viscous flow. For values of the 
stress between dynamic and static yield points, the system was found to exhibit inhomogeneous flow. 
(Reprinted with permission from \cite{varnik}. Copyright: American Institute of Physics.)
}
\label{bandingfigure}
\end{figure}

In Fig.\ref{bandingfigure} we show the simulated flow curve for this system (analogous to that shown in the inset to
Fig.\ref{fig:yielding}) for a glassy statepoint, calculated by applying a fixed rate of strain to 
one of the bounding walls \cite{varnik,varnik_bocquet}.
When the shear stress is plotted as a function of the total strain rate 
(which may differ from the local rate of strain) a dynamic yield stress can clearly be identified. 
It was observed in the simulations that at sufficiently low (total) shear rates, an inhomogeneous 
flow profile develops in which a static layer coexists with a fluidized region exhibiting a linear
flow profile. 
In a complementary set of simulations a lower bound for the static yield stress was identified 
by slowly (stepwise) increasing the shear stress until viscous flow could be detected at long times.
The static yield stress thus obtained was found to provide a criterion for determining the onset 
of inhomogeneous flow.
These observations may be consistent with experiments on PMMA colloids exhibiting inhomogeneous 
flow \cite{expts1} but are apparently at odds with experiments on core-shell particles which do not 
give indications of banding or shear localization effects \cite{zackrisson,crassous,winter}. 
While these discrepancies remain to be understood, it seems possible that the softness of the 
potential interaction in the core-shell systems studied in \cite{zackrisson,crassous,winter} may 
play a role in maintaining homogeneous flow. 

\section{Theoretical approaches to fluid states}\label{flowing}
There currently exist several alternative theoretical 
approaches to first-principles calculation of the microstructure and macroscopic rheology of colloidal 
dispersions subject to externally applied flow. 
Each of the available approximation schemes is tailored to capture the physically relevant 
aspects of the correlated particle motion within a restricted range of volume fractions. 
Theories aiming to treat low and intermediate volume fraction dispersions take as their common 
starting point the pair Smoluchowski equation, which is an exact coarse grained reduction 
of the many-body Smoluchowski equation (\ref{smol_hydro}).
At high volume fractions close to the glass transition the pair Smoluchowski equation no 
longer provides a convenient starting point and an alternative approach capable of capturing  
slow structural relaxation is required. 
This is provided by the recently developed integration through transients mode-coupling theory 
\cite{joeprl_07,joeprl_08,fuchs_cates,fc_jrheol}.
In the following, we will first introduce the pair Smoluchowski equation before proceeding to 
follow the `volume fraction axis' to give an overview of the current 
state of research on the theory of flowing states.

\subsection{Pair Smoluchowski equation}\label{pairsmol}
While Eq.(\ref{smol_hydro}) provides a well defined microscopic dynamics, it has been found useful 
to start from an equivalent coarse grained level of description by integrating 
out uneccessary degrees of freedom from the outset.   
Assuming spatial translational invariance, integration of Eq.(\ref{smol_hydro}) over 
the center-of-mass coordinate of a pair of particles 
and the remaining $N\!-\!2$ particles leads to an equation for 
the flow distorted pair correlation function as a function of ${\bf r}={\bf r}_2-{\bf r}_1$
(see e.g. \cite{brady_morris,wagner89,russel_gast,wagner94})
\begin{eqnarray}\label{pair_smol}
\frac{\partial g({\bf{r}})}{\partial t} + \nabla_r\cdot 
\big[\,
{\bf v}({\bf r})\,g({\bf r}) - {\bf D}({\bf r})\cdot\nabla_r g({\bf r}) 
\,\big] \\
=-\nabla_r\cdot \big[\,
{\bf D}({\bf r})\cdot \beta\,{\bf F}({\bf r})\,g({\bf r})
\,\big]
\notag
\end{eqnarray}
%
where we have suppressed explicit time-dependence in the function arguments for notational
convenience and where we have introduced the gradient operator 
$\nabla_r=\nabla_2 - \nabla_1$. 
The conditional probability to find particles at coordinates ${\bf r}_3\cdots{\bf r}_N$, given that 
the first two are known to be at locations ${\bf r}_1$ and ${\bf r}_2$, respectively, is given by 
$P({\bf r}_3,\cdots,{\bf r}_N | {\bf r}_1,{\bf r}_2) = 
P({\bf r}_1,\cdots,{\bf r}_N)/P({\bf r}_1,{\bf r}_2)$ and is required to calculate the functions 
${\bf F}({\bf r}), {\bf D}({\bf r})$ and ${\bf v}({\bf r})$ entering Eq.(\ref{pair_smol}). 
The first of these functions, ${\bf F}({\bf r})$, describes the force acting between our chosen pair of
particles due to both direct potential interaction $v(r)$, taken here to be pairwise additive, 
and indirect interactions transmitted via the surrounding $N\!-\!2$ particles
\begin{eqnarray}\label{force} 
{\bf F}({\bf r}) = -\nabla_r\, v(r) - \frac{n}{2}\!\int d{\bf r}_3
\frac{g^{(3)}({\bf r}_1,{\bf r}_2,{\bf r}_3)}{g({\bf r}_1,{\bf r}_2)}\times \notag\\
\times\big(
\nabla_2 v(|{\bf r}_2-{\bf r}_3|) - \nabla_1 v(|{\bf r}_1-{\bf r}_3|)
\big),
\end{eqnarray}
where $g^{(3)}({\bf r}_1,{\bf r}_2,{\bf r}_3)$ is the nonequilibrium triplet distribution 
function. 
The diffusion tensor is similarly obtained by conditional averaging and contains details of the
hydrodynamic interactions
\begin{eqnarray}\label{diff_tensor}
{\bf D}({\bf r})=2D_0\left(\, \frac{{\bf r}{\bf r}}{r^2}\,G(r) 
+ \left(\boldsymbol{\delta} - \frac{{\bf r}{\bf r}}{r^2}\right)H(r) \,\right),
\end{eqnarray}
where ${\bf r}{\bf r}$ denotes a dyadic product and the scalar hydrodynamic functions $G(r)$ and 
$H(r)$ remain to be specified. 
Finally, the relative velocity of a pair of particles is given by 
\begin{eqnarray}\label{velocity}
{\bf v}({\bf r}) = \kappa\cdot {\bf r} + {\bf C}({\bf r})\!:\!\bar{\kap}
\end{eqnarray}
where $\bar{\kap}$ is the symmetric rate-of-strain tensor $\bar{\kap}=(\kap+\kap^T)/2$
and ${\bf C}$ is the (third rank) hydrodynamic resistance 
tensor describing the disturbance of the affine flow due to the presence of the particles
\begin{eqnarray}\label{resistance}
{\bf C}({\bf r})\!:\!\bar{\kap}=\!-r\left(
\frac{\rb\rb\cdot\bar{\kap}\cdot\rb}{r^3}A(r) + \left( \boldsymbol{\delta}\!-\!\frac{\rb\rb}{r^2} 
\right)\cdot\frac{\bar{\kap}\cdot\rb}{r}B(r)
\right)\notag\\
\end{eqnarray}
The tensor ${\bf C}$ arises from purely geometrical considerations and is not material 
specific. 
It is interesting to note that the addition of hydrodynamic interactions prevents advection 
leading to unphysical hard core overlap. A pair of approaching particles thus `flow around' each 
other in the solvent flow, an effect taken care of by the second term in Eq.(\ref{velocity}). 

In the dilute limit, much is known about the hydrodynamic functions $A, B, G$ and $H$, as only an 
isolated pair of spheres must be considered. 
For both large and small separations analytical 
expressions for these functions exist \cite{lubrication} and are supplemented by tabulated numerical 
data for intermediate ranges \cite{mifflin}. 
At higher volume fractions approximations are required to obtain the hydrodynamic functions and 
a number of schemes have been developed which aim to incorporate the effects of many-body 
hydrodynamics \cite{noyola,brady_rescale,lio_russ_94,lionberger_rev,brady_bossis}. 
In the absence of hydrodynamic interactions $A=B=0$ and $G=H=1$ leading to considerable 
simplification. 
It should be noted that ${\bf C}=0$ in this limit, with the consequence that affine motion 
alone can lead to hard-core overlaps. 
While an exact treatment of the thermodynamic part of the problem would lend such unphysical
configurations zero statistical weight, care must be exercised in approximate treatments which 
may satisfy only partially this important geometrical constraint.

Although the coarse grained pair Smoluchowski equation (\ref{pair_smol}) is still exact 
(under the assumption of homogeneity), it does not provide a closed expression for the
microstructure, as encoded in $g({\bf r})$. Evaluation of the integral term required to determine the
force (\ref{force}) demands knowledge of the nonequilibrium triplet 
distribution function, which remains unknown and contains the residual 
influence of the surrounding particles which have been integrated out. 
This situation is familiar from the BBGKY heirarchy \cite{hansen} for which the triplet 
correlations must be approximated in terms of the pair correlations (using e.g. the 
Kirkwood superposition approximation) in order to arrive at a closed equation. 
In recent years, accurate approximations for the {\em equilibrium} triplet correlations of certain 
model systems have been developed \cite{attard,joe_triplet}. 
Less is known regarding the nature of the triplet correlations in
nonequilibrium situations. 
Recent simulations \cite{yurko} using accelerated Stokesian dynamics 
\cite{brady_sim,banchio} have revealed the existence of aligned particle 
triplets under shear and it may be hoped that such microstructural insights will eventially lead 
to improved theories by guiding the development of approximate closures for the triplet
correlations.
Some of the approaches to be reported in section \ref{section:integralequations} have 
attempted to make progress in this direction by approximating explicitly the integral 
term on the right hand side of (\ref{pair_smol}).

We note that, although the assumption of spatial homogeneity underlying (\ref{pair_smol}) is 
mathematically convenient (for a translationally invariant system the only physically relevant 
coordinate is the separation vector ${\bf r}={\bf r}_2-{\bf r}_1$) it may not be appropriate 
under all conditions. 
The presence of spatial inhomogeneity induced by either external potential fields, 
shear banded or shear localized states complicates the coarse graining procedure, resulting 
in an inhomogeneous version of (\ref{pair_smol}). 
While these issues should pose no difficulty at low volume fractions, for which the right hand side 
of (\ref{pair_smol}) may be disregarded, caution should be exercised when treating systems 
at higher volume fraction. 

Once $g(\rb)$ is known, calculation of the stress tensor describing the macroscopic rheological 
response becomes possible. Although exact expressions relating the stress tensor to to the 
flow distorted microstructure are known formally, the situtation is complicated by the appearance 
of unknown conditionally averaged hydodynamic functions in the expressions. 
However, reliable approximations for these functions are available and enable 
the stress to be evaluated directly from $g(\rb)$ \cite{lionberger_rev}. 
For the simpler case of a system interacting via a pair potential and in the absence 
of hydrodynamic interactions, the stress tensor
may be completely determined by a simple integral over the pair correlation function 
\cite{kbg} 
\begin{eqnarray}
\sig = -nk_BT{\bf 1} \,+\, 5\eta_s\phi\,\bar{\kap}
\,-\, \frac{\,n^2}{2}\!\int \!d{\bf r}\,\frac{{\bf r}{\bf r}}{r}
v'(r)g({\bf r}),
\label{stress}
\end{eqnarray}
where $v'(r)$ is the derivative of the pair potential, $\eta_s$ the solvent viscosity, 
${\bf 1}$ the identity matrix and ${\bf rr}$ denotes a dyadic product. 
Note that the second term on the right
hand side of Eq.(\ref{stress}) assumes that the particles possess a well defined hard-core  
from which the solvent is excluded. 

In the absence of flow the stress tensor is diagonal with the osmotic pressure given by 
$\Pi=-{\rm Tr}\,\sig/3$. 
For a system of pure hard-spheres the familiar equation of state 
$\beta\,\Pi/n=1 + 4\phi g(d)$ is thus recovered. 
When under shear flow Eq.(\ref{stress}) yields a shear viscosity due to the colloids 
$\eta = 5\,\phi\,\eta_s/2 + \mathcal{O}(\phi^2)$, where the first term corresponds to Einsteins classic 
dilute limit result \cite{einstein} and the corrections to higher order in $\phi$ come from the
anisotropy of $g({\bf r})$ inside the integral term. 
It is clear from Eq.(\ref{stress}) that flow induced microstructural anisotropy can give rise 
to the finite normal stress differences $N_1=\sigma_{\rm xx}\!-\sigma_{\rm yy}$ and 
$N_2=\sigma_{\rm yy}\!-\sigma_{\rm zz}$ characteristic of non-Newtonian rheology. 
The dyadic weight factor entering the integral term has the consequence that if $g({\bf r})$ 
possesses a mirror symmetry about the $x=0$ plane then the integral term will be equal to zero and the 
rheology will thus be Newtonian. 
While this `fore-aft' symmetry of the pair distribution function is an exact mathematical consequence 
of the `pure hydrodynamic limit', in which the motion of the particles is determined by 
Stokes flow alone \cite{batchelor2,batchelor_green}, 
chaotic many-body particle motion and experimental perturbations, such as particle 
surface roughness, present in real colloidal systems break the symmetry and result in a 
non-Newtonian rheology \cite{brady_morris}.

\subsection{Low volume fraction}\label{section:brady}
Efforts to obtain a microscopic understanding of colloid rheology began
with the seminal 1906 work of Einstein in which it was shown how the shear viscosity of a 
dilute dispersion of hard spherical colloids increases with colloidal volume fraction, 
assuming that both the volume fraction and the shear rate remain small 
($\eta=\eta_s(1 + 5\phi/2)$) \cite{einstein}. 
Einstein's study addressed the one-body problem of a single colloid suspended in a Newtonian fluid. 
The next step is naturally to consider the interaction between pairs of colloidal particles, 
thus making possible a discussion of the pair correlation functions and their relation to 
rheological functions at low volume fraction.  
Study of the two-particle dilute limit was initiated by Batchelor 
\cite{batchelor,batchelor_green,batchelor2,batchelor_lowrho} whose fundamental work 
formed the basis for the more recent investigations by Brady and coworkers 
\cite{brady_morris,bergenholtz_brady,brady_vicic}. 

At low volume fraction, Eq.(\ref{pair_smol}) admits analytical solution in the limits 
$Pe\rightarrow 0$ and $Pe\rightarrow\infty$ \cite{brady_morris,brady_vicic} and precise 
numerical results have been obtained for intermediate values of $Pe$, both with and without 
hydrodynamic interactions \cite{bergenholtz_brady,lionberger_thinning}.
Exact results are made possible by the fact that, 
in the dilute limit, triplet correlations in the pair Smoluchowski equation may be neglected 
leading to a closed expression for $g(\rb)$. 
Neglecting the difficult integral term in Eq.(\ref{pair_smol}) incurs an $\mathcal{O}(\phi)$ 
error which becomes irrelevant as $\phi\rightarrow\!0$ and yields a closed equation for 
$g({\bf r})$ which is exact to lowest order in the volume fraction and valid for all $Pe$ 
values. 
For the simple special case of hard-spheres Eq.(\ref{pair_smol}) thus reduces to the equation of
motion 
\begin{eqnarray}
\frac{\partial g({\bf r})}{\partial t} + \nabla_r \cdot\big[{\bf v}(\rb)g({\bf r})
- Pe^{-1}\D(\rb)\cdot\nabla g({\bf r})\big] = 0, 
\label{eq:brady1}
\end{eqnarray}  
where we have scaled distance and time with particle radius and flow rate, respectively, such 
that $Pe$ appears explicitly. 
In order to fully specify the problem Eq.(\ref{eq:brady1}) must be supplemented with 
appropriate boundary conditions enforcing both the requirement that the particles do not 
penetrate, via a no-flux condition at $r=d$, and that $g({\bf r})\rightarrow 1$ as 
$r\rightarrow \infty$ \cite{russel}. The first of these boundary conditions is 
clearly an exact physical requirement and is valid also at higher volume fractions. 
The second condition assumes the decay of `wake' structures which develop in $g({\bf r})$ 
downstream from the reference particle at higher flow rates.  
Detailed analysis of (\ref{eq:brady1}) has shown that the range of the wake scales linearly 
with $Pe$, thus justifying the choice of boundary conditions. 

\begin{figure}
\includegraphics[width=7.5cm]{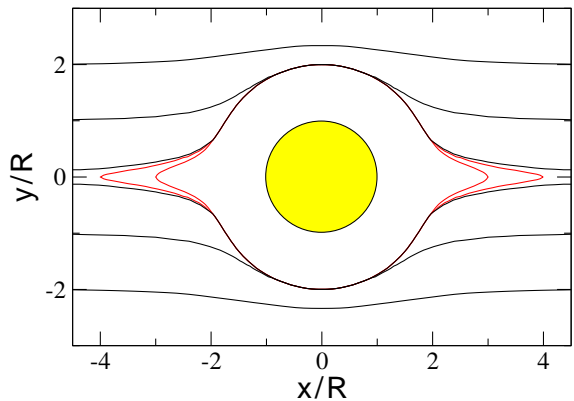}
\caption{
For mathematically perfect hard-spheres with a hydrodynamic radius equal to the radius 
of potential interaction, $R_H\!=\!R$, particle pairs exhibit well defined trajectories. 
Taking a reference frame in which one particle is fixed at the origin (yellow) the 
second particle (red) follows the trajectories shown in the `pure hydrodynamic limit' of 
large $Pe$ \cite{batchelor}. 
The figure shows some sample trajectories in the $z=0$ flow-gradient plane 
(see e.g. \cite{russel}). 
Closed orbits are indicated in red and open in black. 
The apparent fore-aft mirror symmetry gives rise to Newtonian rheology.
}
\label{trajectories}
\end{figure}

For systems interacting via a spherically symmetric pair potential it can be 
shown that, regardless of volume fraction, in the weak flow limit 
$Pe\!\rightarrow\! 0$ a steady flow 
field acts on the spherically symmetric equilibrium distribution $g_{\rm eq}(r)$ to produce an 
$\mathcal{O}(Pe)$ perturbation \cite{batchelor_lowrho,russel}
\begin{eqnarray}
g({\bf r}) = g_{\rm eq}(r)\left[ 1 \,-\, 
Pe\frac{{\bf r}\cdot\hat\kap\cdot{\bf r}}{r^2}f(r) \right],
\label{eq:lowrhosol}
\end{eqnarray}  
where $\hat\kap=\bar{\kap}/\sqrt{\,2\bar{\kap}:\!\bar{\kap}}$, with $\bar\kap$ is defined below
Eq.(\ref{stress}).
In the dilute limit $g_{\rm eq}(r)=\Theta(-r-1)$ and substitution of (\ref{eq:lowrhosol}) into (\ref{eq:brady1}) yields a differential equation for the 
dimensionless function $f(r)$ which has been solved for several interaction potentials 
of interest \cite{russel}. 
Although analytical expressions exist for certain systems, a numerical integration is 
still required to obtain $f(r)$ in the special case of hard-spheres with hydrodynamic
interactions \cite{batchelor_lowrho,russel}.

The extension of Eq.(\ref{eq:lowrhosol}) to higher order in the Peclet number has been 
analyzed in considerable detail \cite{brady_vicic}. 
It can be shown that $g({\bf r})$ has a regular perturbation expansion to 
$\mathcal{O}(Pe^2)$ but that calculation of the next order term requires singular 
perturbation theory, yielding an $\mathcal{O}(Pe^{5/2})$ correction. 
The calculation of higher order terms in the $Pe$ expansion requires use of matched asymptotic 
expansions which rapidly become intractable and make preferable a numerical solution of 
(\ref{eq:brady1}). 
The expansion of the distorted structure is given by \cite{brady_vicic}
\begin{eqnarray}
g({\bf r}) = 1 + f_1Pe + f_2Pe^2 + f_{5/2}Pe^{5/2}+\cdots,
\label{expansion}
\end{eqnarray}
where comparison with (\ref{eq:lowrhosol}), and noting that $g_{\rm eq}(r\!>\!2R)=1$ for 
hard-spheres at low volume fraction, enables identification of the coefficient $f_1$. 
To $\mathcal{O}(Pe)$ the rheology is predicted to be Newtonian with normal stress differences 
identically equal to zero \cite{batchelor}. Non-Newtonian rheology first occurs at 
$\mathcal{O}(Pe^2)$, which is sufficient to capture both non-zero normal stresses and the 
first flow induced correction to the osmotic pressure \cite{brady_vicic}.

Analytic solutions to Eq.(\ref{eq:brady1}) exist also in the `pure hydrodynamic limit' 
of strong flows ($Pe\!\!\rightarrow\!\infty$) and have highlighted the subtle balance between 
hydrodynamic and potential forces in determining the rheological response 
\cite{batchelor,batchelor_green}. For large Peclet number steady flows the solution of 
Eq.(\ref{eq:brady1}) is well approximated by the solution of the the simplified equation 
$\nabla\cdot [{\bf v}(\rb)\,g({\bf r})] = 0$ (subject to the boundary condition $g({\bf r})=1$ at 
$r=\infty$), despite the fact that the approximation neglects the boundary layer and thus 
violates the no-flux condition at contact. 
Subject to certain conditions on the trajectories of particle pairs, Batchelor and Green 
proved the surprising result that the simplified equation predicts a spherically symmetric radial 
distribution function for hard spheres, leading to Newtonian rheology \cite{batchelor_green}.   
This clear prediction is a direct consequence of the fore-aft symmetry of $g({\bf r})$ 
(see subsection \ref{pairsmol}) inherent in the assumed Stokesian solvent flow. 
In Figure.\ref{trajectories} we show sample trajectories of a hard-sphere in shear
flow as it moves around a second sphere held fixed at the coordinate origin. 
Of particular interest is the existence of closed trajectories along which the particles 
become trapped in bound orbits and which are connected to the lubrication force required to
displace solvent from the region between the particles \cite{lubrication}. 
In fact, the lubrication force acting between a pair of perfect spheres at separation $r$ 
shows a divergence, $F_{\rm \,lub}\sim (r/R-2)^{-1}$, corresponding to surface contact. Crucially, the
time-reversibility of the Stokes equations dictating the solvent flow implies that the 
force required to push particles togther is identical to that required to pull them apart, 
with the consequence that particle trajectories exhibit the fore-aft mirror symmetry apparent 
in Fig.\ref{trajectories}.

Despite the sound mathematical evidence provided in \cite{batchelor_green}, serious doubts were cast by 
subsequent experiments on intermediate volume fraction hard-sphere-like colloidal dispersions, 
which seemed to contradict the theoretical predictions by identifying a non-Newtonian rheology at large 
flow rates \cite{gadala}. 
It should be noted that the reversibility of Stokes flow implies that fore-aft symmetry 
in the pure hydrodynamic limit holds also for finite volume fractions and so the value 
$\phi=0.4$ employed in the experiments of \cite{gadala} 
cannot be held responsible for the apparent discrepancy. 
The situation was eventually resolved by Brady and Morris, who analytically 
identified a boundary-layer in the region close to particle contact in which 
Brownian motion balances advection \cite{brady_morris}. 
The analysis of \cite{brady_morris} indeed recovers the findings of \cite{batchelor_green} 
in the case that the hydrodynamic radius is equal to the excluded volume radius, as would 
be the case for mathematically perfect hard-spheres with no surface roughness. 
However, when the excluded volume radius exceeds the hydrodynamic radius, 
even by a very small amount, the residual Brownian motion within the anisotropic boundary layer 
of $g({\bf r})$ leads to a non-Newtonian rheology in the strong flow limit. 
Fig.\ref{coreshell} shows a sketch of the model employed in \cite{brady_morris}. 
The physical origin of these symmetry breaking surface effects remains an open problem and 
is likely to be a function of various system specific parameters (e.g. surface roughness). 
We note that the existence of a boundary-layer structure was originally 
identified in studies of the distorted structure factor of colloids under shear 
\cite{dhont89}.

The analytical results for $Pe\!\rightarrow\! 0$ and $Pe\!\!\rightarrow\!\infty$ for 
hard-spheres under steady shear have been both
confirmed and supplemented by full numerical solutions at all values of $Pe$ \cite{bergenholtz_brady,lionberger_thinning}. 
These accurate numerical studies revealed that the dilute dispersions described by Eq.(\ref{eq:brady1}) demonstrate 
not only shear thinning and finite normal stresses at intermediate flow rates 
(accessible from the expansion (\ref{expansion}) to $\mathcal{O}(Pe^2)$), 
but also shear thickening at high flow rates \cite{bergenholtz_brady}. 
Shear thinning in dilute systems is correlated with a nonvanishing 
distortion of the structure factor in the plane perpendicular to the flow direction 
\cite{ramaswamy}.


\begin{figure}
\includegraphics[width=7.5cm]{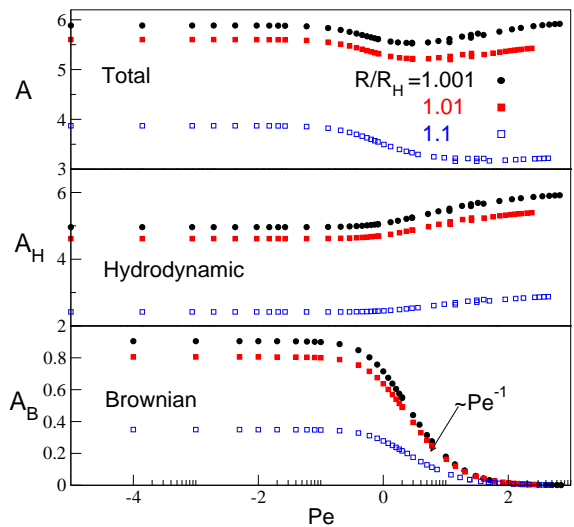}
\caption{
The pair contributions to the relative shear viscosity of a dilute colloidal dispersion under 
steady shear flow for three values of the ratio $R/R_H$ (see Fig.\ref{coreshell}). 
To second order in $\phi$ the relative viscosity is given by 
$\eta_r\equiv\eta/\eta_s = 1 + 5\phi_H/2 + A(Pe,R/R_H)\,\phi^2$, where $\phi_H=(R_H/R)^3\,\phi$, and 
$A(Pe,R/R_H)$ can be divided into hydrodynamic and Brownian contributions $A=A_H+A_B$. 
As $R/R_H\rightarrow 1$ the dispersion shows shear thickening at large $Pe$ values due to 
the increase in the hydrodynamic contribution to the shear stress. 
As the value of $R/R_H$ is increased the increase in $A_H$ becomes balanced by the decrease in 
$A_B$ and only shear thinning remains.  
(Figure adapted from \cite{bergenholtz_brady}) 
}
\label{bergenholtz_pic}
\end{figure}

In Figure \ref{bergenholtz_pic} we show some of the numerical results obtained in 
Ref.\cite{bergenholtz_brady} for the shear viscosity of a dilute dispersion as a function of $Pe$. 
In these calculations the effective sphere model sketched in Fig.\ref{coreshell} was employed and 
results are shown for three values of the ratio of potential to hydrodynamic radius, $R/R_H$.  
To second order in $\phi$ the relative viscosity may be expressed as
\begin{eqnarray}
\eta_r\equiv\eta/\eta_s = 1 + 5\,\phi_H/2 + A(Pe,R/R_H)\,\phi^2, 
\end{eqnarray}
where $\phi_H\!=\!(R_H/R)^3\,\phi$ is 
the volume fraction with respect to the hydrodynamic radius and the function $A(Pe,R/R_H)$ 
contains the effects of microstructural distortion (note that calculation of $g({\bf r})$ to 
$\mathcal{O}(\phi)$ yields the viscosity to $\mathcal{O}(\phi^2)$). 
The numerically determined function $A(Pe,R/R_H)$ may be further split into hydrodynamic and 
Brownian contributions, $A = A_H + A_B$, which are independently accessible from the numerical 
calculations of \cite{bergenholtz_brady}.

The top panel of Fig.\ref{bergenholtz_pic} shows $A$ as a function of $Pe$. For shear rates 
up to $Pe\sim 1$ the qualitative variation of the viscosity is independent of the value of the 
size ratio $R/R_H$, displaying a low-shear Newtonian plateau followed by shear thinning. 
The $R/R_H$ independence of the form of the curves for $Pe\!<\!1$ reflects the fact that hydrodynamic
interactions are not central to the mechanisms underlying shear thinning and only influence the 
absolute value of the viscosity. 
For $Pe\!>\!1$ the viscosity begins to increase as a function of $Pe$ and the dispersion 
shear thickens. In contrast to the shear thinning behaviour, the viscosity increase is 
strongly sensitive to the value of $R/R_H$. In the Limit $R/R_H\rightarrow 1$ the trajectories 
sketched in Fig.\ref{trajectories} are recovered and Batchelor's pure hydrodynamic limit is 
realized with a viscosity independent of $Pe$.  
In \cite{bergenholtz_brady} it is also shown that for $R/R_H\rightarrow 1$ the normal stresses 
also vanish as $Pe\!\rightarrow\!\infty$, indicating a Newtonian response.  
As $R/R_H$ is increased the magnitude of the shear thickening reduces strongly and by 
$R/R_H\!=\!1.1$ is lost entirely. 
This trend strongly indicates the important influence of short range lubrication forces on shear
thickening (see section \ref{thickening}), which can effectively be turned-off by slightly 
reducing the hydrodynamic radius below that of the repulsive potential interaction. 

The central and lower panels of Fig.\ref{bergenholtz_pic} show the individual hydrodynamic and 
Brownian pair contributions to the total stress as a function of $Pe$. 
For $Pe<1$ it is apparent that the shear thinning is due to a reduction in the Brownian 
contribution with increasing $Pe$. 
For values of $R/R_H$ close to unity the reduction in $A_B$ is more than compensated by an 
increase in the hydrodynamic stress afor $Pe>1$, leading to shear thickening. However, 
increasing $R/R_H$ above unity rapidly supresses the influence of lubrication and the 
hydrodynamic contribution is overwhelmed by the strong drop in $A_B$.  

Comparing the first panel of Fig.\ref{bergenholtz_pic} with Fig.\ref{launpic}, it is remarkable 
the extent to which the qualitative rheological response observed in systems at finite volume 
fraction is reproduced by calculations based on the dilute limit. 
However, the fact that $Wi=Pe$ at low volume fractions does not permit investigation of 
potentially interesting interaction effects between shear thinning and thickening. 
Despite the extensive understanding of the response of dilute dispersions 
to steady flows, analogous solutions for time-dependent flows remain to be investigated. 
This leaves open many interesting questions regarding transient response and non-steady states.

\subsection{Intermediate volume fraction}\label{section:integralequations}
The simplest way to extend the dilute-limit results to finite volume fraction 
is via the introduction of empirical volume fraction dependent 
scale factors \cite{brady_vicic,brady_morris,brady_rescale,brady_rescale1}. 
Analysis of Eq.(\ref{eq:brady1}) has provided two important insights. 
Firstly, at finite volume fractions the value of $g(\rb)$ outside the boundary layer close to 
the surface of a reference particle should asymptote to the solution of 
$\nabla\cdot [{\bf v}(\rb)\,g({\bf r})] = 0$, for which the no-flux boundary condition is 
ignored. This amounts to assuming that, outside the thin boundary layer, it is sufficient to 
solve a purely advective problem. Secondly, that the appropriate Peclet number in the presence of
hydrodynamics is $Pe_{\rm H}=\dot\gamma R^2/2D_s(\phi)$, where $D_s(\phi)$ is the short time 
diffusion coefficient (which differs from $D_0$ due to hydrodynamic interactions with neighbouring
particles). 
For weak flows ($Pe\ll 1$) these ideas are manifest in a modified perturbation to the quiescent 
pair correlations
\begin{eqnarray}
g({\bf r}) = g_{\rm eq}(r)\left[ 1 \,-\, 
Pe_{\rm H}\frac{{\bf r}\cdot\hat\kap\cdot{\bf r}}{r^2}f(r) \right].
\label{eq:lowrhosol1}
\end{eqnarray} 
The finite volume fraction equilibrium radial distribution function $g_{\rm eq}(r)$ is an external 
input to the theory and can be calculated using either simulation or equilibrium integral equation 
theory \cite{hansen}. 
The function $f(r)$ is determined by substitution of (\ref{eq:lowrhosol1}) into the {\em dilute
limit} equation (\ref{eq:brady1}), or its generalization for non-hard-sphere potentials. 
Many-body effects are thus included via the equilibrium radial distribution 
function and the short-time diffusion coefficient entering $Pe_{\rm H}$. 
It should be noted that Brady's approach assumes an input $g_{\rm eq}(r)$ which diverges 
at random close packing ($\phi\approx 0.64$). 

Despite considerable success, the scaling approach suffers from two significant drawbacks: 
(i) the mathematical structure of the nonequilibrium part of the theory is that of the dilute 
system. Thus, regardless of rescaling, this approach does not admit the occurrence of possible 
additional physical mechanisms which may only occur as a consequence of cooperative behaviour 
at finite volume fraction, 
(ii) the equilibrium microstructure $g_{\rm eq}(r)$ is required as an external input and does not 
emerge from an approximate treatment of many-body correlation effects within the theory. 
These issues may be addressed by approaches which aim to approximate the triplet distribution 
function entering the pair Smoluchowski equation (\ref{pair_smol}) via the effective force between 
a pair of particles (\ref{force}). 
In order to arrive at a closed theory it is necessary to relate 
$g^{(3)}(\rb_1,\rb_2,\rb_3)$, 
either explicitly or implicitly (by approximating weighted integrals over the triplet 
distribution), to $g(\rb)$ and the pair potential $v(r)$ using an appropriate 
`closure' hypothesis. 

\subsubsection{Superposition approximation}
Guidance in developing an appropriate closure relation to treat non-equilibrium states 
is provided by experience from equilibrium liquid-state integral equation theory. 
One of the earliest approximation schemes aiming to arrive at a closed equation for the 
equilibrium pair correlations was developed by Born and Green \cite{born_green} who employed 
the Kirkwood superposition approximation 
$g^{(3)}(\rb_1,\rb_2,\rb_3)\!\approx\! g(r_{12})g(r_{23})g(r_{13})$ 
in combination with the second member of the exact Yvon-Born-Green heirarchy (Eq.\ref{meanforce}) 
\cite{hansen,kirkwood}. 
Numerical solution of the resulting Born-Green equation yields acceptable results only for 
weak coupling. 
The superposition approximation is asymptotically correct for large particle separations 
but is poor when particles come close to contact, leading to a failure of the Born-Green 
equation at intermediate volume fractions (a breakdown which was erroneously taken as an indicator 
for the first order freezing transition of hard-spheres). 
Subsequent attempts have aimed to systematically improve upon the superposition approximation 
by including additional Mayer cluster diagrams (see e.g. \cite{rice}). 

One of the earliest attempts to close the non-equilibrium Eq.(\ref{pair_smol}) using the superposition approximation, 
albeit in the absence of hydrodynamics, was made by Ohtsuki \cite{ohtsuki1} (self diffusion was 
addressed using an anologous approach in \cite{ohtsuki2}). 
Numerical solution of the closed integro-differential equation resulting from this approximation 
was performed for charged hard-spheres at intermediate volume fraction. 
Although the theoretical results for the zero-shear viscosity were found to be in reasonable 
agreement with those of experiment, the pair correlation $g(\rb)$ was found to be in considerable 
error. 
These findings are supported by the work of Wagner and Russel who investigated a similar
superposition based approach \cite{wagner89}.

The observed discrepancies in $g(\rb)$ arising from superposition are not surprising: 
In equilibrium, the pair flux in Eq.(\ref{pair_smol}) may be set equal to zero, resulting in an 
exact equation for the pair correlation function
\begin{eqnarray}\label{eq_borngreen}
&&k_BT\,\nabla_r \ln g_{\rm eq}(r) = {\bf F}({\bf r}),
\end{eqnarray}  
where ${\bf F}(\rb)$ is given by Eq.(\ref{force}) and contains the unknown triplet distribution 
function. 
Use of superposition to approximate $g^{(3)}(\rb_1,\rb_2,\rb_3)$ in Eq.(\ref{force}) leads 
directly to the Born-Green equation for the equilibrium correlations, the shortcomings of which 
have been noted above. 
It is therefore nontrivial that, despite a relatively poor description of the microstructure, the 
results for the zero-shear viscosity presented in \cite{ohtsuki1} turn out to be rather good agreement with experimental data. 
A similar situation is encountered in a number of the theories to be outlined in this section and 
serves to highlight the fact that good values for integrated quantities (e.g. the viscosity) does not 
neccessarily imply that the underlying correlations are treated adequately. 

When applying superposition to tackle non-equilibrium problems it should be borne in mind that 
the approximation represents an uncontrolled ansatz and only possesses a firm statistical
mechanical basis in the limit of vanishing flow rate. In equilibrium, the superposition approximation 
represents both the exact low volume fraction limit of the triplet correlation function and recovers 
correctly the long range asymptotic behaviour. 
Analogous limiting results for the nonequilibrium triplet correlations which could motivate a more 
appropriate superposition-type approximation are currently lacking and would require a detailed 
analysis of the triplet Smoluchowski equation in the dilute limit. 
Despite these shortcomings, superposition may nevertheless turn out to be useful for some applications. 
As noted at the end of section \ref{thinning}, very recent simulation results investigating the 
microstructure of hard-sphere fluids under shear flow revealed the importance of linear 
triplet configurations \cite{kulk}. 
From equilibrium studies it is known that the error of the superposition approximation is reduced 
for such linear configurations \cite{attard_stell}, 
thus raising the interesting possibility that superposition may be appropriate for treating certain 
non-equilibrium states.  

\subsubsection{Potential of mean force}\label{sec:meanforce}
Inspection of the exact expression for the force (\ref{force}) shows that only a certain weighted
integral of the triplet distribution is required to determine the pair correlations. 
It is thus not strictly neccessary to know the full details of the triplet distribution and
schemes can be developed which aim to approximate directly the integrated quantity. 
The simplest approximation possible is to neglect entirely the integral contibution to the 
effective force (\ref{force}) and set
\begin{eqnarray}\label{first}
{\bf F}({\bf r}) = -\nabla_r\, v(r). 
\end{eqnarray}
Combining this crude approximation with Eq.(\ref{pair_smol}) recovers the dilute limit equation 
of motion for $g(\rb)$ for an arbitrary spherically symmetric pair potential $v(r)$. 
For the special case of hard-spheres this leads to Eq.(\ref{eq:brady1}), but with an additional 
delta function term on the right hand side. 
As demonstrated by Cichocki \cite{cichocki} the resulting equation is completely equivalent to 
(\ref{eq:brady1}), with zero right hand side, supplemented by a no-flux boundary condition at 
contact. 
 
As already noted, in equilibrium the pair Smoluchowski equation (\ref{pair_smol}) reduces to the 
Yvon-Born-Green equation for the pair correlations  
\begin{eqnarray}\label{meanforce}
k_BT\,\nabla_r \ln g_{\rm eq}({\bf r}) \!&=&\!\! -\nabla_r\, v(r) - \frac{n}{2}\!\int d{\bf r}_3
\frac{g^{(3)}_{\rm eq}({\bf r}_1,{\bf r}_2,{\bf r}_3)}{g_{\rm eq}({\bf r}_1,{\bf r}_2)}\!\times \notag\\
&&\hspace*{-0.0cm}\times \big(
\nabla_2 v(|{\bf r}_2-{\bf r}_3|) - \nabla_1 v(|{\bf r}_1-{\bf r}_3|)
\big) \notag\\
&\equiv&-\nabla v_{\rm mf}(\rb),
\end{eqnarray}
where $v_{\rm mf}(\rb)$ is the equilibrium potential of mean force \cite{hansen}, defined by 
$g_{\rm eq}=\exp(-\beta v_{\rm mf})$ in analogy with the low density limit of the pair 
correlations. 
A first step towards improving the zeroth order approximation (\ref{first}) is thus to 
approximate the nonequilibrium force (\ref{force}) by the equilibrium potential of mean force, 
leading to 
\begin{eqnarray}\label{gast_approx}
{\bf F}(\rb) = -k_BT\,\nabla_r \ln g_{\rm eq}(r). 
\end{eqnarray}
This approximation, developed by Russel and Gast \cite{russel_gast}, incorporates equilibrium 
thermodynamic many-body couplings but, as is clear from Eq.(\ref{meanforce}), neglects the 
influence of flow on the triplet correlations. 
It should be noted that the approximation (\ref{gast_approx}) does not provide any information 
regarding the nonequilibrium triplet correlation function. This is in contrast to superposition 
based approaches from which the triplet function can be reconstructed using a product of the
self-consistently determined pair correlation functions. 
It should also be noted that within the Russel-Gast approach the function $g_{\rm eq}(r)$ is 
an input, which can be calculated using either simulation or equilibrium statistical mechanical 
approximations (see subsection \ref{eq_integral}).  
 
The linear equation resulting from combining equations (\ref{pair_smol}) and 
(\ref{gast_approx}) has been solved for hard-spheres in weak shear flow \cite{russel_gast}. 
In these calculations, simple approximations were 
employed to determine the hydrodynamic functions $A, B, G$ and $H$ entering equations (\ref{diff_tensor}) 
and (\ref{resistance}) defining the conditionally averaged hydrodynamic tensors which represent the
effective medium (solvent$\,+\,(N-2)$ colloids) in which the chosen pair of particles are immersed. 
Results were obtained for the zero-shear viscosity, linear response moduli $G'(\omega), G''(\omega)$ 
under small amplitude oscillatory shear and the leading order flow induced distortion of $g(\rb)$ 
(via determination of the function $f(r)$, see Eq.(\ref{eq:lowrhosol})). 
For $\phi<0.3$ good agreement with experiment was obtained for the integrated quantities 
$\eta_0, G'$ and $G''$, despite providing only a poor description of the microstructure 
(see Fig.6 in \cite{szamel}). 

An interesting feature of the Russel-Gast theory is that the predicted zero-shear viscosity becomes 
very large in the vicinity of random-close-packing ($\phi\approx 0.64$), although the precise 
nature of this rapid increase as a function of volume fraction remains to be studied in detail. 
The apparent divergence of $\eta_0$ is a non-trivial output of the theory, given that the 
approximate Verlet-Weiss expression for $g_{\rm eq}(r)$ used as input diverges only at $\phi=1.0$ \cite{verlet_weiss}.  
When viewed within the context of the time-correlation/Green-Kubo formalism 
(see subsection \ref{zero_shear}) it is tempting to infer that the observed growth in 
$\eta_0$ is related to the development of an underlying slow structural relaxation time. 
The solution of Eq.(\ref{gast_approx}) for small amplitude oscillatory shear at finite 
frequencies would enable this issue to be addressed.

\subsubsection{Equilibrium integral equations}\label{eq_integral}
The Russel-Gast theory outlined above neglects the influence of external flow on the triplet 
correlation function, which leads to a force ${\bf F}(\rb)$ generated from the equilibrium potential 
of mean force (Eq.(\ref{gast_approx})). 
In order to go beyond this close-to-equilibrium ansatz it is neccessary to express the force 
${\bf F}(\rb)$ as a functional of $g(\rb)$, such that both functions can be determined 
self-consistently. 
A promising approach in this direction is to generalize equilibrium liquid-state integral equation 
theory \cite{hansen} to treat the nonequilibrium problem. 
It will thus be useful to review briefly some concepts from the equilibrium theory before moving on to 
more unfamiliar territory in the next subsection. 

Integral equation theories generally aim to calculate the equilibrium pair correlation function 
$g_{\rm eq}(r)$ from knowledge of the interaction potential in a non-perturbative fashion.
Fundamental to the integral equation approach is the Ornstein-Zernicke (OZ) equation which, 
for a translationally invariant system, is given by the convolution form \cite{hansen}
\begin{eqnarray}
h_{\rm eq}(r_{12}) = c_{\rm eq}(r_{12}) + n\!\int \!d{\bf r}_3\,c_{\rm eq}(r_{13})h_{\rm eq}(r_{32}), 
\label{oz}
\end{eqnarray}
where $h_{\rm eq}(r)=g_{\rm eq}(r)-1$. The direct correlation function $c_{\rm eq}(r)$ defined 
by (\ref{oz}) is a function of simpler structure than $h_{\rm eq}(r)$ and is thus easier to approximate. 
When supplemented by an independent closure 
relation between $c_{\rm eq}(r)$ and $h_{\rm eq}(r)$, containing details of the interaction potential under 
consideration, (\ref{oz}) provides a closed equation for the pair correlations ( 
note that for pairwise additive potentials the triplet distribution does not enter explicitly). 
The task of the integral equation practitioner is thus to find numerically tractable closures which 
capture the essential physics of the problem under consideration. 

Using diagrammatic techniques \cite{stell_diagram} it can be shown that a formally exact closure 
relation is given by 
\begin{eqnarray}\label{bridge}
g_{\rm eq}(r)=\exp[-\beta v(r) + h_{\rm eq}(r) - c_{\rm eq}(r) + b_{\rm eq}(r)],
\end{eqnarray}
where $b_{\rm eq}(r)$ is the unknown `bridge function' containing the difficult to evaluate 
`irreducible' Mayer cluster diagrams \cite{hansen}. 
Two closures of particular merit are the Hyper-Netted-Chain (HNC) and Percus-Yevick (PY),  
given by 
\begin{eqnarray}
b_{\rm eq} &=& 0\hspace*{4.75cm}{\rm (HNC)} \label{hnc}\\
b_{\rm eq} &=& -\ln(1\!+\!h_{\rm eq}\!-\!c_{\rm eq})-
(h_{\rm eq}\!-\!c_{\rm eq})\hspace*{0.5cm}{\rm (PY)}
\label{py}
\end{eqnarray}
respectively. 
An additional practical advantage of theories based on (\ref{oz}) over superposition-type
approaches is that the convolution form of the integral term enables self consistent solutions 
to be obtained using efficient iterative numerical algorithms \cite{brader_ijtp}. 

Although the majority of integral equation theories focus on the pair correlations, triplet 
correlations can also be handled within the same framework \cite{joe_triplet}. 
In addition to providing a higher level of resolution, the development of triplet-level 
integral equations is motivated by the desire for an improved description of the pair 
correlations. Exact relations, such as the YBG equation (\ref{meanforce}), connect the triplet 
to the pair correlations and the expectation is that errors in an approximate $g^{(3)}_{\rm eq}$ 
may be averaged out by integration to the pair level. 
During the mid-1960s many of the leading liquid-state theorists proposed integral equations 
for the triplet correlations
(e.g. Verlet \cite{verlet3,verlet3a,verlet4,verlet5,verlet6}, Wertheim \cite{wertheim}, 
Baxter \cite{baxter}, Stell \cite{stell1}), all of which showed considerable promise. 
However, the complexity of solving the equations has hindered progress along this route 
and somewhat simpler, numerically tractable, theories now seem preferable 
\cite{barrat,attard_py,joe_triplet}.  

An integral equation which will be of particular relevance for the following section was 
derived by Scherwinski \cite{scherwinski}. 
Within the Scherwinski approximation the triplet correlation function is obtained from 
self-consistent solution of the following linear equation
\begin{eqnarray}\label{triplet}
&&\!\!\!g_{\rm eq}(\rb_1,\rb_2,\rb_3) 
= g_{\rm eq}(r_{12})g_{\rm eq}(r_{13})g_{\rm eq}(r_{23}) + ng_{\rm eq}(r_{12})\times
\notag\\ 
&&\times\int\! d\rb_4 \left(
\frac{g^{(3)}_{\rm eq}(\rb_1,\rb_3,\rb_4)}{g_{\rm eq}(r_{14})} - g_{\rm eq}(r_{13})
\right)
\!h_{\rm eq}(r_{14})h_{\rm eq}(r_{24}). 
\notag\\
\end{eqnarray}
Iteration of Eq.(\ref{triplet}) yields an infinite series expressing the triplet correlation 
function as a functional of the pair correlation function $g_{\rm eq}(r)$. 
Substitution of this series into the exact YBG equation (\ref{meanforce}) yields a diagrammatic
expansion for $g_{\rm eq}(r)$ in perfect agreement with that arising from solution of 
Eqs.(\ref{oz}), (\ref{bridge}) and (\ref{hnc}). 
In this sense, Eq.(\ref{triplet}) represents a triplet generalization of the more 
familiar pair-level HNC theory. 
The more powerful approximations proposed by earlier workers 
\cite{verlet3,verlet3a,verlet4,verlet5,verlet6,wertheim,baxter,stell1} probably provide a 
superior description of the triplet correlations than Eq.(\ref{triplet}) and would, upon 
substitution into the YBG equation, lead to improved (i.e. better than standard HNC) 
estimate of the pair correlations. 
However, the Scherwinski approximation has a number of purely technical advantages which make 
it particularly suitable for application to non-equilibrium situations and which are convenient
for numerical implementation (see also \cite{wagner89}, which predates \cite{scherwinski}, but contains 
several of the key ideas).

\subsubsection{Nonequilibrium integral equations}\label{sec:noneq}
In order to go beyond the Russel-Gast approximation \cite{russel_gast} outlined in subsection 
\ref{sec:meanforce}, 
Lionberger and Russel employed the Scherwinski equation for the triplet correlation function 
(\ref{triplet}) in order to estimate the force (\ref{force}) entering the pair Smoluchowski 
equation \cite{lionberger}. 
Nonequilibrium pair and triplet correlations are thus determined self consistently and are both 
influenced by the externally imposed flow. 
It should be noted that the direct application of an equilibrium relation, such as 
Eq.(\ref{triplet}), to nonequilibrium ignores the fact that nonequilibrium states are intrinsically 
different from equilibrium and thus represents a major approximation. 

The original version of the Lionberger-Russel theory presented in \cite{lionberger} neglected
hydrodynamic interactions and has been implemented numerically for weak flows only. 
No results beyond leading order in $Pe$ have been presented, although in principle the theory remains valid 
also in the nonlinear regime. 
For $\phi<0.4$ the LR theory makes predictions for the zero-shear viscosity, self-diffusion cooefficient and 
distorted microstructure in reasonable agreement with available computer simulation results. 
For $\phi\ge0.45$ significant quantitative deviations appear and the theory becomes unreliable.
We note that the study \cite{lionberger} considered a suspension interacting via a continuous repulsive 
potential which was then mapped onto a hard-sphere system. 
The input equilibrium microstructure was generated using the Rogers-Young integral equation 
\cite{rogers} (an interpolation between PY and HNC), despite the fact that the $Pe\rightarrow 0$ limit 
of the theory reduces to the HNC approximation for $g_{\rm eq}(r)$. 

Although the Lionberger-Russel theory \cite{lionberger} provides a sophisticated 
treatment of the microstructural distortion by incorporating the triplet correlations into the 
self-consistency loop, it is interesting that the results for $\eta_0$ are inferior to those from 
the much simpler Russel-Gast theory \cite{russel_gast} at high volume fractions.  
In particular, the former predicts only a relatively weak growth of $\eta_0$ with volume fraction, 
whereas the latter suggests a divergence. 
Calculations of the linear moduli $G',G''$ as a function of frequency have been performed using 
the LR theory and, perplexingly, reveal a structural relaxation time which 
decreases with increasing volume fraction. 
This unphysical prediction would appear to be at odds with the growth of $\eta_0$ output 
from the theory and represents a weak point of the approach, in need of clarification 
\cite{lionberger2}. 
We note that the only approximation invoked by the Lionberger-Russel theory is the Scherwinski 
closure (\ref{triplet}) for the triplet correlation function. 
It would therefore be interesting, albeit numerically demanding, to see whether any of the more 
sophisticated triplet closures available 
\cite{verlet3,verlet3a,verlet4,verlet5,verlet6,wertheim,baxter,stell1,barrat,attard_py} can improve 
the performance at higher volume fractions. 

The theory developed in \cite{lionberger} omitted hydrodynamic interactions in order to make clearer 
the approximations to the many-body thermodynamic couplings and to facilitate comparision with 
Brownian dynamics simulation results. In \cite{lionberger2} hydrodynamic interactions were 
included into the theory of \cite{lionberger} in order to make closer contact with experiment. 
The hydrodynamic approximations developed were found to be of the correct magnitude but 
errors in the underlying thermodynamic approximation, namely the Scherwinski equation (\ref{triplet}), 
led to an underestimation of the magnitude of the nonequilibrium structure. 
It was identified that the magnitude of the flow induced structural distortion is determined by 
the slow structural relaxation in the system (a finding supported by subsequent theoretical studies 
\cite{henrich}), which, taken together with the results for the volume fraction dependence of the linear
response moduli, suggests that the integral equation approach does not capture the slow dynamics 
characteristic of dense systems. 

Wagner and Russel \cite{wagner89} developed an integral equation approach closely related to that 
of Lionberger and Russel in which the nonequilibrium triplet correlations are approximated 
using a closure motivated by the PY equilibrium theory (\ref{py}). 
Although use of a PY-type closure ensures that excluded-volume packing constraints are treated 
realistically, at least close to equilibrium, the theory of \cite{wagner89} included hydrodynamic interactions by employing only 
the low density limit of the pair Hydrodynamic functions $A, B, G$ and $H$ entering 
Eqs.(\ref{diff_tensor}) and (\ref{resistance}). 
The simultaneous introduction of hydrodynamic and thermodynamic approximations in 
\cite{wagner89} served to obscure the validity of the proposed nonequilibrium PY approximation.

When using integral equation methods to tackle the triplet corelations in nonequilibrium it is 
important to bear in mind that the physical situation is intrinsically different from that in 
equilibrium. Consequently, caution must be exercised when attempting to apply trusted and 
familiar results from equilibrium statistical mechanics to system under flow. 
In order to appreciate more clearly the approximations involved in applying equilibrium triplet 
closures to the pair Smoluchowski equation both Lionberger and Russel \cite{lionberger} 
and Szamel \cite{szamel} have invoked the concept of a `fictitious' flow dependent two-body 
potential $u(\rb,\dot\gamma)$. 
In a study of the kinetic theory of hard-spheres Resibois and Lebowitz \cite{resibois} assumed 
the existence of a two-body potential $u(\rb,\dot\gamma)$ which, if employed in an equilibrium 
calculation, reproduces exactly the nonequilibrium pair correlation function
\begin{eqnarray}
g(\rb)=g_{\rm eq}(\rb;[u\,]),
\label{fict}
\end{eqnarray}
where the square brackets indicate a functional dependence and where $g_{\rm eq}$ is 
anisotropic as a result of the anisotopy of $u$. 
The fictitious potential thus serves as proxy for the flow field acting on the real system. 
It should be made clear that the assumption that an effective two-body potential can yield
the correct $g(\rb)$ is quite distinct from the (erroneous) assumption that an effective 
one-body external potential field can represent the one-body flow induced force acting on 
the particles (see the end of section \ref{formal}).  
Eq.(\ref{fict}) implicitly assumes that a homogeneous one-body density distribution $n$ 
is not a function of the flow rate, thus neglecting possible dilation effects.  
Given equation (\ref{fict}), it is natural to go one step further and assume that the relation 
can be uniquely inverted, such that 
\begin{eqnarray}\label{inv}
u(\rb,\dot\gamma)=u(\rb;[\,g\,]). 
\end{eqnarray}

By construction, $u(\rb,\dot\gamma)$ reproduces the nonequilibrium pair correlations. 
However, if the same fictitious two-body potential is used in an equilibrium statistical 
mechanical calculation of the triplet correlation function, the exact nonequilibrium 
$g^{(3)}$ will not be reproduced. The missing part of the triplet correlation is referred to as 
the `irreducible' term
\begin{eqnarray}
\!\!\!\!\!\!g^{(3)}(\rb_1,\rb_2,\rb_3,\dot\gamma)
\!=\!g^{(3)}_{\rm eq}(\rb_1,\rb_2,\rb_3;[\,g\,]) + g^{(3)}_{\rm irr}(\rb_1,\rb_2,\rb_3;\dot\gamma),
\hspace*{-1cm}
\notag\\
\label{fict_trip}
\end{eqnarray}
where we have assumed that (\ref{inv}) is valid. 
A key approximation in the work of \cite{lionberger} and \cite{szamel} is to set 
$g^{(3)}_{\rm irr}=0$, which essentially amounts to assuming that equilibrium relations 
such as (\ref{triplet}) may be used to connect triplet and pair functions in nonequilibrium.

The nonequilibrium integral equation method considered in \cite{lionberger,lionberger2,wagner89} 
represents a synthesis of the exact dilute-limit results with statistical mechanical 
descriptions of dense systems at equilibrium. 
While this approach is promising for weak flows ($Pe\ll1$), data is lacking for stronger flows 
which would enable the close-to-equilibrium character of the closure approximations to be 
better tested.  
In particular, it has been emphasized by Szamel \cite{szamel} that a major challenge for theories 
of the distorted microstructure is to account for the nonzero distortion of $g(\rb)$ in the 
vorticity-gradient plane perpendicular to an applied shear flow. 
This distortion cannot be detected to leading order in $Pe$ and makes desireable numerical studies 
of the existing closure approximations under strong flow. 
Applications of the nonequilibrium integral equation method have been restricted 
to shear flow and, with the exception of small amplitude oscillatory shear 
\cite{lionberger,lionberger2}, have neglected time-dependent flows entirely. 

\subsubsection{Alternative approaches}
The nonequilibrium pair correlations and rheology of colloidal dispersions under weak shear 
flow was investigated by Szamel \cite{szamel}, who employed functional methods to approximate 
the unknown integral term in Eq.(\ref{force}). 
Hydrodynamic interactions were neglected entirely. 
Assuming that the irreducible term in (\ref{fict_trip}) can be neglected, the triplet correlation 
function can be developed in a functional Taylor expansion about the equilibrium state. 
To first order this is given by
\begin{eqnarray}\label{func_expansion}
&&g^{(3)}(\rb_1,\rb_2,\rb_3;[\,g\,])
= g^{(3)}_{\rm eq}(\rb_1,\rb_2,\rb_3) \\
&&\;\;\;\;\;+ \int \!\!d\rb_4 \!\int\!\!d\rb_5 \frac{\delta g^{(3)}(\rb_1,\rb_2,\rb_3)}{\delta g(\rb_4,\rb_5)}
\left(g(\rb_4,\rb_5)-g_{\rm eq}(r_{45})\right),\notag
\end{eqnarray}
where the functional derivative is evaluated for a constant one-body density. 
Equation (\ref{func_expansion}) is clearly a close-to-equilibrium approximation. 
Insertion of (\ref{func_expansion}) into the expression for the force (\ref{force}) and rearrangement 
of terms leads to a closed equation for $g(\rb)$ which requires both 
$g_{\rm eq}^{(3)}(r_{12},r_{23},r_{13})$ and $c_{\rm eq}^{(4)}(\rb_1,\rb_2,\rb_3,\rb_4)$, a 
higher order equilibrium direct correlation function, as input. 
Following appropriate decoupling approximations for the unknown higher order equilibrium correlations
and linearizing with respect to the shear flow, Szamel obtained a closed equation for $g(\rb)$ which 
only requires $g_{\rm eq}(r)$ as input (for which the Verlet-Weiss approximation was employed
\cite{verlet_weiss}). 

\begin{figure}
\hspace*{-0.3cm}
\includegraphics[width=6.5cm]{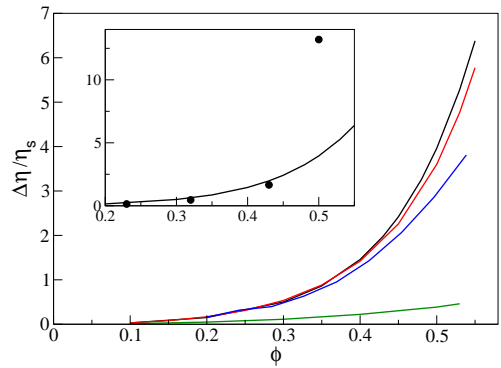}
\caption{
The reduced zero-shear viscosity as a function of volume fraction as predicted by 
various theories based on the pair Smoluchoski equation: 
Szamel (black) \cite{szamel}, Brady (red), Lionberger-Russel (blue) \cite{lionberger}, 
Ronis (green) \cite{ronis}. 
The inset shows the results of the Szamel theory compared with Brownian dynamics 
simulation data (circles). 
For $\phi\!>\!0.43$ the theory strongly underestimates the zero-shear viscosity. 
(Adapted from Ref.\cite{szamel}) 
}
\label{fig:szamel}
\end{figure}

The Szamel theory \cite{szamel} is considerably simpler to implement than the integral equation 
approach of Lionberger and Russel \cite{lionberger} and provides comparable, perhaps even slightly 
better, results for the zero-shear viscosity as a function of volume fraction, 
as is evident from Fig.\ref{fig:szamel}. 
Both $g(\rb)$ and its Fourier transform $S(\kb)$ were found to be qualitatively
similar to those from the Lionberger-Russel theory, substantially underestimating the magnitude 
of the distortion from equilibrium when compared to Brownian dynamics simulation. 
 
In an early study, Ronis took an alternative approach based on fluctuating hydrodynamics, 
in which phenomenological fluctuating terms are added to the macroscopic equations of 
hydrodynamics \cite{ronis}. 
The theory of stochastic processes may then be employed to calculate nonequilibrium time-correlation
functions and the distorted microstucture. 
Of all the theories described in this and the previous section, Ronis theory provides the most accurate 
results for the distortion of the structure factor at low $Pe$ values. 
Nevertheless, it can be shown that the Ronis approximation leads to a vanishing of the microstructural 
distortion in the vorticity-gradient plane at all $Pe$, in contradiction to experiment and simulation
results \cite{blaw}. 
This deficit is related to the fact that the hard-sphere `core condition' $g(|\rb|<1)=0$ is 
violated within this approach. 
The Ronis theory reduces to the closely related theory of Dhont \cite{dhont89} in the low $Pe$, 
linearized limit. 
In another early work, Schwarzl and Hess \cite{schwarzl} postulated a phenomenological equation 
for $g(\rb)$ involving a number of empirical parameters representing the relaxation times in the 
system. 
However, due to the phenomenological nature of both the fluctuating hydrodynamics approach and 
the equation-of-motion for $g(\rb)$ proposed by Schwarzl and Hess, the foundation of the theoretical treatments 
presented in \cite{ronis} and \cite{schwarzl} in nonequilibrium statistical mechanics is unclear. 

Finally, we note that Wagner has assessed approaches based on the pair Smoluchowski equation using 
the GENERIC framework of beyond equilibrium thermodynamics \cite{wagner_generic}. 
This formalism enables any proposed closure of the pair Smoluchowski equation to be checked for 
thermodynamic consistency. 
The study presented in \cite{wagner_generic} identified the thermodynamically admissable expression 
for the stress tensor and clarified the nature of the inconsistencies which can occur when separate
derivations are performed for the equation-of-motion for $g(\rb)$ and the stress tensor.

\subsection{Temporal locality vs. memory functions}
In sections \ref{zero_shear} and \ref{thinning} we introduced briefly the Green-Kubo expressions 
for calculating the shear stress in both the linear (\ref{lr}) and nonlinear 
regimes (\ref{nlr}). Within the Green-Kubo framework, transport coefficients 
are related to integrals over time correlation functions \cite{evans_morriss}. 
The growth of the shear viscosity as a function of volume fraction is thus related to an increasingly 
slow decay of the stress autocorrelation function.  
What is lacking thus far in the discussion of the pair Smoluchowski equation and its approximate 
solutions, is the connection between the temporally local Eqs.(\ref{pair_smol}) and 
(\ref{stress}) and the nonlocal expressions (\ref{lr}) and (\ref{nlr}) fundamental to 
the time-correlation function formalism. 
Moreover, the memory kernels characteristic of the time-correlation approach (and widely employed 
in continuum mechanics approaches, see section \ref{cont}) make no explicit appearance within 
the pair Smoluchowski framework.  

The key to understanding the connection between the nonlocal time-correlation formalism and 
approaches based on the local pair Smoluchowski equation lies in the study of time-dependent external 
flow fields. Indeed, it is the absence of time-dependent data from the pair Smoluchowski approach 
which has served to obscure the relationship between these two methods, despite the fact that they
are formally equivalent. 
The only available Smoluchowski-based time-dependent calculations were performed using the Lionberger-Russel theory 
\cite{lionberger,lionberger2,lionberger_rev} for small amplitude oscillatory shear flow. 
As pointed out in subsection \ref{sec:noneq}, the available results for the volume fraction dependence 
of $G'(\omega)$ and $G''(\omega)$ within the LR theory reveal underlying problems resulting from 
the approximations employed (i.e. a reduction of $\tau_{\alpha}$ with increasing $\phi$) which would 
otherwise have gone unnoticed. 
This serves to emphasize the importance of going beyond steady 
flow calculations in applications of pair Smoluchowski theories.  

The generalization of Eq.(\ref{nlr}) to general time-dependent shear is given by integration 
over the entire flow history \cite{joeprl_07}
\begin{eqnarray}\label{stress_td}
\sigma_{\rm xy}(t) = \int_{-\infty}^{t}\!\!\!dt'\,\dot\gamma(t')\,G(t,t'),
\end{eqnarray}
where $G(t,t')\equiv G(t,t';[\dot\gamma\,])$ is the nonlinear shear modulus.  
The lack of time-translational invariance in the modulus arises from a functional dependence 
on the shear rate. 
The microscopically derived Eq.(\ref{stress_td}) should be contrasted with the more familiar 
phenomenological result (\ref{pre_boltz}). 

Using the generalized Green-Kubo result (\ref{stress_td}) it is instructive to consider a 
simple special case: The stress response of hard-spheres to the onset of steady shear flow in the 
absence of hydrodynamic interactions.
Specifically, we consider a shear field which is switched from zero to a
constant value, $\dot\gamma(t) = \dot\gamma\Theta(-t)$. 
For this choice of shear field Eq.(\ref{stress_td}) reduces to 
\begin{eqnarray}\label{stress_on}
\sigma_{\rm xy}(t) = \dot\gamma\int_{0}^{t}\!\!dt'\,G_{\rm ss}(t'),
\end{eqnarray}
where $G_{\rm ss}(t)$ is the time translationally invariant shear modulus under steady 
shear flow.  
On the other hand, within approaches based on the distorted microstructure the interaction 
contribution to the stress is given by (see Eq.(\ref{stress}))
\begin{eqnarray}\label{dist_on}
\sigma_{\rm xy}(t) = -\frac{\,\,n^2}{2}\!\int \!d{\bf r}\,\frac{{\bf r}{\bf r}}{r}
\,v'(r)\,g({\bf r},t),
\end{eqnarray}
where $g({\bf r},t)$ is calculated from the pair Smoluchowski equation (\ref{pair_smol}) 
subject to the switch-on shear flow under consideration. 
The expressions (\ref{stress_on}) and (\ref{dist_on}) are formally equivalent. 
Equating the time derivatives thus leads to the exact relation
\begin{eqnarray}
G_{\rm ss}(t) = -\frac{\,\,n^2}{2\dot\gamma}\!\int \!d{\bf r}\,\frac{{\bf r}{\bf r}}{r}
\,v'(r)\,\frac{\partial g({\bf r},t)}{\partial t}.
\end{eqnarray}
The quiescent shear modulus is thus recovered in the slow flow limit
\begin{eqnarray}\label{slow}
G_{\rm eq}(t) = -\frac{\,\,n^2}{2}\!\int \!d{\bf r}\,\frac{{\bf r}{\bf r}}{r}
\,v'(r)\,\lim_{\dot\gamma\rightarrow 0}
\left(\frac{1}{\dot\gamma}\frac{\partial g({\bf r},t)}{\partial t}\right).
\end{eqnarray} 
The right hand side of Eq.(\ref{slow}) can be further reduced to the determination of the 
function $f(r,t)$ by substitution of the first order expansion (\ref{eq:lowrhosol}) for 
$g(\rb,t)$. 

As the volume fraction is increased, the time derivative on the r.h.s. of Eq.(\ref{slow}) 
must give rise to a growth in the timescale determining the relaxation of stress fluctuations, 
as described by $G_{\rm eq}(t)$. 
The fact that this is not captured by the Lionberger-Russel theory 
\cite{lionberger,lionberger2,lionberger_rev} simply reflects the failings of the approximate 
closure (\ref{triplet}) relating the nonequilibrium triplet and pair correlation functions. 
However, it is not clear that even an {\em exact} equilibrium expression for the triplet 
correlations would be sufficient to resolve these difficulties. 
If the key source of error lies in the neglect of the irreducible term apprearing in 
(\ref{fict_trip}) then considerable new insight into the nature of nonequilibrium states will 
be required in order to make further progress using integral equation methods. 
Nevertheless, these considerations should serve to motivate time-dependent studies of the 
Russel-Gast \cite{russel_gast} and Szamel \cite{szamel} theories. 

While the particular example chosen (switch-on shear flow) provides access to the steady shear 
rate dependent modulus, other choices of time-dependence may enable connection to be made 
between non-time translationally invariant correlation functions (e.g. $G(t,t')$) and  
time-dependent solutions of the pair Smoluchowski equation. 
The issue of whether temporally local constitutive equations are preferable to nonlocal functionals 
for describing complex fluids has been addressed within the framework of nonequilibrium 
thermodynamics \cite{oettinger_book}. 
The nonequilibrium thermodynamics approach requires identification of an appropriate set 
of structural variables which, in addition to the standard hydrodynamic variables of mass, momentum 
and internal energy density, contain all information about the state of the system at a given time 
that is necessary to determine the macroscopic quantities of interest. 
The pair Smoluchowski approaches discussed in this section essentially introduce $g(\rb)$ as an
additional structural variable \cite{wagner_generic}. 
While this appears to be a valid approach for low and intermediate volume fractions, a correct 
identification of the structural variables appropriate for describing glass formation and 
dynamical arrest remain to be found. 
For this reason the most promising approaches to treating high volume fraction states are based 
on the generalized Green-Kubo relations and mode-coupling theory.

\section{Glass rheology}\label{section:mct}
\begin{figure}
\includegraphics[width=6.2cm]{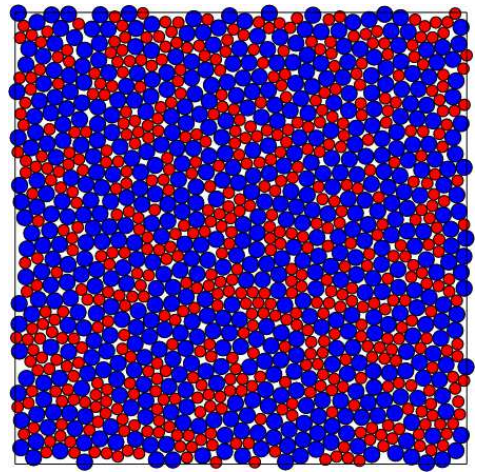}
\caption{
A snapshot from a Brownian dynamics simulation of a quiescent binary hard-disc mixture 
(using a size ratio of $1:1.4$ to supress crystallization). The simulation was performed at a two-dimensional
volume fraction of $\phi_{\rm 2D}=0.81$, which lies above the estimated glass transition packing 
$\phi_{2D}^{(g)}\sim 0.79$, with $50$\% large discs and $50$\% small discs. 
(Figure courtesy of F. Wey\ss er.) 
}
\label{snapshot}
\end{figure}

Assuming that crystallization effects can be suppressed (see section \ref{quiescent}) the volume 
fraction can be increased to the point at which the individual particles are unable to diffuse 
beyond the cage of nearest neighbours and a dynamically arrested glassy state is 
formed. In order to visualize the amorphous cage structure in such a glassy state Figure 
\ref{snapshot} shows a configuration snapshot taken from a Brownian dynamics simulation of a 
binary hard-sphere mixture in two dimensions (hard-discs) \cite{weysser}.
The two-dimensional volume fraction $\phi_{\rm 2D}=0.81$ of the simulation is above the estimated 
glass transition point of $\phi_{\rm 2D}\approx 0.79$ and the size ratio of large to small disc 
radii is $1.4$, a value empirically found to frustrate crystallization in two dimensional systems 
(which occurs at $\phi_{\rm 2D}=0.69$ for monodisperse discs). In both two- and three-dimensional 
systems the physics of the glass transition becomes important for determination of both the rheology and flow distorted microstructure 
of high volume fraction systems. 

The response to externally applied flow of states close to, or beyond, the glass
transition is only beginning to be understood and establishing the basic principles of glass rheology 
remains a challenging task. 
At present, the only truly microscopic theories available are provided by recent extensions of the quiescent 
MCT which enable the effects of external flow to be incorporated into the formalism and thus make possible 
a theoretical investigation of the complex interaction between arrest and flow. 

\subsection{MCT inspired approaches}\label{sec:inspired}
Extending earlier work on the low volume fraction self diffusion of colloidal dispersions 
\cite{indrani}, Miyazaki and Reichman constructed a self-consistent mode-coupling-type approach 
to describe collective density fluctuations for dense colloidal fluids under shear below the 
glass transition \cite{miyazaki1,miyazaki2,miyazaki3}. 
The Miyazaki-Reichman theory considers time-dependent fluctuations about the steady state and 
thus requires the (unknown) flow distorted structure factor $S(\kb)$ as an input quantity. 
Approximating $S(\kb)$ by the quiescent correlator, results have been presented for colloidal
dispersions in two-dimensions under steady shear \cite{miyazaki1,miyazaki2} and in three
dimensions (subject to additional isotropic approximations) under oscillatory shear
\cite{miyazaki3}. 
Applications to glassy states have been avoided as the theory relies upon an ergodic 
fluctuation-dissipation theorem. 
An alternative extended-MCT approach has been proposed by Koblev and Schweizer \cite{koblev} 
and Saltzman {\em et al} \cite{saltzman} which is built on the idea that entropic barrier hopping 
is the key physical process driving the microscopic dynamics and rheology of glassy colloidal 
suspensions. 
Due to the activated nature of the barrier hopping process the ideal glass transition described 
by quiescent MCT (see section \ref{quiescent}) plays no role. 
A nonlinear rheological response results from a stress induced modification of the barrier heights. 
 
A currently promising method of extending quiescent MCT to treat dense systems under flow 
involves integration through the transient dynamics, starting from an equilibrium Boltzmann 
distribution in the infinite past. In contrast to \cite{miyazaki1,miyazaki2,miyazaki3} the 
distorted microstructure is an output of this approach. 
The initial form of the theory was outlined by Fuchs and Cates for steady shear flow 
\cite{fuchs_cates} and 
presented two essential developments: Firstly, that integration through the transient dynamics 
leads directly to exact generalized Green-Kubo formulas, relating average quantities to integrals over 
microscopic time-correlation functions. 
Secondly, that MCT-type projection operator approximations reduce the formal Green-Kubo expressions 
to closed equations involving transient correlators (which can be calculated self-consistently). 
A strong prediction of the ITT-MCT theory resulting from combining these two steps is that the 
macroscopic flow curves exhibit a dynamic yield stress (see section \ref{yieldstress})
in the limit $Pe\!\rightarrow\!0$, for states which would be glasses or gels in the 
absence of flow. 
Moreover, the yield stress appears discontinuously as a function of volume fraction, in contrast 
to mesoscopic approaches \cite{mes1,mes2,mes4}. 
The ITT-MCT thus provides a scenario for a nonequilibrium transition between a 
shear-thinning fluid and a yielding amorphous solid which is supported by considerable 
evidence from both colloidal experiments \cite{petekidis_yield2,crassous,winter,zackrisson} and 
Brownian dynamics simulation \cite{varnik1,weysser}.

Due to the numerical intractability of the microscopic theory of \cite{fuchs_cates}, subsequent 
work focused on the construction of both isotropically averaged approximations to the full 
anisotropic equations 
and simplified schematic models inspired by these \cite{faraday}.   
Comparison of the theoretical predictions with experimental data for thermosensitive core-shell 
particles (see Fig.\ref{coreshell}) has proved highly successful 
\cite{fuchs_cates2003,fuchs_ballauff,crassous0,crassous,winter}. 
The original formulation of the ITT-MCT (more details of which can be found in \cite{fc_jpcm}) 
has subsequently been superseded by a more elegant version \cite{fc_jrheol}. 
It is interesting to note that the significant technical changes to the ITT-MCT formalism 
introduced in \cite{fc_jrheol} lead to expressions which resemble more closely those of Miyazaki 
and Reichman \cite{miyazaki1,miyazaki2,miyazaki3}. 
Given the very different nature of the formal derivation (fluctuating hydrodynamics vs. 
projection operator 
methods) and approximations employed, the similarity of the final expressions is reassuring and 
serves to highlight the robustness of MCT-based approaches. 
For a comprehensive overview of the status of the steady shear theory we refer the reader 
to the recent review \cite{fuchs_review}.

Going beyond steady shear, the original formulation of ITT-MCT \cite{fuchs_cates,faraday,fc_jpcm} 
has been generalized to treat arbitrary time-dependent shear \cite{joeprl_07}. 
These developments not only enable shear fields of particular experimental relevance to be investigated 
(e.g. large amplitude oscillatory shear flow), but have also revealed an underlying 
mathematical structure which is not apparent from consideration of steady flows alone. 
The theory has been applied (albeit subject to various simplifying approximations) to 
investigate the build-up of stress and corresponding microscopic particle motion, 
as encoded in the mean-squared-displacement, following the onset of shear \cite{zausch}. 
More recently, the modern version of ITT-MCT \cite{fc_jrheol} has been extended to describe 
time-dependent flow of arbitrary geometry \cite{joeprl_08}, thus making possible 
the study of non-shear flow and enabling the full tensorial structure of the theory to 
be identified. 
The developments presented in \cite{joeprl_08} elevate the ITT-MCT approach to the level 
of a full constitutive theory for dispersion rheology and may be regarded as the most up-to-date 
formulation of the theory.
While the development of numerical algorithms to efficiently solve the fully 
microscopic theory \cite{joeprl_08} is currently in progress, this task is made 
computationally demanding by the combination of spatial anisotropy and lack of 
time-translational invariance presented by many flows of interest. 
In \cite{pnas} a simplified theory was presented which contains the 
essential physics of the full microscopic equations, including the tensorial 
structure, but which is much more convenient for numerical solution (see subsection 
\ref{sec:schematic}). 

\subsection{Integration through transients}\label{section:itt}
The integration through transients (ITT) approach originally developed by Fuchs and Cates 
\cite{fuchs_cates} and subsequently developed in \cite{joeprl_07,joeprl_08} provides a formal 
expression for the nonequilibrium distribution function $\Psi(t)$ required to calculate 
average quantities under flow. 
In essence, ITT provides a very elegant method of deriving generalized 
(i.e. nonlinear in $\kap(t)$) Green-Kubo relations which invite mode-coupling-type closure 
approximations. 
The current formulation of the theory neglects hydrodynamic interactions (HI) entirely. 
On one hand this omission is made for purely technical reasons, but it is also hoped that 
HI will prove unimportant for the microscopic dynamics of the dense states to which the 
final theory will be applied. 
In the following we will briefly outline the key points of ITT, employing throughout the 
modern formulation of \cite{joeprl_08}. 

The starting point for ITT is to re-express the Smoluchowski equation (\ref{smol_hydro}) 
in the form
\begin{eqnarray}
\frac{\partial \Psi(t)}{\partial t} = \Omega(t) \Psi(t) ,
\label{smol}
\end{eqnarray}
where, in the absence of HI, the Smoluchowski operator controlling the dynamical evolution of 
the system is given by \cite{dhont}
\begin{eqnarray}
\Omega(t) = \sum_{i} \bp_i\cdot[\,D_0(\bp_i - \beta{\bf F}_i) - \kap(t)\cdot\rb_i\,].
\label{smol_op}
\end{eqnarray}
Equation (\ref{smol}) may be formally solved using a time-ordered exponential function (which arises 
because $\Omega(t)$ does not commute with itself for different times \cite{vankampen}) 
\begin{eqnarray}
\Psi(t) = \left[ \exp_{+} \int_{-\infty}^t\!\!\!ds\,\Omega(s)\right] \Psi_{\rm eq},
\label{formal1}
\end{eqnarray}
where $\Psi_{\rm eq}$ is the initial distribution function in the infinite past, which is 
taken to be the equilibrium Boltzmann-Gibbs distribution corresponding to the thermodynamic 
state-point under consideration. The assumption of an equilibrium distribution is clearly acceptable 
in situations for which the quiescent state is one of ergodic equilibrium. 
However, the role of the initial state is less clear for statepoints in the glass and the dependence, if any, 
on the initial condition may depend upon the details of the flow between $t=-\infty$ and the present time $t$. 
The absence of a general proof that $\Psi(t)$ is independent of $\Psi(-\infty)$ leaves open the possibility 
that certain flow histories do not restore ergodicity and that the system thus retains a dependence on the
initial state. 

The solution (\ref{formal1}) is formally correct, but not particularly useful in its present form. 
A partial integration yields an alternative solution of (\ref{smol}) which is exactly equivalent to 
(\ref{formal1}), but more suited to approximation
\begin{eqnarray}
\Psi(t) &=& \Psi_e + \int_{-\infty}^{t}\!\!\!\!\! dt_1\,
\Psi_e \,\kap(t_1)\!:\!\hat{\sig}\,e_-^{\int_{t_1}^{t}ds\,\Omega^{\dagger}(s)}, 
\label{itt}
\end{eqnarray}  
where $\hat{\sigma}_{\alpha\beta}\!=\! -\sum_iF_i^{\alpha}r_i^{\beta}$ and the `double dot' 
notation familiar from continuum mechanics, 
${\bf A}\!:\!{\bf B}={\rm Tr}\{{\bf A}\cdot{\bf B}\}$ \cite{mcquarrie}, has been employed. 
As a result of the partial integration the dynamical evolution in Eq.(\ref{itt}) is dictated 
by the adjoint Smoluchowski operator 
\begin{eqnarray}
\Omega^{\dagger}(t) = \sum_{i} [\,D_0(\bp_i + \beta{\bf F}_i) + \rb_i\cdot\kap^{T}(t)\,]\cdot\bp_i.
\label{adj_smol}
\end{eqnarray}

Equation (\ref{itt}) is the fundamental formula of the ITT approach and expresses the nonequilibrium
distribution function as an integral over the entire transient flow history. 
Both solutions (\ref{formal1}) and (\ref{itt}) are valid for arbitrary flow geometries and 
time-dependence. 
The relation between the two formal solutions is analogous to the Heisenberg and Schr\"odinger pictures of quantum mechanics
in which the time evolution of the system is attributed to either the wavefunction 
(equation (\ref{formal1})) or the operators representing physical observables (equation (\ref{itt})).  
It should be understood that the ITT form (\ref{itt}) is an operator expression to be used with the understanding that a function 
to be averaged is placed on the right of the operators and integrated over the particle coordinates. 
A technical point to note is that in cases for which phase space decomposes into disoint pockets 
(`nonmixing' dynamics) the distribution (\ref{itt}) averages over all compartments.  
A general function of the phase-space coordinates $f(t,\{ {\bf r}_i \})$ thus has the average
\begin{eqnarray}
\langle f\rangle^{ne} = \langle f \rangle
+ \int_{-\infty}^{t}\!\!\!\!\! dt_1\,
\langle\,
\kap(t_1)\!:\!\hat{\sig}\,
e_-^{\int_{t_1}^{t}ds\,\Omega^{\dagger}(s)} f\,\rangle,
\label{average2}
\end{eqnarray}
where $\langle f\rangle^{ne}$ denotes an average over the non-equilibrium distribution (\ref{itt}). 
Equation (\ref{itt}) generalizes the original formulation of ITT \cite{fuchs_cates} to treat 
arbitrary time-dependence \cite{joeprl_08}. 

\subsection{Translational invariance}\label{sec:advection}
Before applying mode-coupling-type approximations to the exact result (\ref{average2}) we first 
address an important consequence of assuming homogeneous flow (reflected in the spatial constancy 
$\kap(t)$ appearing in Eq.(\ref{smol})). 
On purely physical grounds, it seems reasonable that for an infinite system the assumed translational invarance of the 
equilibrium state (crystallization is neglected) will be preserved by the Smoluchowski dynamics. 
However, proving this for a general time-dependent flow is mathematically nontrivial, due to the 
fact that the Smoluchowski operator (\ref{smol_op}) is itself not translationally invariant. 
By considering a constant vectorial shift of all particle coordinates, $\rb_i'=\rb_i+{\bf a}$,  
Brader {\em et al} have shown that the nonequilibrium distribution function $\Psi(t)$ 
is translationally invariant (but anisotropic) for any homogeneous velocity gradient 
$\kap(t)$ \cite{joeprl_08}. 

Given the translational invariance of $\Psi(t)$ it becomes possible to investigate the more
interesting invariance properties of the two-time correlation functions. 
The correlation between two arbitrary wavevector-dependent 
fluctuations 
$\delta f_{\bf q}=f_{\bf q}-\langle f_{\bf q}\rangle^{ne}$ 
and 
$\delta g_{\bf k}=g_{\bf k}-\langle g_{\bf k}\rangle^{ne}$ 
occuring at times $t$ and $t'$ is given by 
\begin{eqnarray}\label{correlation}
C_{f^{}_{\qb}g^{}_{\kb}}(t,t')
=\langle\delta f^*_{\qb}(t)\delta g^{}_{\kb}(t')\rangle^{ne}.
\end{eqnarray}
It is clear that in a homogeneous system the correlation function (\ref{correlation}) must 
be translationally invariant. 
However, in this case, shifting the particle coordinates by a constant vector ${\bf a}$ yields
\begin{equation}
C_{f^{}_{\qb}g^{}_{\kb}}(t,t') = e^{-i (\,\bar{\qb}(t,t')-\kb\,)\cdot {\bf a}}
\,C_{f^{}_{\qb}g^{}_{\kb}}(t,t'),
\label{invariance}
\end{equation}
where 
\vspace*{-0.2cm}
\begin{eqnarray}
\bar{\qb}(t,t')=\qb\cdot e_-^{-\int_{t'}^{t}\!ds\,\kap(s)}.
\label{advection1}
\end{eqnarray}
The only way in which the required translational invariance of the correlation function can 
be preserved is if the exponential factor in (\ref{invariance}) is equal to unity. 
This requirement has the consequence that a fluctuation at wavevector 
$\kb=\bar{\qb}(t,t')$ at time $t'$ is correlated with a fluctuation with wavevector 
$\qb$ at time $t$ as a result of the affine solvent flow.  
Eq.(\ref{advection1}) thus defines the {\em advected} wavevector which is central to the ITT-MCT 
approach and which captures the affine evolution of the system in approaches focused on Fourier 
components of fluctuating quantities (e.g. the density $\rho^{}_{\bf k}$) rather than particle 
coordinates directly. 
The wavevector $\bar{\qb}(t,t')$ at time $t'$ evolves due to flow-induced advection to become 
${\bf q}$ at later time $t$. 
It should be noted that various definitions and notations for the advected wavevector have been 
employed in the literature documenting the development of ITT-MCT and which could provide a source 
of confusion. In the present work we exclusively employ the modern definition used in 
\cite{joeprl_08,fc_jrheol,pnas}. 

Although Eq.(\ref{advection1}) arises from microscopic considerations it is nevertheless fully 
consistent with the continuum mechanics approaches outlined in section \ref{cont}, despite the very different 
mindset underlying the two methods. 
Eq.(\ref{advection1}) simply describes the affine deformation of material lines in Fourier 
space and can thus be used to define the inverse deformation gradient tensor via 
$\bar{\qb}(t,t')=\qb\cdot {\bf E}^{-1}(t,t')$ in complete accord with continuum approaches. 
Doing so leads to the identification
\begin{eqnarray}
{\bf E}^{-1}(t,t') = e_-^{-\int_{t'}^{t}\!ds\,\kap(s)}.
\end{eqnarray}
As the deformation gradient tensor simply describes the affine distortion of a material line under
flow, it is natural to define also a {\em reverse-advected} wavevector resulting from the inverse 
transformation 
$\qb(t,t')=\qb\cdot {\bf E}(t,t')$, where 
\begin{eqnarray}\label{def}
{\bf E}(t,t') = e_{+}^{\int_{t'}^{t}\!ds\,\kap(s)}.
\end{eqnarray}
The choice of using either advected or reverse-advected wavevectors in treating the effects of affine motion 
within a microscopic theory has parallels with the choice between Lagrangian and Eulerian
specifications of the flow field in continuum fluid dynamics approaches \cite{batchelor}. 
Within a continuum mechanics framework the deformation gradient would simply be {\em defined} 
as the solution of the equation
\begin{eqnarray}\label{defdef}
\frac{\partial}{\partial t}\E(t,t')=\kap(t)\E(t,t'),
\end{eqnarray}
for a given flow $\kap(t)$. 
According to the rules of time-ordered exponential algebra \cite{vankampen}, Eq.(\ref{def}) 
is the formal solution of (\ref{defdef}), thus demonstrating the consistency between the 
Fourier-space microscopic approach of \cite{joeprl_08} and traditional real-space continuum 
mechanics. 

The advected wavevector introduced above provides a convenient way to keep track of the affine 
deformation in a particulate system. 
Mode-coupling-type approximations (to be discussed below) seek to factorize the average entering Eq.(\ref{average2}) by 
projecting the dynamics onto the subspace of density fluctuations $\rho_{\bf q}$ 
\cite{goetze_leshouches}. 
For a flowing system a fluctuation at wavevector $\bar{q}(t,t')$ at time $t'$ evolves (in the absence 
of interactions and Brownian motion) to one at ${\bf q}$ at time $t$. 
It thus becomes essential to project onto density fluctuations at the correct advected wavevectors in 
order to avoid spurious decorrelation effects in the resulting approximations.

\subsection{Microscopic constitutive equation}\label{sec:constitutive}
In order to address dispersion rheology the special choice $f=\hat{\sig}/V$ is made in 
(\ref{average2}), leading to an exact generalized Green-Kubo relation for the time-dependent 
shear stress tensor \cite{joeprl_08}
\begin{eqnarray}
\sig(t) = 
\frac{1}{V}\int_{-\infty}^{t}\!\!\!\!\! dt_1\,
\langle
\kap(t_1)\!:\!\hat{\sig}\,
e_-^{\int_{t_1}^{t}ds\,\Omega^{\dagger}(s)} \hat{\sig}\rangle,
\label{gk_stress}
\end{eqnarray} 
noting that there are no `frozen in' stresses in the equilibrium state, $\langle \sig \rangle=0$. 
The adjoint Smoluchowski operator $\Omega^{\dagger}(t)$ has a linear dependence on $\kap(t)$ and equation
(\ref{gk_stress}) is thus nonlinear in the velocity gradient tensor. 
Equation (\ref{gk_stress}) is a formal constitutive equation expressing the stress tensor 
at the present time as a nonlinear functional of the flow history. 
Although the result (\ref{gk_stress}) does not provide an exact description of 
the physical system under consideration (particle momenta are assumed to have 
relaxed and hydrodynamic interactions are absent), it has a formal status equivalent to that of 
Eq.(\ref{itt}). 
For the special case of steady shear flow (\ref{gk_stress}) is consistent with (\ref{nlr}) with a 
shear modulus given by (\ref{exact_mod_nonlin}).

Application of MCT-type projection operator factorizations \cite{joeprl_08} to the average in (\ref{gk_stress}) leads 
to a complicated, but closed, constitutive equation expressing the deviatoric stress in terms of the 
strain history \cite{joeprl_08,pnas}
\begin{eqnarray}
\sig(t) = -\int_{-\infty}^{t} \!\!\!\!\!\!dt'\!\int\!\!\!
\frac{d{\bf k}}{32\pi^3} \left[\frac{\partial}{\partial t'}(\kb\!\cdot\!\Finger(t,t')\!\cdot\!\kb)
\kb\kb\right]\times\notag\\
\times
\left[\left(\!\frac{S'_k S'_{k(t,t')}}{k k(t,t')S^2_k}\!\right)\Phi_{{\bf k}(t,t')}^2(t,t')\right],
\label{nonlinear}
\end{eqnarray}
where $S_k$ and $S'_k$ are the equilibrium static structure factor and its derivative, respectively. 
The influence of external flow enters both explicitly, via the Finger tensor ${\bf B}(t,t')$ (see
subsection \ref{sec:lodge}), and implicitly through the reverse-advected wavevector. 
As noted above, the reverse-advected wavevector, which provides an important source of nonlinearity
in (\ref{nonlinear}), enters as a result of judicious projection of the dynamics onto 
appropriately advected density fluctuations $\rho_{\bf k(t,t')}$. 
The normalized transient density correlator describes the decay under flow of
thermal density fluctuations and is defined by
\begin{eqnarray}\label{corr_def}
\Phi_{\kb}(t,t')=\frac{1}{NS_k}\langle\, \rho^*_{\kb} e_{-}^{\int_{t'}^{t}\!ds\,\Omega^{\dagger}(s)}
\rho^{}_{\bar{\kb}(t,t')} \rangle.
\end{eqnarray}
The occurance of the advected wavevector in (\ref{corr_def}) ensures that trivial decorrelation
effects are removed (i.e. that in the absence of Brownian motion and potential interactions 
$\Phi_{\kb}=1$ for all times). 

In order to close the constitutive equation (\ref{nonlinear}) we require an explicit expression
for the transient correlator (\ref{corr_def}). 
Time-dependent projection operator manipulations combined with the theory of Volterra integral
equations yield an exact equation of motion for the time evolution of the transient correlator 
containing a generalized friction kernel - a memory function formed from the autocorrelation 
of fluctuating stresses. 
Mode-coupling type approximations to this kernel yield the nonlinear integro-differential 
equation \cite{joeprl_07,joeprl_08,pnas}
\begin{eqnarray}
\dot\Phi_{\bf q}(t,t_0)
&+& \Gamma_{\bf q}(t,t_0)\bigg(
\Phi_{\qb}(t,t_0)
\label{equom}
\\
&+&
\int_{t_0}^t dt' m_{\qb}(t,t',t_0) \dot\Phi_{\qb}(t',t_0)
\bigg) =0
\notag
\end{eqnarray}
where the overdots denote partial differentiation with respect to the first time argument.
Here the `initial decay rate' obeys
$\Gamma_{\bf q}(t,t_0)=D_0\bar{q}^2(t,t_0)/S_{\bar{q}(t,t_0)}$ with $D_0$ the diffusion constant of 
an isolated particle. 
The formal manipulations presented in \cite{joeprl_07,joeprl_08} have revealed that imposing a 
time-dependent flow results in a memory kernel which depends upon three time arguments. 
The presence of a third time argument, which would have been difficult to guess on the basis 
of quiescent MCT intuitition, turns out to have important consequences for certain rapidly varying 
flows (e.g. step strain \cite{joeprl_07}) and is essential to obtain physically sensible results 
in such cases.  
The memory kernel $m_{\bf q}(t,t',t_0)$ entering (\ref{equom}) is given by the factorized 
expression \cite{joeprl_08,pnas}
\begin{eqnarray}
\label{approxmemory}
&& \hspace*{-1cm}
m_{\qb}(t,t'\!,t_0) \!\!= \!\!
\frac{\rho}{16\pi^3} \!\!\int \!\! d\kb
\frac{S_{\bar{q}(t,t_0)} S_{\bar{k}(t',t_0)} S_{\bar{p}(t',t_0)} }
{\bar{q}^2(t',t_0) \bar{q}^2(t,t_0)}\\
&\times&
V_{\qb\kb\pb}(t',t_0)\,V_{\qb\kb\pb}(t,t_0)\Phi_{\bar{\kb}(t',t_0)}(t,t')
\Phi_{\bar{\pb}(t',t_0)}(t,t'),
\notag
\end{eqnarray}
where $\pb=\qb-\kb$, and the vertex function obeys
\begin{eqnarray}
\!\!\!\!\!V_{\qb\kb\pb}(t,t_0) \!=\! \bar\qb(t,t_0)\cdot(
\bar\kb(t,t_0) c_{\bar{k}(t,t_0)} \!+
\bar\pb(t,t_0) c_{\bar{p}(t,t_0)})\notag\\
\label{vertex}
\end{eqnarray}
with Ornstein-Zernicke direct correlation function $c_k\!=\!1\!-\!1/S_k$ (see Eq.(\ref{oz})).
In the linear regime Eqs.(\ref{nonlinear}) and (\ref{equom}) reduce to 
the standard quiescent MCT forms (\ref{lrmodulus}) and (\ref{eom_quiescent}), respectively.

An important feature of Eqs.(\ref{nonlinear})--(\ref{vertex}) is that they offer a closed constitutive
equation requiring only the static structure factor and velocity gradient tensor $\kap(t)$ 
as input to calculating the stress tensor. 
The equilibrium $S_{q}$ is determined by the interaction potential and thermodynamic statepoint 
and, as in quiescent MCT, serves as proxy for the pair potential (an interpretation arising from 
field-theoretical approaches to MCT \cite{mike_poincare}). 
The role of $S_q$ within the ITT-MCT should be contrasted with that within the Miyazaki-Reichman theory 
\cite{miyazaki1,miyazaki2,miyazaki3}, discussed in subsection \ref{sec:inspired}, where it enters as an 
approximation to the flow-distorted structure factor $S(\kb)$.  

In subsection \ref{matob} we introduced the principle of material objectivity; an approximate 
symmetry requiring that a valid constitutive relation be rotationally invariant. 
While verification of rotational invariance is straightforward for the phenomenlogical Lodge equation 
introduced in subsection \ref{sec:lodge}, proof becomes more demanding for the microscopic constitutive 
theory given by Eqs.(\ref{nonlinear})--(\ref{vertex}). Nevertheless, substitution of 
Eqs.(\ref{rotationE}) and (\ref{finger_rotation}) into (\ref{nonlinear})--(\ref{vertex}) yields the 
result (\ref{requirement}), thus verifying that the ITT-MCT constitutive equation is indeed material
objective as desired \cite{joeprl_08}. 
Material objectivity is an important consistency check for constitutive theories based on overdamped
Smoluchowski dynamics, for which it represents an exact symmetry constraint.

Possibly the most exciting feature of the ITT-MCT constitutive equation (\ref{nonlinear})--(\ref{vertex}) is 
that it incorporates a mechanism for describing the slow structural relaxation leading to dynamical 
arrest. The predicted rheological response thus goes from that of a viscous fluid to that of an 
amorphous solid, characterised by an elastic constant, upon variation of the thermodynamic control 
parameters. 
The ability of the theory to unify the description of fluid and glassy states stems from the 
underlying mode-coupling-type approximations which are tailored to capture the cooperative particle
motion in dense colloidal dispersions, ultimately leading to particle caging and arrest. 
Mathematically, this scenario arises from a bifurcation in the solution of the nonlinear
integro-differential equation (\ref{equom}) at suffiently high volume fraction/attraction strength
associated with a diverging relaxation time of the transient density correlator. 
Glass formation within the MCT-ITT approach is a purely dynamical phenomenon, as the equilibrium 
$S_q$ used as input varies smoothly accross the transition. 
The fluid-solid transition contained within the ITT-MCT equations can be better appreciated by
considering the small strain limit of Eq.(\ref{nonlinear}) which yields the linear 
response result \cite{joeprl_08}
\begin{eqnarray}\label{linear}
\sig^{l}(t)\!=\!\!\int_{-\infty}^{t} \!\!\!\!\!\!dt'\!\int\!\!\!
\frac{d{\bf k}}{16\pi^3} \{(\kb\!\cdot\!\bar{\kap}(t')\!\cdot\!\kb)\kb\kb\}\!
\left(\!\frac{S'_k\Phi_k(t\!-\!t')}{k S_k}\!\right)^{\!2}
\notag\\
\end{eqnarray}
where $\Phi_k(t)$ is the correlator from quiescent MCT \cite{goetze_leshouches}. 
In the glass the correlator does not relax to zero for long times and a partial integration 
of (\ref{linear}) followed by taking the limit of small strain leads to the result
\begin{eqnarray}\label{hooke}
\sig(t)=2G(t\!\rightarrow\!\infty)\boldsymbol{\epsilon}(t),
\end{eqnarray}
where $\boldsymbol{\epsilon}(t)$ is the infinitessimal strain tensor and 
$G(t\!\rightarrow\!\infty)$ 
is an elastic modulus obtained from Eq.(\ref{lrmodulus}) (also known as Lam\'e's second coefficient, 
the first being zero here due to incompressibility). 
Eq.(\ref{hooke}) is essentially Hookes law, describing the small strain response of a glassy solid. 
Going beyond linear response, Eq.(\ref{nonlinear}) incorporates the fluidizing effect of flow 
and thus makes possible investigation of a large number of time-dependent rheological situations 
in which externally applied flow fields compete with glass formation and slow structural 
relaxation (see subsection \ref{apps}).

\subsection{Distorted structure factor}
The microscopic ITT-MCT constitutive equation discussed above enables comparisons to be made 
with traditional continuum rheological modelling (section \ref{cont}) for which the macroscopic stress
tensor is the fundamental quantity of interest. 
However, the formal ITT result (\ref{average2}) also enables calculation of the distorted structure 
factor, $S({\bf k},t)\!=\!1 + n\int d\rb\,(g(\rb,t)\!-\!1)\,e^{i\,\kb\cdot\rb}$, which makes possible 
a comparison with the microstructure obtained from approaches based on the pair Smoluchowski equation 
(section \ref{pairsmol}). 
In particular, setting $f=\Delta\rho^*_{\bf k}\rho^{}_{\bf k}\equiv\rho^*_{\bf k}\rho^{}_{\bf k} 
- \langle\,\rho^*_{\bf k}\rho^{}_{\bf k}\rangle$ in Eq.(\ref{average2}) yields the formal result
\begin{eqnarray}\label{formal_dist}
S({\bf k},t)
\!=\! \langle\, \rho^*_{\bf k}\rho^{}_{\bf k} \rangle \!+\! \int_{-\infty}^{t}\!\!dt' 
\langle \kap(t')\!:\!\hat{\boldsymbol{\sigma}}\,
e_-^{\int_{t'}^t ds\,\Omega^{\dagger}(s)}
\Delta\rho^*_{\bf k}\rho^{}_{\bf k} \,\rangle.
\notag\\
\end{eqnarray}
Mode-coupling projection operator steps analagous to those leading to (\ref{nonlinear}) 
yield the ITT-MCT expression for the distorted structure factor
\begin{eqnarray}\label{distorted_structure}
S({\bf k},t) = S_{k}\, -
\int_{-\infty}^t \!\!\!\!\!dt'\, \frac{\partial
S_{k(t,t')}}{\partial t'}\,\Phi^2_{\kb(t,t')}(t,t')
\end{eqnarray}
where the transient density correlator is given by solution of Eqs.(\ref{equom})--(\ref{vertex}) for 
given $S_k$ and $\kap(t)$ and where an isotropic term has been suppressed. 
Eq.(\ref{distorted_structure}) has the appealing interpretation that flow-induced microstructural changes 
are built-up by integration of the affinely shifted equilibrium structure factor over the entire flow
history, weighted by the transient density correlator describing the fading memory of the system. 
The temporally non-local character of Eq.(\ref{distorted_structure}) is in striking contrast 
to the local approximations based on the pair Smoluchowski equation. 
The former consists of a history integral over a memory function which is itself determined 
by solution of a nonlocal integro-differential equation (\ref{equom}) whereas the latter are purely
Markovian approximations.

Within the pair Smoluchowski approach, in the absence of HI the stress tensor is exactly 
related to the distorted pair correlation function by Eq.(\ref{stress}). 
It is therefore of interest to inquire whether a similar connection holds within the 
approximate ITT-MCT approach. 
It is a relatively simple exercise to show that, subject to a certain contraint to be 
discussed below, Eqs.(\ref{nonlinear}) and (\ref{distorted_structure}) are connected 
by the relation \cite{joeprl_08}
\begin{eqnarray}
\sig(t)=-\Pi\,{\bf 1} -nk_BT\int \!\!\frac{d\kb}{16\pi^3}
\,\frac{\kb\kb}{k}
\,c'_k
\,\delta S_{\kb}(t),
\label{fl_equation}
\end{eqnarray}
where $\delta S_{\kb}(t;\kap) = S_{\kb}(t;\kap) - S_k$ and $\Pi$ is the equilibrium 
osmotic pressure.
For shear flow $\sigma_{xy}(t)$ from (\ref{fl_equation}) coincides with a result of 
Fredrickson and Larson \cite{fredrickson} for sheared copolymers, reflecting the Gaussian 
statistics underlying both the field-theory approach of \cite{fredrickson} and the ITT-MCT 
factorization approximations.
Equation (\ref{fl_equation}) thus connects stresses to microstructural distortions, which 
build up over time via the affine stretching of density fluctuations competing with structural 
rearrangements encoded in $\Phi_{\bf k}(t,t')$. 

For the off-diagonal stress tensor elements Eq.(\ref{fl_equation}) connects 
Eqs.(\ref{nonlinear}) and (\ref{distorted_structure}) directly. 
For the diagonal elements contibuting to the osmotic pressure Eq.(\ref{fl_equation}) is 
also valid, providing that the following approximate `sum-rule' is obeyed
\begin{eqnarray}
\frac{\partial \Pi}{\partial \phi}=\frac{1}{6\pi}\int \!d\kb 
\left( \frac{\partial \ln S_k}{\partial \ln k} \right)
\left( \frac{\partial \ln S_k}{\partial \ln\phi} \right).
\end{eqnarray}
Although it is not at all obvious that the above relation should hold, numerical calculations for hard
spheres (using e.g. the Percus-Yevick approximation for $S_k$ \cite{hansen}) show that it represents 
a rather good approximation. 
It should be emphasized that the application of projection operator approximations to 
(\ref{gk_stress}) and (\ref{formal_dist}) yield approximate expressions for the stress and structure
factor, respectively, which are not neccessarily self-consistent, in the sense that integration 
of the approximate $S(\kb,t)$ leads to the approximate $\sig(t)$. 
The fact that the ITT-MCT $S(\kb,t)$ is {\em almost} consistent with the direct ITT-MCT approximation to 
the stress is a testament to the underlying robustness of the method.

Although the caging mechanism is expected to be most important for statepoints close to 
the glass transition, ITT-MCT calculations of $S(\kb)$ for hard-spheres under weak shear flow 
\cite{henrich} suggest that the (truncated) divergence of the structural relaxation timescale at the point of
dynamical arrest (be it at the idealized glass transition point, as predicted by MCT, or some higher 
volume fraction \cite{russel_zero}) remains relevant for volume fractions well removed from the 
singularity and has a range of influence which extends back to dense equilibrium fluid states. 
More recently the MCT-ITT $S(\kb)$ has been evaluated numerically for two-dimensional hard-discs 
under shear at finite values of $Pe$ \cite{weysser}, without invoking any additional isotropic 
approximations (see subsection \ref{apps}). 
These calculations show only qualitative agreement with Brownian dynamics simulation results 
and overestimate the magnitude of the distortion by around an order of magnitude, a failing which 
is attributed to the fact that ITT-MCT apparently underestimates the speeding up of structural 
relaxations induced by the shear flow. 
This is to be contrasted with pair Smoluchowski-based approaches (e.g. \cite{szamel,lionberger}) 
which underestimate the magnitude of the low-shear distortion for dense fluid states 
($\phi\sim 0.5$) by around an order of magnitude. 
It therefore appears that neither the pair Smoluchowski nor the ITT-MCT approach can account
adequately for the shear induced distortion of the microstructure.

\subsection{Applications}\label{apps}
Explicit numerical solution of the ITT-MCT constitutive equation (Eqs.(\ref{equom})--(\ref{vertex})) 
has been performed for a one-component system of hard spheres under steady flows of various 
geometry \cite{joeprl_08}. 
However, the computational resources required to solve the anisotropic 
Eqs.(\ref{equom})--(\ref{vertex}) in three-dimensions over many decades of time are daunting and the efficient numerical
algorithms required to reduce the computational load are still under development. 
Nevertheless, it is hoped that much of the essential physics may be captured by solving a simplified 
set of equations in which the advected wavevectors are approximated by an isotropic average 
$\kb(t,t')\rightarrow k_{\rm is}(t,t')$, where
\begin{eqnarray}\label{isotropic}
k^2_{\rm is}(t,t')=\frac{1}{4\pi}\int \!d\Omega\, k^2(t,t'). 
\end{eqnarray}
This technical approximation has been successfully applied to the case of simple shear 
\cite{faraday,miyazaki3} and enables the angular integrals entering 
Eqs.(\ref{equom})--(\ref{vertex}) to be performed analytically. 
For two-dimensional systems algorithms have been developed which enable accurate numerical 
solution of the ITT-MCT equations without additional isotropic approximation \cite{weysser}. 

\begin{figure}
\includegraphics[width=8.2cm,angle=0]{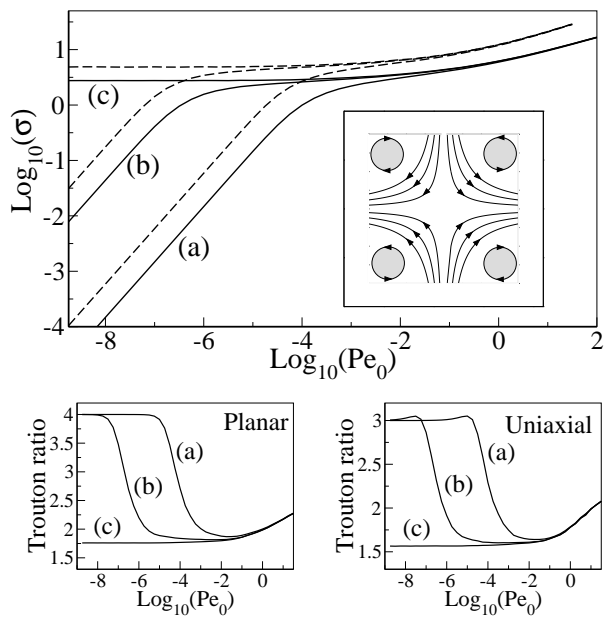}
\caption{
Flow curves from the microscopic ITT-MCT approach of \cite{joeprl_08} 
for hard-spheres at three different volume fractions close to the glass transition 
$\phi_c$. 
Full lines show the steady shear stress, $\sigma\!=\!\sigma_{\rm xy}$ in units of 
$k_BT/(2R)^3$ under shear flow as a function of $Pe$.  
Broken lines show the stress difference $\sigma\!=\!\sigma_{xx}\!-\!\sigma_{yy}$ 
(related to the extensional viscosity) under planar extensional flow.
Each curve is labelled according to the distance in volume fraction from the glass
transition, $\Delta\phi = \phi - \phi_{c}$. 
(a) and (b) are fluid states with $\Delta\phi=-10^{-4}$ and $\!-10^{-3}$, respectively. 
State (c) is in the glass, $\Delta\phi=\!10^{-4}$, and exhibits a dynamic yield stress for 
both flow geometries.
The inset shows a possible realization of planar extensional flow. 
The lower two panels show the Trouton ratio 
$\sigma_{\rm xx}\!-\!\sigma_{\rm yy}/\sigma_{\rm xy}$ 
as a function of $Pe$ for both uniaxial and planar extensional flow.
} 
\label{dynamic}
\end{figure}

In Fig.\ref{dynamic} flow curves for hard-spheres resulting from solution of 
Eqs.(\ref{equom})--(\ref{vertex}) subject to the isotropic approximation (\ref{isotropic}) are shown 
for both steady shear and steady planar extensional flow. 
For these two choices of flow the defining velocity gradient tensors are given by 
\begin{eqnarray}
     \kap_{\rm s}=\left(
     \begin{array}{ccc}
     0 & \dot\gamma & 0 \\
     0 & 0 & 0 \\
     0 & 0 & 0 
     \end{array}
     \right)
     \;\;\;\;\;
     \kap_{\rm e}=\left(
     \begin{array}{ccc}
     \dot\gamma  & 0 & 0 \\
     0 & -\dot\gamma  & 0 \\
     0 & 0 & 0 
     \end{array}
     \right).
\end{eqnarray}
Specifically, Fig.\ref{dynamic} shows $\sigma_{\rm xy}$ under shear flow and 
$\Delta\sigma\equiv\sigma_{\rm xx}\!-\!\sigma_{\rm yy}$ under planar extension 
for hard-spheres as a function of $Pe$, for various volume fractions around the glass transition. 
The equilibrium structure factors used as input for 
these calculations were provided by the monodisperse Percus-Yevick theory \cite{hansen}.  
For extensional flows it is natural to plot the stress difference $\Delta\sigma $ as a function 
of $Pe$, as this is simply related to the extensional viscosity 
$\eta_{\rm e}=\Delta\sigma/\dot\gamma$. 
Flow curves below the glass transition show a regime of linear response, characterized by a constant 
viscosity, followed by shear thinning as $Pe$ is increased. 
On approaching $\phi_{\rm g}$ from below the linear regime moves to lower values of $Pe$ and disappears 
entirely on crossing the glass transition. 
The resulting plateau in the flow curves identifies a dynamical yield stress 
(see subsection \ref{yieldstress}) for both of the considered flow geometries. 

In the lower panels of Fig.\ref{dynamic} the Trouton ratio 
$(\sigma_{\rm xx}-\sigma_{\rm yy})/\sigma_{\rm xy}$ \cite{hassager} is shown as a function of 
$Pe$ for both planar and uniaxial extensional flow \cite{joeprl_08}. 
For fluid states in the linear regime {\em Trouton's rules} assert that the ratio of extensional to 
shear viscosity $\eta_{\rm e}/\eta_{\rm s}$ takes the values $4$ and $3$ for planar and uniaxial
extension, respectively. 
These characteristic ratios arise from purely geometrical considerations and emerge naturally from 
the ITT-MCT approach as the $Pe\rightarrow 0$ limiting values of the Trouton ratio. 
For glassy states (curves labelled (c) in Fig.\ref{dynamic}) the structural relaxation time 
diverges and the linear response regime vanishes. 
As a consequence, the classical Trouton ratios are not recovered for glassy states in the limit 
$Pe\rightarrow 0$ and non-trivial values dictated by the dynamical yield stress may be identified. 
The results for extensional flow presented in \cite{joeprl_08} alongside those for simple shear 
\cite{fuchs_cates} thus provide the first steps towards the prediction from first-principles theory 
of a dynamic yield stress surface for glasses \cite{hill}. 
Calculation of the yield surface from a simplified schematic version of 
Eqs.(\ref{equom})--(\ref{vertex}) will be discussed in subsection \ref{sec:yield} below. 

Going beyond steady flow, in \cite{zausch} experiments on PMMA colloidal dispersions, molecular 
dynamics simulation and the ITT-MCT approach of \cite{joeprl_07} were combined to study the 
evolution of stresses during start-up shear flow for high volume fraction fluids close to 
glassy arrest. 
The sudden onset of a steady shear flow leads to the build-up of stresses in the systems as a 
function of the accumulated strain $\gamma\equiv\dot\gamma t$. 
For small values of $\gamma$ the response is elastic (as described by Eq.(\ref{hooke})), whereas 
for large strains system enters steady state viscous flow with a stress independent of $\gamma$. 
In between these two limits, typically at strains around $10$\%, the stress exhibits a maximum 
as the local microstructure is broken up by the external flow \cite{varnik}. 
Although a stress `overshoot' in response to start-up shear flow is a rather generic 
feature of the rheology of complex fluids, its microscopic origins remain poorly understood. 
A central novel aspect of \cite{zausch} was thus to connect the stress overshoot to 
anomolous behaviour in the mean-squared-displacement (`superdiffusion') identified in 
both simulation and confocal microscopy experiment. 
From the exact Eq.(\ref{stress_on}) it is clear that the only way in which the shear stress 
can exhibit a maximum is if the modulus under steady shear, $G_{\rm ss}(t)$, becomes negative 
at long times. 
Fig.\ref{fig:nonlinear_mods} shows that $G_{\rm ss}(t)$ from the ITT-MCT approach 
(employing approximation (\ref{isotropic})) indeed predicts negative values at long times. 
The inset to Fig.\ref{fig:nonlinear_mods} shows the same data as a function of strain and demonstrates
that the negative region of $G_{\rm ss}(t)$ occurs at around $10$\% strain, consistent with the 
position of the stress overshoot.



\subsection{Schematic model}\label{sec:schematic}
The microscopic ITT-MCT constitutive equation outlined in subsection \ref{sec:constitutive} 
provides a route to first-principles prediction of the rheological behaviour of arrested colloidal 
states. However, the anisotropic, wavevector dependent expressions are rather intractable for 
three dimensional flows, hindering both their practical use and interpretation. 
In order to facilitate numerical calculations for flows of interest a simplified 
`schematic' version of the tensorial microscopic theory has very recently been proposed 
\cite{pnas}. 
Such schematic models have proved invaluable in the analysis of mode-coupling 
theories and provide a simpler set of equations which retain 
the essential mathematical structure of the microscopic theory \cite{faraday,goetze_leshouches}. 

Applying the isotropic approximation (\ref{isotropic}) to the microscopic ITT-MCT expression for 
the stress (\ref{nonlinear}) enables the angular integrals to be performed explicitly, 
leading to the simplified form
\begin{eqnarray}\label{schematic_stress}
\sig(t)= \int_{-\infty}^{t} \!\!\!\!dt'
\left[
-\frac{\partial}{\partial t'}
\Finger(t,t')
\right]
G(t,t'),
\end{eqnarray}
where ${\bf B}$ is the Finger tensor and an explicit expression for $G(t,t')$ may be found in 
\cite{pnas}. 
By disregarding all wavevector dependence the modulus can be expressed in terms 
of a single-mode transient density correlator
\begin{eqnarray}
G(t,t')=v_{\sigma}\Phi^2(t,t')
\end{eqnarray}
where $v_{\sigma}=G(t,t)$ is a parameter measuring the strength 
of stress fluctuations (typically taking values of the order $100 k_BT/R^3$ for hard-sphere-like 
colloids). 
A schematic equation-of-motion for $\Phi(t,t')$ may be obtained by neglecting the wavevector 
dependence of the microscopic expression (\ref{equom}), leading to 
\begin{eqnarray}\label{schematic_eom}
\dot\Phi(t,t') 
+ \Gamma\bigg(
\Phi(t,t') + \int_{t'}^{t}ds\, m(t,s,t')
\dot\Phi(s,t')
\bigg) =0,
\notag\\
\end{eqnarray}
where the single decay rate $\Gamma$ simply sets the time scale and may thus be set to unity. 
Experience with the construction of schematic MCT models both in the quiescent 
\cite{goetze_leshouches} and steady shear cases \cite{faraday} combined with analysis of the 
way in which strain enters the microscopic memory (\ref{approxmemory}) lead to the 
following schematic ansatz for the memory function
\begin{eqnarray}\label{schematic_memory}
m(t,t',t_0)=h(t,t_0)\,h(t,t')\,[\,\nu_1\Phi(t,t')+\nu_2\Phi^2(t,t')\,].
\end{eqnarray}  
The parameters $\nu_1$ and $\nu_2$ represent in an unspecified way the role of $S_q$ in the 
microscopic theory and, following standard MCT practice, are given by 
$v_2=2$ and $v_1=2(\sqrt{2} -1) + \epsilon/(\sqrt{2}-1)$. The separation parameter 
$\epsilon$ encodes the distance from the glass transition, with negative values corresponding 
to fluid states and positive values corresponding to glass states. 
Finally, the $h-$function is given by
\begin{eqnarray}\label{hfunc}
h(t,t_0)=\frac{\gamma_c^2}{\gamma_c^2 +  \nu(I_1(t,t_0)-3) + (1-\nu)(I_2(t,t_0)-3)},
\notag\\
\end{eqnarray}
where $\gamma_c$ sets the strain scale (typically $\gamma_c\approx 10$\%) and $\nu$ is a mixing
parameter ($0\!<\!\nu\!<\!1$). 
The invariants of the Finger tensor, $I_1={\rm Tr}\,{\bf B}$ and $I_2={\rm Tr}\,{\bf B^{-1}}$ 
incorporate the fluidizing influence of flow into the memory function. 
Requiring that flow enter via the Finger tensor alone guarantees that the resulting schematic theory 
is material objective (see subsection \ref{matob}), consistent with the fully microscopic theory.  

\begin{figure}[!t]
\includegraphics[width=7.3cm,angle=0]{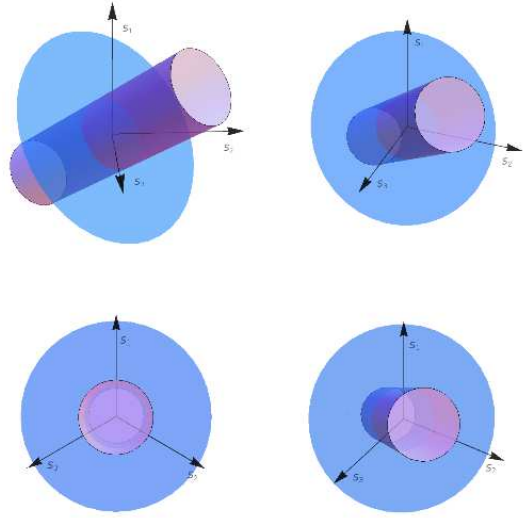}
\caption{
The state of stress of a material under applied force can be represented by 
a point in the three dimensional space of principal stresses. 
The cylinder shown here is the surface defined by the von Mises criterion 
(\ref{vonmises_yield}). Stress states lying within the cylinder are deformed 
by the applied stress, but do not yield, whereas states outside the cylinder 
exhibit plastic flow. 
Due to invariance along the `hydrostatic' axis $s_1=s_2=s_3$ (arising from incompressibility) 
the yield surface is usually viewed in the `deviatoric' plane perpendicular to this 
(the lower left representation in this figure).
}
\label{vm}
\end{figure}

The schematic expression for the stress (\ref{schematic_stress}) is closely related to the Lodge 
equation of continuum mechanics (see subsection \ref{sec:lodge}). 
Integrating Eq.(\ref{schematic_stress}) by parts 
yields 
\begin{eqnarray}\label{parts}
\sig(t) = \int_{-\infty}^{t}\!\!dt'\; \Finger(t,t') 
\left( \frac{\partial}{\partial t'}\Phi^2(t,t') \right).
\end{eqnarray} 
Comparison of this expression with Eq.(\ref{lodge1}) shows that the present schematic theory goes 
considerably beyond the standard Lodge equation by incorporating memory which is both nonexponential 
and a function of two time arguments, reflecting the loss of time-translational invariance under 
time-dependent flow. 
Replacing the correlator with a simple exponential trivially recovers the Lodge equation (\ref{lodge1}). 

The tensorial schematic theory given by Eqs.(\ref{schematic_stress})--(\ref{hfunc}) 
has been applied to predict flow curves under shear and extensional flow, 
normal stresses under shear and the transient stress response to step strain. 
In all cases tested so far the predictions of the schematic model are in good qualitative 
agreement with those available from the microscopic theory using the isotropic approximation 
(\ref{isotropic}) in three-dimensions \cite{fuchs_cates,faraday,joeprl_07,joeprl_08} 
and exact numerical solution in two-dimensions \cite{weysser}. 
The schematic theory predicts a positive value for the first normal stress difference 
$N_1\!=\!\sigma_{\rm xx}\!-\!\sigma_{\rm yy}$ under shear flow in accord with microscopic ITT-MCT
calculations. 
On the other hand, the second normal stress difference 
$N_2\!=\!\sigma_{\rm yy}\!-\!\sigma_{\rm zz}$ is from Eq.(\ref{schematic_stress}) identically 
zero, in disagreement with both analytical low $Pe$ analysis of the pair Smoluchowski equation 
\cite{brady_vicic} and colloidal experiments \cite{frank} (finite negative values are found). 
The disappearance of $N_2$ from the schematic theory can be traced back to the isotropic
approximation leading to Eq.(\ref{schematic_stress}) which effectively kills off this feature 
of the fully microscopic theory. 

\begin{figure}
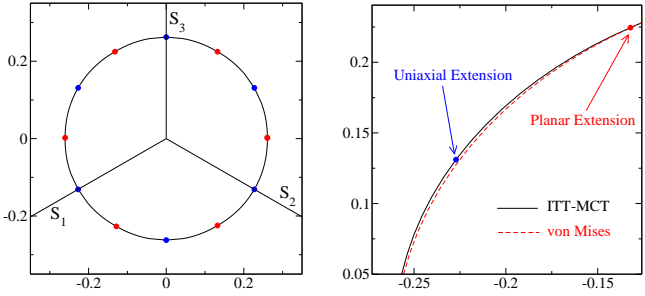

\hspace*{-0.5cm}
\includegraphics[width=4.cm]{vonmises.eps}
\hspace*{0.3cm}
\includegraphics[width=4.cm]{vonmises_zoom.eps}
\caption{
Left Panel: The dynamical yield surface from ITT-MCT \cite{pnas} for a glass with 
$\epsilon\approx 3(\phi-\phi_{\rm gt})=10^{-3}$, where $\phi_{\rm gt}$ is the volume 
fraction at the hard-sphere glass transition, 
in the space of  principal stresses $(s_1,s_2,s_3)$ as  
viewed along the hydrostatic axis $s_1=s_2=s_3$ (stress in units of $k_BT/(2R)^3$). 
The red points correspond to planar  
extensional flow and the blue points to uniaxial extensional flow. 
Right Panel: A closer view reveals that the  
surface is not a perfectly circular cylinder (as predicted by the von Mises criterion for static
yielding) and that maximal deviation from  
circularity occurs at points of pure uniaxial extension. 
These deviations are connected with the existence of finite normal stress differences. 
\label{vonmises}
}
\end{figure}  

\subsection{Yield stress surface}\label{sec:yield}
A striking feature of the theory developed in \cite{pnas} is that it 
permits direct calculation of a dynamic yield stress surface for glasses.
Related {\em static} yield surfaces have been empirically postulated and employed for 
over a century in the engineering community to study the yielding of amorphous solids 
(see also subsection \ref{yieldstress}). 
The two classical criteria for determining the onset of plastic yield  are due to 
Tresca \cite{tresca} and von Mises \cite{vonmises}. 
The Tresca criterion asserts that a material will yield when the maximum shear 
stress due to the deformation exceeds a critical value. 
Recalling that an external force imposed on a material can be represented as a stress 
tensor which can be diagonalized to obtain values for the three principal stresses 
$s_1, s_2$ and $s_3$, the Tresca criterion can be compactly stated in the form \cite{hill}
\begin{eqnarray}
{\rm Max}\left(\,|\,s_1-s_2|,|s_2-s_3|,|\,s_1-s_3|\,\right)= \sqrt{3}\,\sigma_{ss}^{y}\,,
\label{tresca_yield}
\end{eqnarray} 
where $\sigma_{ss}^{y}$ is the shear stress at yield under a simple shear deformation. 
According to Eq.(\ref{tresca_yield}) knowledge of $\sigma_{ss}^{y}$ is thus sufficient to 
determine the mechanical stability of a material under an arbitrary applied force.  
Alternatively, the von Mises critierion requires that the distortion strain energy exceeds a 
critical value at yield \cite{hill} 
\begin{eqnarray}		
\frac{1}{6}\left( (s_1-s_2)^2 + 
(s_2-s_3)^2 + (s_1-s_3)^2\right)= (\sigma_{ss}^{y})^2\,.
\label{vonmises_yield}
\end{eqnarray}
Both the Tresca and von Mises criteria have proven to be in reasonable qualitative agreement 
with yield experiments on crystalline metals \cite{schuh}. 
Two main assumptions underly Eqs.(\ref{tresca_yield}) and (\ref{vonmises_yield}):
(i) the microscopic rearrangements leading to plastic deformation do not lead to 
significant dilation of the material, (ii) that residual stresses arising from the 
deformation history of the sample do not influence the yielding (i.e. there is no 
Bauschinger effect).

It is useful to interpret equation (\ref{vonmises_yield}) geometrically in the space of 
principal stresses, where it describes a surface separating elastically deformed states 
from states of plastic flow. 
Equation (\ref{vonmises_yield}) defines a circular cylinder with axis along the line 
$s_1=s_2=s_3$ and radius $\sqrt{2}\,\sigma_{ss}^{y}$. 
The symmetry about this `hydrostatic' axis is a geometrical reflection 
of the fact that the yield condition (\ref{vonmises_yield}) is independent of hydrostatic
pressure.
The plane which passes through the origin and which lies perpendicular to the cylinder axis 
is the so-called deviatoric plane. 
All yield stress surfaces which are independent of hydrostatic pressure may be projected 
without loss of information onto the deviatoric plane. In the case of the von Mises and 
Tresca criteria this generates a circle (see Fig.\ref{vm}) and a hexagon, respectively.  

\begin{figure}[!b]
\hspace*{-0.8cm}\includegraphics[width=6.3cm]{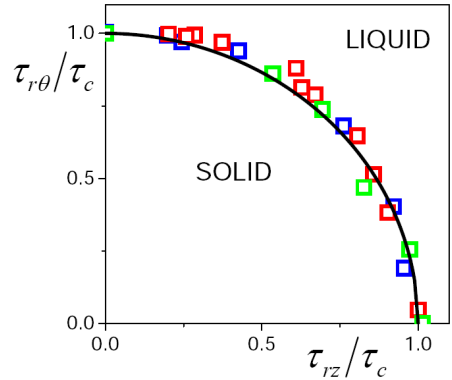}
\caption{
The yield stress surface in the $\tau_{{\rm r}\theta}$ (rotational shear stress) 
$\tau_{\rm rz}$ (squeeze shear stress) plane, where the stresses have been scaled by the yield 
stress $\tau_c$ in simple shear. 
The blue and red squares show results for two different emulsions with differing values 
of $\tau_c$ ($28Pa$ and $52Pa$). 
The green squares are data taken using a carbopol gel ($\tau_c=70Pa$). 
In this representation the von Mises criterion (\ref{vonmises_yield}) becomes a circle 
and is indicated by the solid line.   
Reprinted by permission from MacMillan Publishers Ltd: 
Nature Materials {\bf 9} 115, copyright (2010). 
}
\label{ovarlez_fig}
\end{figure}

The results presented in \cite{pnas} reveal intriguing connections between 
the static yielding discussed above and dynamic yield, as determined from the 
$Pe\rightarrow 0$ values of the flow curves. 
Within the ITT-MCT approach, for any given steady flow field 
(e.g. shear, uniaxial extension) there exists for glassy states a finite stress 
tensor in the limit of vanishing flow rate. 
This stress tensor at yield may be diagonalized to obtain three eigenvalues 
and plotted as a point in the cartesian space of principal stresses. 
By considering all possible nondegenerate flows (made possible by a suitable parameterization 
of the velocity gradient tensor \cite{pnas}) a closed locus of points may be 
constructed in the space: The dynamic yield stress surface. 
In \cite{pnas} this procedure was realized using the schematic model 
Eqs.(\ref{schematic_stress})--(\ref{hfunc}).

In Fig.\ref{vonmises} we show the deviatoric projection of the dynamical yield stress 
surface of a colloidal glass, as predicted by the schematic model. 
To a first approximation the numerically calculated dynamical yield surface from 
the theory agrees well with the empirical von Mises criterion for static yield 
(\ref{vonmises_yield}). 
However, closer inspection (see Fig.\ref{vonmises}, right panel) reveals that deviations 
at around the percent level occur, with the maximal deviation located at points of pure 
uniaxial extension. 
A careful analysis of the schematic equations reveals that this fine structure of 
the surface can be attributed to the existence of a finite first normal stress difference 
\cite{pnas}. 
Expanding the schematic model result to first order in $N^y_1$, the first normal stress difference
at yield, provides an explicit expression 
for the schematic model yield surface \cite{pnas}
\begin{eqnarray}		
\frac{1}{6}\big( (s_1\!-\!s_2)^2 &+& 
(s_2\!-\!s_3)^2 + (s_1\!-\!s_3)^2 \big) = (\sigma_{\rm ss}^y)^2 
\notag\\
&+&\frac{1}{12}(N_1^y)^2
+\frac{3(1-A^2)}{(3+A^2)^{3/2}}N_1^y\sigma_{\rm ss}^y,
\label{vonmises_yield2}
\end{eqnarray} 
where $0\!<\!A\!<\!1$ parameterizes the geometry of the imposed flow. Eq.(\ref{vonmises_yield2}) 
thus describes a noncircular cylinder with a radius which varies according to the value of the
parameter $A$, each value of which corresponds to a given azimuthal angle about the hydrostatic 
axis.  
Higher order terms in the expansion (\ref{vonmises_yield2}) exist (and can be 
explicitly calculated) but remain numerically negligable due to the smallness of 
$N_1/\sigma_{\rm ss}^y$.

Very recent experiments on yielding soft materials have attempted to determine
the shape of the yield surface \cite{ovarlez}. The novel rheometer employed in \cite{ovarlez} applies a combined squeeze and rotational shear flow to a material sample loaded between two parallel plates. The claim is that by independently varying the rotation and squeeze rate it may become possible to explore the entire family of flows in a way analogous to the mathematical parameterization of the velocity gradient tensor employed in \cite{pnas} to calculate the schematic model yield surface. However, it remains uncertain whether the superposition of two shear flows (radial Poiseuille and tangential homogeneous shear) is really sufficient to map the yield surface.
Fig.\ref{ovarlez_fig} shows experimental data taken using three different yielding materials: a Carbopol gel and two different emulsions \cite{ovarlez}.

The yield data are shown in the rotational shear stress ($\tau_{{\rm r}\theta}$), squeeze shear stress ($\tau_{\rm rz}$) plane.
While the data shown in Fig.\ref{ovarlez_fig} are not inconsistent with the von Mises criteria, it may well be that the chosen superposition flow actually constrains the surface to be spherical, regardless of the true form of the yield surface. It would certainly be remarkable if the soft materials considered in \cite{ovarlez} obey yield criteria developed for crystalline solids, despite the very different underlying microscopic plasticity mechanisms \cite{schall}.

Whether the microscopic theory (Eqs.(\ref{equom})--(\ref{vertex})) 
indeed predicts a similar dynamic yield surface, as expected, 
and its relationship to static yielding in glassy materials remain 
important open problems. 
Nevertheless, the results from the schematic model are promising and represent a 
considerable step towards a microscopic derivation of material specific yield surfaces 
from first principles.

\section{Outlook}\label{section:outlook}
In this review we have attempted to provide an overview of the rheological phenomenology 
presented by colloidal dispersions and to outline some of the leading theoretical approaches 
aiming to rationalize this. 
It is clear that much remains to be done and that exisiting theories have met with only partial 
success in solving the complex many-body problem of driven, strongly interacting colloidal 
systems.  
We hope that both the presentation and choice of topics contained within the present work 
serve to emphasize the common ground between different theories (continuum mechanics, pair 
Smoluchowski treatments and Green-Kubo based approaches), as well as to highlight where progress 
still needs to be made. 

One of the clear deficiencies of approximate closures of the pair Smoluchowski equation is 
their apparent inability to describe, even qualitatively, the slow structural relaxation time 
present in colloidal dispersions at finite volume fraction. 
This failing is inherent in the Markovian character of the approximate closures, which follows 
as a natural consequence of applying equilibrium statisitical mechanical relations to connect 
pair and triplet correlation functions. 
A clear challenge to future theories which attempt to improve this situation is thus to tackle 
directly the intinsic difference between nonequlibrium and equlibrium by confronting 
the irreducible term (see Eq.\ref{fict_trip}) containing the missing physics. 
Moreover, the existing numerical data in the literature is restricted, largely for technical 
reasons, to the low $Pe$ limit. 
While this enables interesting analysis of the zero-shear viscosity and leading order 
microstructural distortion, there is an absence of data for the nonlinear rheology as a function 
of $Pe$. 

Despite the success of the ITT-MCT approach in describing the nonlinear response of states close 
to dynamic arrest, there remain gaps in the theoretical formulation 
which should be filled and many fundamental questions are yet to be addressed. 
A notable omission is that the present formulation of the theory does not enable incorporation of hydrodynamic interactions. 
The influence of hydrodynamics on the rheology of densely packed glassy states 
is largely unexplored and their incorporation into the theory, even at a crude approximate 
level, would therefore be of considerable interest. 
Moreover, due to the recent nature of the ITT-MCT theory and the complexity of numerically solving
the equations, many flows of rheological interest remain to be explored in detail. 
Of particular importance is to assess the predictions of the theory for nonsteady flows where 
interesting interaction effects between yielding and the time-dependent strain field may be 
envisaged (e.g. large amplitude oscillatory shear, where higher harmonics will contribute to 
the stress response).

\section*{Acknowledgments}
I would particularly like to thank M. Fuchs, M.E. Cates, Th. Voigtmann and M.~Kr\"u{}ger for many 
stimulating discussions. 
Support was provided by the SFB TR6 and the Swiss National 
Science Foundation.


\begin{thebibliography}{}
\bibitem{capillary}
For example, in Refs.\cite{vanmegen2} and \cite{vanmegen3} dynamic light scattering 
was employed to measure the coherent transient density correlator of 
spherical hard-sphere-like colloids. The data confirmed the factorization properties 
of the $\alpha$ and $\beta$ relaxation, as predicted by the mode-coupling theory of 
the glass transition \cite{sjoegren,goetze_review}. 
In \cite{aarts} confocal microscopy experiments on a mixture of PMMA colloids 
and non-adsorbing polymer were used to confirm the capillary wave theory of 
the fluctuating interface between demixed fluid phases \cite{mandelstam}.
\bibitem{banding_note}
In \cite{expts2} a confocal microscopy study of polymethylmethacrylate (PMMA) 
colloids suspended in a mixture of decalin and tetralin revealed  
inhomogeneous plug flow for glassy states. In contrast, the rheological experiments 
performed in \cite{crassous} and \cite{winter} using thermosensitive PNIPAM core-shell 
particles are fully consistent with homogeneous flow, both above and below the glass 
transition.
\bibitem{poly_note}
In order to observe glassy behaviour the timescale of crystal nucleation $\tau_{\rm fr}$ 
has to be extended beyond the accessible experimental or simulation measurement time. 
Fortunately, $\tau_{\rm fr}$ is a sensitive function of the degree of polydispersity and 
for hard-sphere-like systems a polydispersity $\sim 5$\% is usually sufficient to effectively 
supress crystallization.
\bibitem{div_note}
In general, the initial value of the shear modulus is determined by near-field 
hydrodynamic lubrication forces. The commonly studied case of Brownian hard-spheres 
in the absence of hydrodynamic interactions is a pathological special case for which 
$G(t\rightarrow 0)\sim t^{-\frac{1}{2}}$ as a consequence of the discontinuous potential 
interaction. Nevertheless, the Brownian hard-sphere 
$G(t)$ rapidly becomes physical for later times.     
We refer the reader to \cite{lio_russ_94} for more details on this issue.
\bibitem{creepnote}
We note that there exist numerous alternative definitions of the `yield stress' in the 
literature. 
Two common choices are: (i) the maximum value of the stress overshoot which occurs 
in response to the onset of steady shear flow, 
(ii) the point for which $G'=G''$ as a function of amplitude for fixed frequency in 
oscillatory experiments (strain sweeps).
\bibitem{creepnote1}
This expecation is supported by the simulations performed in \cite{varnik} which 
found that the maximum of the stress-strain curve during start-up shear flow, 
often identified as the static yield-stress, exhibits a logarithmic dependence on 
the age of the system.
\end{thebibliography}
\end{document}